\documentclass[11pt,dvipsnames]{article}
\usepackage{caption}
\usepackage{scalerel}
\usepackage{amsthm,latexsym,amsxtra,graphicx,appendix,epstopdf,feynmf,hyperref,setspace,fix-cm}
\usepackage[normalem]{ulem}
\usepackage{makeidx}
\usepackage{fullpage}
\usepackage{amsmath}
\usepackage{amssymb}
\usepackage{setspace}
\usepackage{bbm}
\usepackage{dsfont}
\usepackage{epsfig}
\usepackage[font=footnotesize,labelfont=bf,justification=centerlast,width=.94\textwidth]{caption}
\usepackage{cite}
 \usepackage{multirow}
\usepackage{array,booktabs,tabularx}
\usepackage{subcaption}
\usepackage{mathtools}
\usepackage{color}
\usepackage[makeroom]{cancel}
\usepackage{tikz}
\usepackage{pgfplots}
\usepackage{bm}

\newcommand{\defeq}{\vcentcolon=}

\usepackage{tensor}
\usepackage{simplewick}
\usepackage{frcursive}

\usepackage{pgfornament,multicol}

\usepackage{caption}
\usepackage{hyperref}

\usepackage[skip=0.2cm, indent=0pt]{parskip}

\hypersetup{
    colorlinks,%
    citecolor=blue,%
    filecolor=blue,%
    linkcolor=blue,%
    urlcolor=blue
}
\usepackage{float}
\usepackage{slashed}
\usepackage{tikz}
\usetikzlibrary{decorations.pathmorphing}

\hypersetup{bookmarksdepth=subsection}
\usepackage{etoolbox}
\pretocmd{\appendix}{
  \addtocontents{toc}{\protect\setcounter{tocdepth}{1}}
  \hypersetup{bookmarksdepth=section}
}{}{}

\usepackage{tcolorbox}
\usepackage{tabularx}
\usepackage{array}
\usepackage{colortbl}
\tcbuselibrary{skins}

\tcbset{tab2/.style={enhanced,fonttitle=\bfseries,fontupper=\normalsize\sffamily,
colback=Goldenrod!20!white,colframe=Blue!50!black,colbacktitle=Goldenrod!50!white,
coltitle=black,center title}}

\def\r{\rightarrow}

\def\a{\alpha}
\def\b{\beta}

\newcommand{\bi}{\begin{itemize}}
\newcommand{\ei}{\end{itemize}}
\newcommand{\bea}{\begin{eqnarray}}
\newcommand{\eea}{\end{eqnarray}}

\def\={\, = \,}
\def\rh{r_{\text{h}}}

\def\frakr{\mathfrak{r}}

\def\bomega{\bm{\omega}}
\def\r{\mathfrak{r}}

\def\b{\tilde\beta}
\def\L{\mathfrak{L}}

\def\r{\mathfrak{r}}
\def\frakr{\mathfrak{r}}
\def\bomega{{\boldsymbol{\omega}}}

\def\XXint#1#2#3{{\setbox0=\hbox{$#1{#2#3}{\int}$}
     \vcenter{\hbox{$#2#3$}}\kern-.5\wd0}}

\def\={\, = \,}

\newcommand{\beq}{\begin{equation}}

\newcommand{\eeq}{\end{equation}}

\usepackage{tikz}
\usetikzlibrary{positioning}
\usetikzlibrary{intersections}
\usetikzlibrary{fadings} 
\usetikzlibrary{arrows.meta} 
\usetikzlibrary{arrows}

\tikzfading[name=fade out,
inner color=transparent!0,
outer color=transparent!100]

\definecolor{cherryblossompink}{rgb}{1.0, 0.72, 0.77}
\definecolor{lightblue}{rgb}{0.68, 0.85, 0.9}

\usetikzlibrary{decorations.pathmorphing}
\usetikzlibrary{decorations.pathreplacing,decorations.markings}

\usetikzlibrary{backgrounds,automata}

\newsavebox\CBox
\newcommand\hcancel[2][0.5pt]{%
  \ifmmode\sbox\CBox{$#2$}\else\sbox\CBox{#2}\fi%
  \makebox[0pt][l]{\usebox\CBox}%
  \rule[0.75\ht\CBox-#1/2]{\wd\CBox}{#1}}

\numberwithin{equation}{section}
\allowdisplaybreaks  

\begin{document}

\vspace*{2.5cm}
\begin{center}
{ \huge {The Stretched Horizon Limit} }\\

\vspace*{1cm}
\end{center}

\renewcommand{\thefootnote}{\fnsymbol{footnote}}

\begin{center}
{\small Dionysios Anninos,$^{1,2}$ Dami\'an A. Galante,$^{1}$ Silvia Georgescu,$^{1}$ Chawakorn Maneerat,$^{1}$ and Andrew Svesko$^{1}$}
\end{center}
\begin{center}
{
\footnotesize
\vspace{0.2cm}
$^1$ Department of Mathematics, King's College London, Strand, London WC2R 2LS, UK
\\
$^2$ Instituut voor Theoretische Fysica, KU Leuven, Celestijnenlaan 200D, B-3001 Leuven, Belgium} 
\end{center}
\begin{center}
{\textsf{\footnotesize{
dionysios.anninos@kcl.ac.uk, damian.galante@kcl.ac.uk, silvia.georgescu@kcl.ac.uk, \\ chawakorn.maneerat@kcl.ac.uk, andrew.svesko@kcl.ac.uk}} } 
\end{center}

\renewcommand{\thefootnote}{\arabic{footnote}}

\vspace*{0.5cm}

\vspace*{1.5cm}
\begin{abstract}
\noindent  We consider four-dimensional general relativity with a positive cosmological constant, $\Lambda$, in the presence of a boundary, $\Gamma$,  of finite spatial size. The boundary is located near a cosmological event horizon, and is subject to boundary conditions that fix the conformal class of the induced metric, and, $K$, the trace of the extrinsic curvature along $\Gamma$. The proximity of $\Gamma$ to the horizon is controlled by the dimensionless parameter ${K}{{\Lambda}^{-\frac{1}{2}}}$. We provide an exhaustive analysis of linearised gravitational perturbations for the setup. This is performed both for a $\Gamma$ encasing a portion of the static patch that ends just before the cosmological horizon (pole patch), as well as a $\Gamma$ containing only the region near the cosmological horizon (cosmic patch). In the pole patch, we uncover a layered hierarchy of modes: ordinary normal modes, a novel type of boundary gapless mode, and boundary soft modes of frequency $\omega \approx \pm  2\pi i T_{\text{dS}}$, with $T_{\text{dS}}$ the horizon temperature.  Minkowskian behaviour is recovered only for angular momenta $l \gtrsim {K}{{\Lambda}^{-\frac{1}{2}}}$ which can be made parametrically large, thus attenuating previously found growing modes. In the cosmic patch, we uncover sound and shear fluid-dynamical modes that we interpret in terms of a conformal fluid with shear viscosity over entropy density ratio $\tfrac{\eta}{s} = \tfrac{1}{4\pi}$ and vanishing bulk viscosity $\zeta=0$. The fluid dynamical sector is shown to admit a non-linear treatment. We describe a scaling regime in which the stretched horizon gravitational dynamics is dictated by a universal Rindler geometry, independent to the details of the infilling horizon. We briefly discuss quantitative features that distinguish cosmological and black hole horizons away from the Rindler regime.

\end{abstract}

\newpage

\setcounter{tocdepth}{2}
\tableofcontents

\section{Introduction} \label{sec:intro}

The idea of placing general relativity on a manifold endowed with a finite size boundary has a long and interesting history. This is as true in Lorentzian, as it is in Euclidean signature.  

It may be of some use for the reader to offer a brief and incomplete overview. Already in the sixties, Penrose \cite{Penrose:1964wq} introduced the notion of a trapped surface to describe the dynamical properties of black holes. Making sense of the variational problem for the Einstein-Hilbert action in general relativity on a manifold with a boundary led, in the seventies, to the necessity of boundary terms in the action \cite{York:1972sj,Gibbons:1976ue}. These were later found to play a key role for Euclidean black hole thermodynamics, and the construction of the quasilocal stress-energy tensor of Brown and York \cite{Brown:1992br}. In the late seventies and eighties, considerations of dynamical phenomena near astrophysical black hole horizons led to the introduction of a stretched horizon \cite{Damour:1978cg,Damour:1979wya,Znajek:1978nwp,Price:1986yy,Parikh:1997ma} yielding equations resembling the fluid dynamical behaviour for an effective membrane. Also in the eighties appeared the bubble of nothing \cite{Witten:1981gj}, itself a dynamical timelike boundary of finite size. Attempts to clarify the nature of quantum information across a black hole invoked notions of brick walls, and entangling surfaces \cite{tHooft:1984kcu,Bombelli:1986rw,Srednicki:1993im,Susskind:1993if,Freidel:2020xyx} all the way into the nineties. The onset of the AdS/CFT correspondence witnessed a revived interest in the membrane paradigm, as a physical hydrodynamic state of the dual CFT \cite{Policastro:2002se,Bhattacharyya:2007vjd,Iqbal:2008by}. Accompanied by this were general attempts to relate the renormalisation group flow to physics to AdS spacetimes with boundaries of more finite size \cite{deBoer:1999tgo,Heemskerk:2010hk,McGough:2016lol,Skenderis:2002wp}. Attempts to generalise such ideas to more general horizons, irrespective of the spacetime asymptotia, \cite{Bredberg:2010ky,Bredberg:2011xw,Anninos:2011zn,Eling:2011ct,Donnay:2019jiz} followed suit. More recently, a finite size timelike boundary surrounding the horizon of near-extremal black holes has played an important role in characterising the nature of the approximately AdS$_2$ near horizon physics \cite{Maldacena:2016upp}. In the last few years, the mathematical work of \cite{Friedrich:1998xt,Anderson:2006lqb,Witten:2018lgb,Fournodavlos:2020wde,Fournodavlos:2021eye,An:2021fcq,An:2025rlw,An:2025gvr} accompanied by \cite{Odak:2021axr,Anninos:2023epi,Liu:2025xij} has attempted to sharpen notions of well-posedness, or lack thereof, for general relativity in the presence of finite size boundaries.

Although the question of stretched horizons and membranes has been traditionally tied to black hole horizons, a few works also considered the possibility of porting such ideas to cosmological horizons  \cite{Banks:2003cg,Banks:2006rx,fischler2000taking,Banks:2016taq,Anninos:2011zn,Anninos:2020hfj,Shaghoulian:2021cef,Shaghoulian:2022fop,Law:2025ktz,Fischler:2024idi}. The main inspiration stems from ongoing attempts to make sense of the microphysical structure of the de Sitter horizon originally envisioned by Gibbons and Hawking \cite{Gibbons:1977mu}. This has been accompanied by broader considerations of timelike surfaces \cite{Wang:2001gt,Anninos:2011zn,Anninos:2024wpy,Coleman:2021nor,Batra:2024kjl,Silverstein:2024xnr,Banihashemi:2022jys} and infinitesimal worldlines \cite{Anninos:2011af,Anninos:2017hhn,Chandrasekaran:2022cip,Loganayagam:2023pfb,Loganayagam:2025jmw,Maldacena:2024spf} in a de Sitter world. From a Euclidean perspective, manifolds with finite size boundaries are an essential ingredient to the construction of a cosmological wavefunction proposed by Hartle and Hawking \cite{Hartle:1983ai}, at least in the regime where the spatial size of the world is small. 

The aim of this paper is to further elucidate the notion of a stretched horizon living parametrically close to the cosmological horizon of a de Sitter Universe (see \cite{Spradlin:2001pw,Witten:2001kn,Anninos:2012qw, Galante:2023uyf} for reviews covering different perspectives of the de Sitter problem). Much of our treatment generalises to other horizons as well.  To proceed, we will endow the static patch of a de Sitter spacetime with a timelike hypersurface, $\Gamma$, located parametrically close to the true de Sitter horizon. The presence of $\Gamma$ necessitates a specification of boundary conditions. Here, following \cite{Anninos:2024wpy,Banihashemi:2025qqi}, we will consider boundary conditions whereby the conformal class of the metric on $\Gamma$ is fixed, along with the trace of the extrinsic curvature, which we denote by $K$. We will take our conformal class to contain the representative metric of a round two-sphere times a time coordinate, and we will further take $K$ to be spacetime independent. Additionally, one must specify data along a spacelike Cauchy surface $\Sigma$ that intersects $\Gamma$ on a two dimensional spatial boundary. The initial data will include the specification of initial conditions for a corner degree of freedom which is part of the gravitational phase space \cite{An:2025rlw}. 
Other choices of boundary conditions are also viable; one could, for instance, consider the Dirichlet problem. 

There are two essential reasons for our choice. Firstly, it was noted in \cite{Bredberg:2011xw,Anninos:2011zn} that in the strict stretched horizon limit, the Dirichlet problem significantly restricts the space of dynamical solutions on top of the unperturbed horizon. Moreover, as explained in appendix B of \cite{Anninos:2024xhc}, in the strict horizon limit, whereby the spacetime is replaced with a Rindler geometry, the Dirichlet problem further suffers from uniqueness issues. These issues are an artifact of the strict horizon limit in a classical (as opposed to the more complete quantum mechanical) approximation, and it is interesting to understand how they become regulated upon slightly moving away a Dirichlet boundary from the strict horizon or otherwise \cite{Anninos:2023epi,An:2025gvr}. On the other hand, the conformal boundary conditions appear to have properties that are non-singular all the way into the strict horizon limit. 

We will consider the problem of a stretched horizon for both the case of a surface encasing the de Sitter horizon, which we denote as the cosmic patch, as well as the case of a surface shielding the de Sitter horizon, which we denote as the pole patch. The former is closer in spirit to more recent work on the fluid gravity correspondence \cite{Policastro:2002se,Bhattacharyya:2007vjd,Iqbal:2008by,Bredberg:2010ky}, whilst the latter is closer in spirit to the original membrane paradigm \cite{Damour:1979wya,Znajek:1978nwp,Price:1986yy,Parikh:1997ma}. We will work in four spacetime dimensions. As we shall explore in great detail, the stretched horizon is intimately connected to the large-$K\ell$ limit.

\subsection{The basic setup}

We now discuss in some more detail the setup under consideration. As mentioned, we must specify the conformal class $[g]_{mn}$ of the induced metric along $\Gamma$ along with the trace of the extrinsic curvature $K$. We take $[g]_{mn}$ to have the round metric on $S^2$ times time as a representative. To give a flavor of what $\Gamma$ looks like, let us consider the empty static patch of four-dimensional de Sitter space
\begin{equation}
\frac{ds^2}{\ell^2} = d\psi^2 -dt^2 \cos^2\psi + \sin^2\psi d\Omega_2^2~, \quad\quad \psi \in (0,\pi/2)~.
\end{equation}
The cosmological constant is related to the de Sitter length, $\ell$, as $\Lambda = + \tfrac{3}{\ell^2}$. Constant $\psi = \psi_0$ surfaces, $\Gamma_0$, have an induced metric given by $S^2\times \mathbb{R}$. When $\Gamma_0$ points toward the de Sitter horizon at $\psi = \tfrac{\pi}{2}$ one has $K\ell = \tan \psi_0 -2\cot\psi_0 \in \mathbb{R}$. Similarly, constant $\psi=\psi_0$ surfaces pointing toward the de Sitter worldline at $\psi = 0$ have $K\ell = 2\cot\psi_0-\tan \psi_0 \in \mathbb{R}$. Thus, the stretched horizon limit resides in the limit $K\ell \to \pm \infty$, depending on whether we are in the cosmic patch ($+$) or the pole patch ($-$). There will generally be many additional infillings satisfying our boundary conditions, even in the stretched horizon limit. For instance, the timelike surface pointing toward ($-$) or away from ($+$) the horizon with $\psi_0 \to 0$, which is a thin worldline limit, has $K\ell \to \pm \infty$.

In fact, even within a spherically symmetric sector, we will also have timelike surfaces in the Schwarzschild de Sitter geometry that will also have such values for $K\ell$. A systematic analysis of their properties was given in \cite{Anninos:2024wpy} (see appendix D in particular), and it follows from the results in that paper that in the limit $K\ell\to -\infty$, the energy (as measured by the appropriately defined boundary stress-tensor \cite{Odak:2021axr}) of the empty pole patch is the lowest of all static and spherically symmetric configurations. In this sense, the stretched horizon limit of the pole patch plays a distinguished role among the static and spherically symmetric solutions. It has lower energy, for instance, than the $K\ell\to-\infty$ thin worldline surface whose properties resemble those of a timelike hypersurface in Minkowski spacetime. That the stretched horizon limit of the pole patch is energetically favored over the approximately Minkwowksi timelike surface may tie, in an interesting way, to the fact that there are exponentially growing modes at the linear level around the Minkowski worldtube \cite{Anninos:2023epi,Liu:2024ymn,Liu:2025xij}.

In addition to the static spherically symmetric configurations, one also has time-dependent hypersurfaces that satisfy the same boundary conditions. For instance, a three-dimensional de Sitter (dS$_3$) slice of four-dimensional de Sitter obeys our conformal boundary conditions also. For $K\ell\to-\infty$, the dS$_3$ hypersurface will become very thin at the time-symmetric point, and just surround the de Sitter worldline. There will be again a family of such time-dependent hypersurfaces, which can be viewed as deformations of the maximally symmetric dS$_3$ hypersurface, and some of these may have lower energy than the stretched horizon of the pole patch (see \cite{Galante:2025tnt} and (\ref{Econfzero}) below). Nevertheless, if we wish to consider the static configurations as elements of a thermodynamic ensemble, we may be able to separate out the time dependent ones from our consideration. This last point merits a clearer understanding.  

For $K\ell \to \infty$,  one has a timelike surface that just encases the de Sitter horizon. In addition, there are again many additional configurations allowed by the same boundary conditions. Restricting to the spherically symmetric sector, there will again be a family of near cosmological horizon surfaces that shield the family of Schwarszchild de Sitter solutions, or even naked timelike singularities. Furthermore, there is again a family of dS$_3$ slices and deformations thereof that are permitted, but do not have an immediate thermodynamic interpretation. In addition there is the inward pointing thin worldline limit. It is less clear how to distinguish the stretched horizon of the empty static patch from the remaining $K\ell\to\infty$ phase space. However, it is a particular configuration that admits a thermodynamic interpretation containing horizon entropy at a specific temperature $\beta_{\text{dS}}$. At this specific temperature, it was shown in \cite{Anninos:2024wpy} that the static patch stretched horizon hypersurface encasing the de Sitter horizon has positive specific heat, and is the thermodynamically favoured state within the class of static and spherically symmetric configurations.

In our work, we provide an exhaustive analysis of linearised gravitational perturbations of the stretched de Sitter horizon for $K\ell\to\pm \infty$.

\subsection{Brief summary of results}

The paper is rather lengthy and detailed, so for the sake of the reader we will summarise here the main results that we have uncovered throughout our work. 

The general setup and perturbative framework is reviewed in section \ref{sec:gravdynamics}. Following \cite{Odak:2021axr}, we discuss the appropriate notion of boundary stress tensor, $T_{mn}$ given in (\ref{eq: tmn}),  for our choice of conformal boundary conditions and note that it is traceless and conserved. The gravitational perturbations are decomposed into those built from vector and scalar spherical harmonics, and are labeled by the  momentum parameters $(l ,m)$. The frequency of the modes, $\omega$, is measured with respect to the inertial clock at the origin of the static patch.  Part of the perturbative solution-space includes a boundary mode that can be expressed locally as a diffeomorphism, but is physical due to the presence of a boundary. The boundary mode sector can be treated non-linearly, and is governed by equation (\ref{eq: 2.12d3}).  Properties of the perturbative solutions for general $K\ell$ are given in sections \ref{sec:gravdynamics} and \ref{sec: large l}, where we recover and extend the results of previous analyses \cite{Anninos:2023epi,Liu:2024ymn,Liu:2025xij}. 

We now summarise the properties of the linearised solution space in the stretched horizon limit. 

\subsection*{Pole patch stretched horizon} 

For the stretched horizon limit, $K\ell\to -\infty$, of the pole patch one has to ensure that the perturbations are smooth in the static patch interior whilst obeying the boundary conditions at the boundary. Upon taking the stretched horizon limit, we find that the solution space decomposes into three main classes. This is the subject of section \ref{sec: large K pole}.

\textbf{Normal modes.} A discretuum of normal type modes in both the vector and scalar sectors. These are  modes one physically expects when placing a field theory in a box. To leading order at large-$|K\ell|$, the spacing between their frequencies goes as $\Delta\omega\ell \approx \tfrac{\pi}{\log |K\ell|}$ (\ref{eq:spacingscalnormmodes}), which tends to zero as the boundary goes toward the de Sitter horizon. Scattering these modes from the boundary, into the bulk, and back yields a scattering phase (\ref{QNMssmatrix}) whose pole structure is approximately given by the collection of quasinormal modes in the static patch, with slight corrections of order $\mathcal{O}((K\ell)^{-2})$. Configurations that have compact support in the static patch interior are captured by these modes. There is a volume worth of normal modes --- the solutions are labeled by three quantum numbers, namely, the $(l ,m)$ associated to $S^2$, and a bulk quantum number $n \in \mathbb{Z}$. These perturbations leave the Weyl factor of the boundary metric, $\boldsymbol{\omega}(t,\Omega)$, unchanged.

\textbf{Gapless modes.}  A discretuum of real frequency modes, in the scalar sector, exhibiting a massless type of dispersion relation. To leading order at large-$|K\ell|$, these modes have $\omega \ell  \approx \pm \tfrac{\sqrt{l (l +1)}}{\sqrt{2} K\ell}$ (\ref{eq:gaplessfreqsPP}). They are labeled by each $(l, m)$, so there is an area (rather than volume) worth of such modes. These modes are not present in the absence of a timelike boundary, and their gapless dispersion relation is analogous to a novel set of modes that appear when bringing in the infinite AdS$_4$ boundary subject to conformal boundary conditions \cite{Anninos:2024xhc,Allameh:2025gsa}. The radial profile of these modes is increasingly localised near the timelike boundary (cf. figure \ref{fig:bulkprofsandDom0}) as we take $K\ell \to -\infty$. Their characteristic time evolution is parametrically slower than that of the normal modes, so they appear to be very slowly evolving. Perhaps the gapless nature of their dispersion is related to some form of spontaneous breaking in the presence of a timelike boundary. These perturbations modify the Weyl factor,  $\boldsymbol{\omega}(t,\Omega)$, of the boundary metric.

\textbf{Soft and growing modes.} A discretuum of complex frequency modes, in the scalar sector, that is partitioned into two subsectors depending on whether $l \lesssim |K\ell|$ or $l \gtrsim |K\ell|$. In the strict $K\ell \to -\infty$ limit only the former are present. To leading order in the stretched horizon limit, the first subsector has $\omega \ell \approx \pm i$ for each value of $l$, and the modes are parametrically localised near the boundary, see figure \ref{fig:bulkprofsandDomi}. Remarkably, their boundary stress tensor to leading order, (\ref{eq: leading tmn w=i}), is subleading  as compared to the normal mode boundary stress tensor (\ref{stresstensor tower}). They are somewhat reminiscent of the infinite number of soft horizon modes discussed in \cite{Maldacena:2016upp,Hawking:2016msc,Knysh:2024asf}, and perhaps should be viewed as pure gauge configurations of an emergent gauge symmetry. Thus, we refer to these as soft modes.  The second subsector appears at parametrically large $l \gtrsim |K\ell|$ and their frequency scales as $\omega\ell \approx \tfrac{l}{|K\ell|} + {\nu} \left(\tfrac{l}{2|K\ell|}\right)^{\frac{1}{3}}$ where $\nu \approx 0.0674\pm0.4279i$.\footnote{There are other allowed values for $\nu$, but they are always real, see \eqref{eqn: real nus}.} These modes are what remains of the Minkowskian growing modes uncovered in \cite{Anninos:2023epi,Liu:2024ymn} in the stretched horizon limit. Again, there is an area worth of these modes and they are exponentially localised near the boundary (\ref{widthpole}). It is interesting to note that the large $K\ell$ limit creates a scale separation in $l$. As such, at sufficiently large-$K\ell$ the second subsector will happen only for Planckian scales, at which point the surface will be a Planckian distance away from the physical horizon. 
\newline\newline
Beyond the linearised regime we expect all the above modes to interact with each other, bringing about a rich but layered dynamical problem. Part of the dynamics will be captured by the evolution of the boundary Weyl mode $\boldsymbol{\omega}(t,\Omega)$ whose interplay with the bulk dynamics is governed by the Lichnerowicz type constraint equation (\ref{eq: 2.12}).

\subsection*{Cosmic patch stretched horizon} 

For the stretched horizon limit, $K\ell\to +\infty$, of the cosmic patch we take the perturbations to be purely outgoing at the de Sitter horizon whilst obeying the boundary conditions at the boundary. Upon taking the stretched horizon limit, we find that the solution space decomposes into three main classes.  This is the subject of section \ref{sec: large K cosmic}.

\textbf{Quasinormal modes.} A discretuum of quasinormal mode type excitations. We find that $K\ell$ deforms the original quasinormal mode spectrum of the empty static patch \cite{Lopez-Ortega:2006aal} to a new set of quasinormal modes labeled by three quantum numbers $(l,m,n)$. Thus, there is a volume worth of quasinormal modes. Their frequencies are complex and go as $\omega\ell \approx -in+\frac{i\omega_{1}}{(K\ell)^{2n}}$,  (\ref{eq:QNMlikemodsscal}) in the scalar sector where $n=2,3,\dots,l$ and (\ref{eq:QNMlikemodsvec}) in the vector sector where $n=1,2,\dots,l$. These perturbations leave the Weyl mode  $\boldsymbol{\omega}(t,\Omega)$ of the boundary metric unchanged. They encode the dissipative nature of the de Sitter horizon.

\textbf{Fluid dynamical modes.} A discretuum of fluid dynamical type modes that come in two classes. The first class, which appears for the vector perturbations, takes the form of linearised shear modes for a Navier-Stokes equation. Their dispersion relation goes as $\omega\ell \approx - i\tfrac{ l(l+1)-2}{2(K\ell)^2}$ in the stretched horizon limit (\ref{eq:shearmodes}). The non-linear completion of this sector in terms of an incompressible Navier-Stokes equation was obtained in \cite{Anninos:2011zn}. The second class, which appears for the scalar perturbations, take the form of linearised sounds modes of the Navier-Stokes equation. The fluid type modes are labeled by each $(l,m)$, so there is an area (rather than volume) worth of such modes. Their frequency goes as $\omega\ell \approx \pm \tfrac{\sqrt{l(l+1)}}{\sqrt{2}K\ell}$ (\ref{cosmicsound}). We can read off the shear viscosity, $\eta$, and bulk viscosity, $\zeta$, directly from the fluid dynamical modes. We find that the shear viscosity to entropy density ratio is $\tfrac{\eta}{s} = \tfrac{1}{4\pi}$, confirming the classic result  \cite{Damour:1979wya,Price:1986yy,Policastro:2001yc}. We also find that the bulk viscosity  vanishes, $\zeta=0$, reminiscent of the behaviour of a conformally invariant fluid \cite{Policastro:2002se}. The speed of sound is found to be $c_s = \tfrac{1}{\sqrt{2}}$ as measured by the clock at the conformal boundary. In our context there is no dual conformal field theory or AdS$_4$ boundary, and our result $\zeta=0$ contrasts in a sharp and interesting way the negative bulk viscosity , $\zeta_{\text{m.p.}} =-\tfrac{1}{16\pi}$ found in the original membrane paradigm! 

\textbf{Soft and growing modes.} As before, a discretuum of complex frequency modes that is partitioned into two subsectors depending on whether $l \lesssim K\ell$ or $l \gtrsim K\ell$. In the strict $K\ell \to \infty$ limit only the former are present. To leading order in the stretched horizon limit, the first subsector has $\omega \ell \approx \pm i$, found in (\ref{eq:cosmicplusIexp}) and (\ref{cosmicminusIexp}), for each value of $l$.  Again, they are  reminiscent of the infinite number of soft horizon modes discussed in \cite{Maldacena:2016upp,Hawking:2016msc,Knysh:2024asf}, and perhaps should be viewed as pure gauge modes of an emergent gauge symmetry. Relative to the quasinormal mode and sound mode boundary stress tensors, (\ref{eq:stresstenpimodesQNM}) and (\ref{eqn: Tmn sound}), the soft mode stress tensor (\ref{eq:stresstenpimodesCP}) is subleading. In terms of the de Sitter temperature measured by an inertial observer, $T_{\text{dS}} = (2\pi\ell)^{-1}$, one has $\omega \approx \pm 2\pi i T_{\text{dS}}$. The second subsector appears at parametrically large $l \gtrsim K\ell$ and their frequency scales as $\omega\ell \approx \tfrac{l}{K\ell} \pm {\nu} \left(\tfrac{l}{2K\ell}\right)^{\frac{1}{3}}$ where $\nu \approx -0.4043+0.1556i$. Again there is an area worth of these modes and their radial profile decays exponentially away from the boundary. It is interesting to note that the large-$K\ell$ limit again creates a scale separation in $l$. 

\subsection*{Towards a non-linear completion of the fluid dynamical sector}

As before, beyond the linearised regime we expect all the above modes to interact with each other, bringing about a rich dynamical problem. In section \ref{sec: non-linear fluid}, we offer some insight into this for the fluid dynamical sector. We show that in the large-$K\ell$ limit, the solution space to Einstein's equations can be expressed in terms of a non-linear compressible Navier-Stokes equation (\ref{eqn: fluid eom}). The linearised approximation of this equation reduces to both the shear and sound modes discussed previously. The fluid dynamical sector will, in turn, interact with the remaining dynamical modes. We view the two types of dynamical fluid modes, as accounting for the two propagating degrees of freedom of the gravitational field.  Our approach may pave the way toward a complete picture for the relationship between the dynamics of fluids and general relativity.

\subsection*{Rindler viewpoint \& the disparity of horizons}

In order to gain a clearer understanding of the generality of our results, we have dedicated  section \ref{sec: rindler} to an analysis of conformal boundaries in a Rindler spacetime. We view the Rindler region as the universal geometry near the horizon and we consider both an introspective and extraspective perspective. In this regime, the spherical horizon is replaced with a planar region, and the discretuum of angular momentum modes $(l,m)$ with a continuum of spatial two-momenta $\bold{k}\in\mathbb{R}^2$. We identify a Rindler avatar for each of the linearised modes in both the pole patch and cosmic patch to leading order in the large $|K\ell|$ limit. More concretely, taking the concurrent limit $l\to \infty$ and $|K\ell|\to\infty$, with $\kappa \equiv \tfrac{\sqrt{l(l+1)}}{K\ell}$ fixed we find that to all orders in the small-$\kappa$ expansion the de Sitter features are mapped to Rindler features under the replacement $\tfrac{l(l+1)}{\ell^2} \to \bold{k}\cdot\bold{k}$.

So the dynamics of a cosmological and black hole stretched horizon are likely to be indistinguishable in a suitable Rindler regime. Indeed, in this regime the de Sitter length, $\ell$,  drops out from our expressions altogether. Nonetheless, once we keep track of the finiteness of the horizon and geometry, we find qualitative and quantitative differences (see \cite{Anninos:2018svg,Anninos:2022ujl} for an exposition contrasting black hole and de Sitter horizons). As a simple but explicit example, in the stretched horizon limit the first correction to the soft mode frequencies of the de Sitter horizon, (\ref{eqn: +i dS}), and those of a Schwarzschild black hole with horizon radius $\rh$, (\ref{bhsoft}), differ by a relative sign. This effect is amplified in the $l=0$ sector governed by (\ref{eqn: brane eqn}), where one finds a more immediate distinction  between the black hole and de Sitter horizon behaviour.

\subsection*{Emergence of stretched horizon locality?}

A rather striking feature of our findings is that in the stretched horizon limit $K\ell\to \pm \infty$ several structures appear to enjoy a type of boundary locality. The low lying modes we have identified, and more particularly the fluid dynamical type modes and gapless modes, organise themselves into a $(2+1)$-dimensional local structure. One could also consider the $\omega\ell = \pm i$ modes as being ultralocal at the boundary. The remaining modes do not obey any form of boundary locality, but all the modes will interact non-trivially among each other. It is also striking that the stretched horizon limit creates a large hierarchy in $l$, limiting the Minkowskian type behaviour to parametrically large values $l \gtrsim |K\ell|$. Also notable is the relation between the proper distance between the timelike boundary to the actual horizon and the large-$l$ cutoff. They both scale with $|K\ell|$ --- the closer we push the timelike surface to the horizon, the sharper the resolution of the horizon sphere can be made before hitting the growing Minkowskian modes. This has the flavour of coarse graining.

The approximate local boundary structure is perhaps related to the observation of \cite{Anninos:2023epi,Banihashemi:2024yye,Banihashemi:2025qqi,Allameh:2025gsa} that subject to conformal boundary conditions, the high temperature scaling of the horizon entropy, $S$, in a $(d+1)$-dimensional gravitational theory scales as $S \approx N_{\text{d.o.f.}} \beta^{-(d-1)}$,  characteristic of a (local) $d$-dimensional conformal field theory. Here $N_{\text{d.o.f.}}$ is a dimensionless quantity built from $K$, $\ell$, and $G_N$. For a four-dimensional de Sitter horizon, it was found in \cite{Anninos:2024wpy} that 
\begin{equation}
N_{\text{d.o.f.}} = \frac{32\pi^3 \ell^2}{81 G_N} \left(\sqrt{K^2\ell^2 + 9} - K \ell \right)^2~.
\end{equation}
In the cosmic patch stretched horizon limit, $K\ell\to \infty$, we find that $N_{\text{d.o.f.}} \approx \tfrac{8\pi^3}{G_N K^2}$, which recovers the Minkowskian black hole expression (see (6.17) of \cite{Anninos:2023epi}) and is independent of $\ell$. In the pole patch stretched horizon limit, $K\ell\to - \infty$, we instead find $N_{\text{d.o.f.}} \approx  \tfrac{128 \pi ^2 }{81}(K \ell)^2 \mathcal{S}_{\text{dS}}$, where $\mathcal{S}_{\text{dS}}\equiv \tfrac{3\pi}{\Lambda G_N}$ is the tree-level Gibbons-Hawking de Sitter horizon entropy \cite{Gibbons:1977mu}. Amusingly, in the stretched horizon limit,  the number of pixelated points on an $S^2$ with an angular momentum cutoff at $l=|K\ell|$ scales as $(K\ell)^2$.

Viewed from the perspective of effective field theory, our treatment will require a completion due to the presence of irrelevant operators (see \cite{Chapman:2012my,Galante:2025emz} for relevant analyses), suitably renormalised quantum field excitations near sharp boundaries \cite{Deutsch:1978sc,Philcox:2025faf,Draper:2025kcr}, or other potential boundary phenomena \cite{Witten:1981gj,Horava:1996ma,Ahmadain:2024uyo}. In judicious circumstances, we may have a hierarchy of scales, $\ell \gg K^{-1} \gg \ell_{\text{EFT}}\gg \ell_{\text{Planck}}$,  allowing us to take the large-$K\ell$ limit whilst still keeping the additional effects suppressed. Ultimately, however, these effects must be incorporated.  

At any rate, we hope that the pieces we have assembled will aid in the microscopic (or in the least mesoscopic) construction of a quasilocal piece of quantum cosmology.

\section{Gravitational dynamics with conformal boundaries}  \label{sec:gravdynamics}

We consider the theory of general relativity with positive cosmological constant $\Lambda$ in four spacetime dimensions, whose Lorentzian action is given by
\begin{equation}
   I = \frac{1}{16 \pi G_N} \int_{\mathcal{M}} d^4 x \sqrt{-\text{det} \, g_{\mu\nu}} \,\left(R-2\Lambda\right) + \frac{1}{24 \pi G_N} \int_\Gamma d^{3}x \sqrt{-\text{det}\,g_{mn}} \, K  \, , 
    \label{eq: action}
\end{equation}
with $\Lambda=+\tfrac{3}{\ell^2}$, for dS$_{4}$ length scale $\ell$. The timelike boundary is denoted by $\Gamma$, and is endowed with the induced metric $g_{mn}$. Further, $K = g^{mn} K_{mn}$ denotes the trace of the extrinsic curvature, $K_{mn} = \tfrac{1}{2} \mathcal{L}_{n^\mu} g_{mn}$ with respect to the outward-pointing unit normal $n^\mu$. The boundary term is such that the variational problem is well-posed when conformal boundary conditions are imposed at the boundary $\Gamma$. These boundary conditions fix both the conformal class of the induced metric at the boundary, denoted $[g_{mn}]|_{\Gamma}$, and the trace of the extrinsic curvature $K$ at $\Gamma$. 

Solutions obey the Einstein field equations of motion stemming from action \eqref{eq: action},
\begin{equation}\label{eqn: Einstein tensor}
    G_{\mu\nu} \equiv R_{\mu\nu} - \tfrac{1}{2}g_{\mu\nu} R + \Lambda g_{\mu\nu} = 0 \,.
\end{equation}

Varying the action \eqref{eq: action} on-shell with respect to the boundary conformal metric $[g]_{mn}|_{\Gamma}$ (a conformal representative of the conformal class $[g_{mn}]$) defines the conformal version of the Brown-York stress tensor density \cite{An:2021fcq,Odak:2021axr},
\begin{equation}
    T_{m n} \equiv \left. -\frac{e^{\boldsymbol{\omega}}}{8\pi G_N}\left( K_{m n} - \frac{1}{3} K g_{mn} \right) \right|_\Gamma  \,,
    \label{eq: tmn}
\end{equation}
where the metric at the boundary is written as follows,
\begin{equation}
    ds^2|_\Gamma = g_{mn} dx^m dx^n\equiv e^{2\bomega}[g]_{mn} dx^m dx^n \,.
\label{eq:bdrymet}\end{equation} 
It is straightforward to verify that $T_{mn}$ is traceless with respect to the conformal representative of the induced metric at the boundary. Note that $T_{mn}$ is not invariant under Weyl transformations of the conformal representative. In fact, it transforms with conformal weight $-1$. We will define a quantity that is invariant under Weyl transformations in the next subsection.

Interestingly, the projection of the Einstein equations onto the boundary can be written in terms of $T_{mn}$. For boundary metrics of the form  (\ref{eq:bdrymet}), the analogue of the momentum constraint gives,
\begin{equation}
\mathcal{D}^m T_{mn} =-\frac{1}{12\pi G_{N}}e^{3\bomega} \mathcal{D}_n K\;.
\label{eq:momconst}\end{equation}
That is, provided $K$ is constant, it follows that  $T_{mn}$ is covariantly conserved with respect to the conformal metric. Meanwhile, the analogue of the Hamiltonian constraint for the timelike boundary $\Gamma$ characterises the dynamics of the Weyl factor,
\begin{equation}
    \mathcal{D}_m\mathcal{D}^m \bomega + \frac{1}{2}\mathcal{D}_m \bomega \mathcal{D}^m \bomega - \frac{R^{(\Gamma)}}{4} - 16\pi^2G_N^2 T^{mn}T_{mn} e^{-4\bomega} + \left(K^2 + 3\Lambda\right)\frac{e^{2\bomega}}{6}=0 \, ,
    \label{eq: 2.12d3}
\end{equation}
where
$R^{(\Gamma)}$ refers to the Ricci scalar of the boundary conformal metric with $\bomega=0$. Thus, in general, the equation of motion for the Weyl factor  is a second order (in boundary coordinates), non-linear differential equation. Together, equations (\ref{eq:momconst}) and (\ref{eq: 2.12d3}) are the hyperbolic version of the Lichnerowicz-York equations \cite{LichKing,York:1972sj,Choquet-Bruhat:2009xil}.

For a general $d$-dimensional boundary, the Weyl equation of motion equation (\ref{eq: 2.12d3}) generalises to
\begin{equation}
    \mathcal{D}_m\mathcal{D}^m \bomega + \frac{d-2}{2}\mathcal{D}_m \bomega \mathcal{D}^m \bomega - \frac{R^{(\Gamma)}}{2(d-1)} - \frac{64 \pi^2G_N^2}{2(d-1)}T^{mn}T_{mn} e^{-2(d-1)\bomega} + \left(K^2 + \frac{2d}{d-1}\Lambda\right)\frac{e^{2\bomega}}{2d}=0 \, ,
    \label{eq: 2.12}
\end{equation}
with
\begin{equation}
    T_{mn} = \left.- \frac{e^{(d-2)\bomega}}{8\pi G_N}\left(K_{mn}-\frac{1}{d}K g_{mn}\right)\right|_{\Gamma}\,.
\end{equation}

For the rest of the paper, we will focus mostly on $d=3$.

\subsection{Background solution}

We consider a finite region of the static patch of de Sitter space,
\begin{equation}
    ds^2 = - f(r) dt^2 + \frac{dr^2}{f(r)} +r^2 d\Omega_2^2 \quad \,, \, \quad f(r) \equiv \left(1-\tfrac{r^2}{\ell^2}\right) \quad \,, \, \quad d\Omega_2^2 \equiv d\theta^2 + \sin^2 \theta d\varphi^2 \,, \label{eq:dSmetric}
\end{equation}
where $\ell$ denotes the de Sitter length scale, and the coordinate ranges are $t\in\mathbb{R}$, $r\in(0,\ell)$, $\theta\in(0,\pi)$, and $\varphi\in(0,2\pi)$. A static patch observer is located at the pole $r=0$ and encounters a cosmological horizon at $r=\ell$. We position a codimension-one timelike boundary $\Gamma$ at $r=\r$, dividing the static patch into two regions of interest: the pole patch, with $r\in [0,\r]$, and the cosmic patch, with $r\in[\r,\ell]$, see figure~\ref{fig:SPbdry}. 
Quantities with respect to the background geometry (\ref{eq:dSmetric}) are denoted using a barred notation, e.g., the background metric is $\bar{g}_{\mu\nu}$ where Greek indices run over all coordinates.

\begin{figure}[t!]
\centering
\begin{tikzpicture}[scale=1]
	\pgfmathsetmacro\myunit{4}
	\draw[dashed]	(0,0)			coordinate (a)
		--++(90:\myunit)	coordinate (b);
	\draw (b) --++(0:\myunit)		coordinate (c)
							node[pos=.5, above] {};
	\draw[dashed] (c) --++(-90:\myunit)	coordinate (d);
    \fill[fill=magenta, fill opacity=.4] (a) to[bend right=35]  (b) -- (2,2) -- (a);
    \fill[fill=blue, fill opacity=.4] (a) to[bend right=35]  (b) -- (0,0) -- (a);
    \path (b) to[bend left=35] node[pos=.65,above,sloped] {$r=\mathfrak{r}$} (a);
    \draw[dashed, name path=rB] (a) to[bend right=35] (b);
    \draw (d) -- (a) 		node[pos=.5, below] {};
    \draw (b) -- (d) node[pos=.3, above, sloped] {$r = \ell$} -- (c) -- (a);
    \draw (a) -- (b) node[pos=.7, above, sloped] {$r = 0$};
    \draw (c) -- (d);
    \filldraw (2,2) circle (0.025cm);
\end{tikzpicture}
\begin{caption}{Static patch with a timelike boundary. The boundary delineates the static patch into the ``cosmic patch'' (magenta) and ``pole patch'' (blue). In the worldline limit ($\bar{K}\ell\to\sigma\infty$), the pole patch occupies a small bulk region near the pole, while the cosmic patch fills nearly the entire static patch. Alternatively, in the stretched horizon limit ($\bar{K}\ell\to -\sigma\infty$), the pole patch fills nearly the entire static patch, while the cosmic patch is a small bulk region near the horizon.
\label{fig:SPbdry}}
\end{caption}
\end{figure}

The induced metric at the boundary has the form
\begin{equation}
    ds^2|_\Gamma = \bar{g}_{mn} dx^m dx^n = -f(\r) dt^2+\r^2 d\Omega_2^2 \,,
    \label{eq: induced}
\end{equation}
where indices $m,n$ range over $(t,\theta,\varphi)$. The extrinsic curvature at $\Gamma$ is
\begin{equation}
    \left. \bar{K}_{mn}dx^mdx^n\right|_\Gamma=\sigma \frac{\r}{\ell^2}\sqrt{1-\frac{\r^2}{\ell^2}}\left(dt^2+\ell^2 d\Omega_2^2\right) \,,
\end{equation}
where $\sigma = \pm 1$, for the pole and cosmic patch, respectively. Meanwhile, the trace of the extrinsic curvature is 
\begin{equation}
    K \ell = \sigma \frac{2\ell^2- 3\frakr^2}{\frakr\sqrt{\ell^2-\frakr^2}} \, .
\label{eq:KdS}\end{equation}
Our main interest will be the \emph{stretched  horizon limit}, where the boundary approaches the horizon, $\mathfrak{r}\to\ell$. Note in particular this means, when restricted to the pole patch, the stretched horizon limit leads to a bulk region occupying nearly the entire static patch. Also of interest is the \emph{worldline limit}, where the finite boundary approaches the pole, $\mathfrak{r}\to0$. The worldline limit produces a small portion of the bulk spacetime (and vice versa for the cosmic patch). Note further that as the boundary moves from the worldline towards the cosmic horizon, ${K} \ell$ spans from $-\infty$ to $+\infty$ in the case of the cosmic patch, and vice-versa for the pole patch. Inverting (\ref{eq:KdS}) yields 
\beq \left(\frac{\mathfrak{r}}{\ell}\right)^{2}=\frac{12+(K\ell)^{2}-\sigma K\ell\sqrt{(K\ell)^{2}+8}}{2(9+(K\ell)^{2})}\;,\label{eqn: r as Kl}\eeq
such that the stretched horizon and worldline limits are, respectively, 
\beq \frac{\mathfrak{r}}{\ell}\biggr|_{K\ell\to- \sigma \infty}=1-\frac{1}{2| K\ell|^{2}}+\mathcal{O}(|K\ell|^{-4})\;,\quad \frac{\mathfrak{r}}{\ell}\biggr|_{K\ell\to \sigma\infty}=\frac{2}{|K\ell|}-\frac{8}{|K\ell|^{3}}+\mathcal{O}(|K\ell|^{-4})\;.\label{eq:limitsbackK}\eeq

For the de Sitter background (\ref{eq:dSmetric}), the conformal stress tensor (\ref{eq: tmn}) reads,
\begin{equation}
     \frac{\bar{T}_{tt}}{1-\r^2/\ell^2} = \frac{2}{\r^2} \bar{T}_{\theta\theta} =\frac{ 2}{\r^2 \sin^2\theta} \bar{T}_{\varphi\varphi} = -\frac{\sigma}{8\pi G_N} \frac{2}{3\r \sqrt{1-\tfrac{\r^2}{\ell^2}}} \,,
\end{equation}
which is manifestly traceless and transverse with respect to the induced metric \eqref{eq: induced}. In terms of the trace of the boundary extrinsic curvature (\ref{eq:KdS}), this reads
\begin{equation}
   8\pi G_N \bar{T}_{tt} = -\frac{\sigma}{3 \ell} \sqrt{ 2 + \frac{( K\ell)^2}{2} + \sigma\frac{ K \ell}{2} \sqrt{( K\ell)^2+8}} = \frac{2}{3 K \ell^2} + \mathcal{O}((K\ell)^{-3})\,.
\end{equation}

From the conformal stress tensor, it is possible to define the conformal energy as \cite{Anninos:2024xhc}
\begin{equation}
    E_{\text{conf}} \equiv \frac{\r^2}{\sqrt{f(\r)}} \int T_{tt} \, d\Omega_2 \,. \label{eqn: def Econf}
\end{equation}
This quantity is invariant under Weyl transformations of the conformal representative of the metric at the boundary and agrees with the energy computed using the Euclidean partition function \cite{Anninos:2024wpy}.\footnote{Note that the time chosen here can be related to the one in \cite{Anninos:2024wpy} by  $\tau \, \r = \sqrt{f(\r)} t$. Given this identification in the time coordinate, the expressions agree.} From the conservation of $T_{mn}$, it follows that $E_{\text{conf}}$ is time independent.

For the background solution (\ref{eq:dSmetric}), and in terms of the static patch time, we obtain 
\begin{equation}
    \bar{E}_{\text{conf}} = -\frac{\sigma \r}{3G_N} \,,
\end{equation}
such that in the stretched horizon limit, $\bar{E}_{\text{conf}} \to -\frac{\sigma \ell}{3 G_N}$, while in the worldline limit $E_{\text{conf}}$ vanishes. We observe that the energy is negative in the pole patch ($\sigma=+1$) and positive in the cosmic patch ($\sigma=-1$). An analogous behaviour can be observed in the case of a Dirichlet boundary \cite{Banihashemi:2022htw}.

To make contact with the Weyl factor equation of motion \eqref{eq: 2.12}, it is convenient to compute
\begin{equation}
    (8\pi G_N)^2 \bar{T}_{mn} \bar{T}^{mn} = \frac{2}{3 \r^2 \left(1- \frac{\r^2}{\ell^2}\right)}= \frac{3}{\ell^2} + \frac{5 (K\ell)^2}{12 \ell^2} -\sigma \frac{K\ell \sqrt{(K \ell)^2+8}}{4 \ell^2} \,,
    \end{equation}
which in the stretched horizon limit becomes,
\begin{equation}
     (8\pi G_N)^2 \bar{T}_{mn} \bar{T}^{mn} =
         \frac{2 K^2}{3}+\frac{4}{\ell^2}+\mathcal{O}\left((K \ell)^{-2}\right) \,.
         \label{eq: large K Tmn}
\end{equation}
Note that the leading contribution becomes independent of $\ell$. It is now easy to check that this background solution with a constant $\bomega=0$ satisfies \eqref{eq: 2.12}. Indeed, 
\begin{equation}
- \frac{1}{4}\bar{R} - 16 \pi^2G_N^2 \bar{T}^{mn}\bar{T}_{mn} + \frac{1}{6}\left(K^2 + \frac{9}{\ell^{2}}\right)=0\;,
\end{equation}
where we used that the Ricci scalar of the boundary metric $\bar{g}_{mn}$ is $\bar{R}=\frac{2}{\mathfrak{r}^{2}}$.

\subsection{Linearised dynamics}\label{section 2.2:linearised dynamics}
We would like to study the dynamics of small perturbations around the background metric that preserve the conformal boundary conditions. In particular, we consider metric perturbations,
\begin{equation}
    g_{\mu\nu} = \bar{g}_{\mu\nu} + \varepsilon \, h_{\mu\nu} \,, \;\;|\varepsilon| \ll1 \,,
\end{equation}
where $\bar{g}_{\mu\nu}$ is given by \eqref{eq:dSmetric}, and we have the boundary conditions
\begin{equation}
     ds^2|_\Gamma = e^{2\bomega(x^{m})} \left( -f(\r) dt^2+\r^2 d\Omega_2^2 \right) \,, \quad\quad \varepsilon \delta K \equiv K (g_{mn}+\varepsilon h_{mn}) - K (g_{mn}) = 0 \,.
\label{eq:linCBCs}\end{equation}
Here $\bomega(x^{m})$ is a dynamical conformal factor that is not fixed as a boundary condition, and we adopt the notation $\varepsilon \, \delta X \equiv X(g+\varepsilon h) - X(g) $ to denote the linearised variation of any object $X$.

Studying the linearised dynamics of finite conformal boundaries in dS$_{4}$ was performed in \cite{Anninos:2024wpy}, where the spherical symmetry of the background metric was used to organize the metric perturbations via the Kodama-Ishibashi formalism \cite{Kodama:2000fa,Kodama:2003kk}. Details about the method and our conventions can be also found in appendix C of \cite{Anninos:2023epi}. In the absence of a boundary, this method provides a gauge-invariant way of studying gravitational perturbations. Here, we use a convenient gauge such that the metric perturbations about the dS$_{4}$ background (\ref{eq:dSmetric}) take the form \cite{Anninos:2024wpy} 
\begin{eqnarray}
\begin{cases}
	h_{mn} &= - \bar{g}_{mn} \frac{\ell}{2r}\left[ l\left(l+1\right)\left(1- \frac{2r^2}{\ell^2}\right)+2 r^2 \partial_t^2 + 2  \left(1-\frac{r^2}{\ell^2}\right)^2  r \partial_r\right]\Phi^{(S)} \mathbb{S}  \\
 & \quad + \ell\left(\delta^i_m\delta^t_n + \delta^i_n \delta^t_m\right) \left(1-\frac{r^2}{\ell^2}\right)\partial_r \left(r \Phi^{(V)}\right) \mathbb{V}_i \, , \\
    h_{rr} &= -\frac{\ell}{r\left(1-\frac{r^2}{\ell^2}\right)^2}\Bigg[\frac{l(l+1)}{2}\left(3-\frac{7r^2}{\ell^2}+\frac{4r^4}{\ell^4}\right)+ \left(3-\frac{2 r^2}{\ell^2}\right)r^2\partial_t^2 \\
    		&\quad + \left(1-\frac{r^2}{\ell^2}\right)\left(\left(1-\frac{r^2}{\ell^2}\right)\left(l(l+1) + 1-2\frac{r^2}{\ell^2} \right)+ r^2\partial_t^2\right) r\partial_r\Bigg]\Phi^{(S)} \mathbb{S} \, , \\
    h_{tr} &=  -\frac{\ell}{2\left(1-\frac{r^2}{\ell^2}\right)}\partial_t \left[l(l+1)\left(1-\frac{r^2}{\ell^2}\right) - 2 + r^2\partial_t^2 - \left(1-\frac{r^2}{\ell^2}\right)\frac{r^2}{\ell^2}r\partial_r\right]\Phi^{(S)} \mathbb{S} \, ,  \\
    h_{ri} &=  \frac{\ell\sqrt{l(l+1)}}{2\left(1-\frac{r^2}{\ell^2}\right)}\left[l(l+1)\left(1-\frac{r^2}{\ell^2}\right)+r^2\partial_t^2+\left(2-\left(3-\frac{r^2}{\ell^2}\right)\frac{r^2}{\ell^2}\right)r\partial_r\right]\Phi^{(S)} \mathbb{S}_i + \frac{\ell r}{1-\frac{r^2}{\ell^2}} \partial_t \Phi^{(V)} \mathbb{V}_i \, ,
    \end{cases}\label{eqn: spherical l>2 ansatz}
\end{eqnarray}
where $\bar{g}_{mn}$ has line element (\ref{eq:linCBCs}) and indices $i,j$ range over coordinates $\{\theta,\varphi\}$. 

Due to the spherical symmetry and time-translation invariance, the metric perturbation $h_{\mu\nu}$ is uniquely decomposed into a vector and scalar perturbation, $h_{\mu\nu}=h^{(V)}_{\mu\nu}+h_{\mu\nu}^{(S)}$. The content of $h^{(V)}_{\mu\nu}$ is captured by vector spherical harmonics, i.e., transverse (vectorial) eigenfunctions $\mathbb{V}_{i}$ of the unit two-sphere Laplacian with eigenvalues $k^{2}_{V}=l(l+1)-1$ for $l=1,2,\ldots$. The content of $h^{(S)}_{\mu\nu}$ is encoded in transverse eigenfunctions $\mathbb{S}$ of the unit two-sphere Laplacian with eigenvalues of $k^{2}_{S}=l(l+1)$ for $l=0,1,2,...$ (the $l=0$ and $l=1$ modes require a different treatment -- see Appendix E of \cite{Anninos:2024wpy}). In particular, here $\mathbb{S}=Y_{ml}(\theta,\varphi)$ (the usual spherical harmonics), $\mathbb{S}_{i}=-\frac{1}{\sqrt{\ell(\ell+1)}}\tilde{\nabla}_{i}\mathbb{S}$ for covariant derivative $\tilde{\nabla}_{i}$ with respect to the two-sphere, and $\mathbb{V}_{i}=(\star d\mathbb{S})_{i}$ for Hodge operator $\star$.

Further, the vector and scalar metric perturbations
 each depend solely on a master field $\Phi^{(S/V)}$ that satisfies, 
 \begin{equation}
     \nabla^2_{\text{dS}}\Phi^{(S/V)}(t,r)-\frac{l(l+1)}{r^2}\Phi^{(S/V)}(t,r) = 0 \,,
 \end{equation}
where $\nabla^2_{\text{dS}}$ is the Laplacian on a two-dimensional de Sitter space with curvature $+\tfrac{2}{\ell^2}$. Explicitly, 
\begin{equation}
\left(1-\frac{r^{2}}{\ell^{2}}\right)^{2}\partial_{r}^{2} \Phi^{(S/V)}(t,r) -\frac{2r}{\ell^{2}}\left(1-\frac{r^{2}}{\ell^{2}}\right)\partial_{r} \Phi^{(S/V)}(t,r)-\left[\frac{l(l+1)}{r^{2}}\left(1-\frac{r^{2}}{\ell^{2}}\right)+ \partial_t^2 \right] \Phi^{(S/V)}(t,r) =  0~.
\label{eq: master equation}
\end{equation}
It is straightforward to verify the metric satisfies the perturbed Einstein field equations to leading order provided the master fields satisfy \eqref{eq: master equation}. The choice of gauge is such that a generic metric perturbation purely in the scalar sector already preserves the conformal class of the background metric, while a metric perturbation purely in the vector sector preserves the trace of the extrinsic curvature of the background induced metric at the boundary. 

Given the background solution satisfies \eqref{eq: 2.12}, at the linearised level, we are left with a linear equation for the variation of the conformal factor $\delta \bomega$,
\beq 
\mathcal{D}_{m} \mathcal{D}^{m}\delta\bomega+\left( 64\pi^{2}G_{N}^{2}\bar{T}_{mn}\bar{T}^{mn}
+\frac{K^{2}}{3}+\frac{3}{\ell^{2}} \right)\delta\bomega = \frac{64\pi^{2}G_{N}^{2}}{2}\bar{T}^{mn}\delta T_{mn}\;,
\label{eq: linearised 2.12}
\eeq
where $\mathcal{D}_{m}$ is the covariant derivative with respect to the conformal class of the boundary metric, we used $\delta K|_{\Gamma}=0$, and $\varepsilon \delta T_{mn} \equiv T_{mn} (g+h) - T_{mn}(g)$. Here the conformal stress tensor encodes information about the bulk dynamics and explicitly contains $\delta \bomega$ in its definition (\ref{eq: tmn}). Notice that the conformal energy associated to $\delta T_{mn}$ will always vanish at the linearised level for $l\geq2$. This is because the integral over the two-sphere of any spherical harmonic with $l\geq 2$ will vanish.

Writing $\delta\bomega = \delta \bomega(t) \mathbb{S}$, 
it is straightforward to show \eqref{eq: linearised 2.12} becomes 
\begin{align} \label{linearized Weyl eq}
    \frac{2\mathbb{S}}{\mathfrak{r}^2(1-\frac{\mathfrak{r}^2}{\ell^2}) }\bigg( -\mathfrak{r}^2 \partial_t^2 \delta\bomega(t) +  \left(1+\left(1-\frac{\mathfrak{r}^2}{\ell^2}\right)(1-l(l+1))\right)\delta\bomega(t)\bigg)&=  64\pi^{2}G_{N}^{2}\bar{T}^{mn}\delta T_{mn} \,.
\end{align}

\textbf{Vector perturbations.} With our choice of gauge, a purely vector perturbation already satisfies by definition $\delta K|_{\Gamma}=0$, so we only need to impose that the conformal class of the metric is preserved at the linearised level. This implies
    \begin{equation}
    \left. \Phi^{(V)} + r \, \partial_r \Phi^{(V)}\right|_{\Gamma} =0 \, . 
    \label{eq: vector bdy condition}
\end{equation}
Moreover, it is direct to verify that a vector perturbation does not affect the Weyl factor at the boundary, that is, $\delta \bomega = 0$ for all vector perturbations. Consistency with \eqref{eq: linearised 2.12}, then, implies  $\bar{T}^{mn} \delta T_{mn}^{(V)}$ should vanish (to linear order in the metric perturbation). However, not all of the individual components of the on-shell stress tensor in the vector sector vanish. Indeed, a direct calculation shows they are given by
\begin{equation}
\begin{cases}
    8\pi G_N\delta T_{t i}= -\frac{\ell}{2\r} \sqrt{1-\frac{\mathfrak{r}^2}{\ell^2}} (l-1)(l+2) \Phi^{(V)}(t,\mathfrak{r}) \mathbb{V}_{i}\;,\\
    8\pi G_N\delta T_{ij}= \frac{\mathfrak{r}\ell}{2\sqrt{1-\frac{\mathfrak{r}^2}{\ell^2}}}  \partial_t \Phi^{(V)}(t, \mathfrak{r})(\tilde{\nabla}_i \mathbb{V}_j+\tilde{\nabla}_j \mathbb{V}_i)\;.
\end{cases}
\label{eq:deltaTmnvecgen}\end{equation}
Provided $\tilde{\nabla}_{i}\mathbb{V}^{i}=0$,\footnote{The vectorial spherical harmonics $\mathbb{V}_{i}$ of angular momentum $l$ satisfy $(\tilde{\nabla}^{2}+l(l+1)-1)\mathbb{V}_{i}=0$ for $\tilde{\nabla}^{2}=\frac{1}{\sin(\theta)}\partial_{\theta}(\sin(\theta)\partial_{\theta})+\frac{1}{\sin^{2}(\theta)}\partial^{2}_{\varphi}$ being the unit two-sphere Laplacian, and $\tilde{\nabla}_{i}\mathbb{V}^{i}=0$.} it is clear now that, at this order, the stress-tensor is traceless, and, as expected,
\beq 
64\pi^{2}G_{N}^{2}\bar{T}^{mn}\delta T^{(V)}_{mn}=-\frac{2}{3\mathfrak{r}^{2}\left(1-\frac{\mathfrak{r}^{2}}{\ell^{2}}\right)}\tilde{\nabla}_{i}\mathbb{V}^{i}\partial_{t}\Phi^{(V)} (t,\r) = 0 \;.
\eeq

\textbf{Scalar perturbations.} From the boundary point of view, the scalar perturbations are more interesting since they modify the Weyl factor at the boundary. Indeed comparing the boundary condition (\ref{eq:linCBCs}) with the metric perturbations (\ref{eqn: spherical l>2 ansatz}) evaluated at the boundary yields
\begin{equation}
\delta\bomega=-\frac{\ell}{4r}\left[l(l+1)\left(1-\frac{2r^{2}}{\ell^{2}}\right)+2r^{2}\partial_{t}^{2}+2\left(1-\frac{r^{2}}{\ell^{2}}\right)^{2}r\partial_{r}\right]\Phi^{(S)}\mathbb{S}\biggr|_{r=\mathfrak{r}}\;. 
\label{eq: weyl scalar}
\end{equation}
In this case, the conformal class of the boundary metric is preserved by construction, and so in order to keep the trace of the extrinsic curvature unchanged, scalar perturbations must satisfy,
\begin{eqnarray}\label{eq: deltaK scalar}
	\mathcal{F}_l(K\ell\,,  \omega\ell) \,\equiv\,\left.\left(\frac{a_1}{r^4}+\frac{a_2}{r^2}\partial_t^2 - 2 \partial_t^4\right)\Phi^{(S)}  + \left(\frac{a_3}{r^2} -2 \partial_t^2\right)\left(1-\frac{r^2}{\ell^2}\right)^2\frac{\partial_r \Phi^{(S)}}{r} \right|_{r=\frakr} = 0 \, ,
 \end{eqnarray}
 where
 \begin{eqnarray}
     \begin{cases}
	& a_1  \= l(l+1)\left(1-\frac{r^2}{\ell^2}\right)\left(3-\frac{2r^2}{\ell^2}-2l(l+1)\left(1-\frac{r^2}{\ell^2}\right)\right) \, , \\
	& a_2  \= 4-\frac{2r^2}{\ell^2} - 4 l(l+1)\left(1-\frac{r^2}{\ell^2}\right) \, , \\
	& a_3  \= 4-\frac{2r^2}{\ell^2}-l(l+1)\left(3-\frac{2r^2}{\ell^2}\right) \, .
    \end{cases}
\end{eqnarray}
Given the non-trivial Weyl factor at the boundary, in order to satisfy \eqref{eq: linearised 2.12}, scalar perturbations must have a non-vanishing conformal stress tensor. Assuming \eqref{eq: deltaK scalar}, the components of the stress tensor can be written in terms of Weyl factor $\delta\bomega$ as 
\beq
\begin{cases} 
8\pi G_{N}\delta T_{tt}=-\frac{2\sqrt{1-\frac{\mathfrak{r}^{2}}{\ell^{2}}}}{l(l+1)\mathfrak{r}}\mathcal{G}\;\tilde{\nabla}^{2}\delta\bomega\;,\\
8\pi G_{N}\delta T_{ti}=-\frac{2\mathfrak{r}}{l(l+1)\sqrt{1-\frac{\mathfrak{r}^{2}}{\ell^{2}}}}\mathcal{G}\;\partial_t \tilde{\nabla}_{i}\delta\bomega\;,\\
8\pi G_{N}\delta T_{ij}=\frac{2\mathfrak{r}\left[l(l+1)\left(1-\frac{\mathfrak{r}^{2}}{\ell^{2}}\right)+2\mathfrak{r}^{2}\partial_{t}^{2}\right]\mathcal{G}}{l(l+1)(l-1)(l+2)\left(1-\frac{\mathfrak{r}^{2}}{\ell^{2}}\right)^{3/2}}\left(\tilde{\nabla}_{i}\tilde{\nabla}_{j}-\frac{1}{2}\tilde{g}_{ij}\tilde{\nabla}^{2}\right)\delta\bomega-\frac{\mathfrak{r}\;\mathcal{G}}{l(l+1)\sqrt{1-\frac{\mathfrak{r}^{2}}{\ell^{2}}}}\tilde{g}_{ij}\tilde{\nabla}^{2}\delta \bomega\;,
\end{cases}
\label{eq:compsdeltaTweyl}\eeq
where we have defined the differential operator $\mathcal{G}\equiv (l(l+1)-1)\left(1-\frac{\mathfrak{r}^{2}}{\ell^{2}}\right)-(1-\mathfrak{r}^{2}\partial_{t}^{2})$.
It is easy to verify the linearised stress-tensor is traceless and transverse with respect to the background $\bar{g}_{mn}$.
Further, 
\beq 64\pi^{2}G_{N}\bar{T}^{mn}\delta T_{mn}=-\frac{2\mathcal{G}}{\mathfrak{r}^{2}\left(1-\frac{\mathfrak{r}^{2}}{\ell^{2}}\right)}\delta\bomega\;,\label{eq:TdeltaTweyl}\eeq
which self-consistently recovers the equation of motion for the linearised Weyl factor \eqref{eq: linearised 2.12}.

\textbf{Gauge-invariant observables.} Given these metric perturbations, it is useful to introduce a bulk geometric quantity which is gauge-invariant at the linearised level. Consider the following quantity,
\begin{equation}
    \hat R_{\mu\nu\rho\sigma}[g_{\mu\nu}]\equiv R_{\mu\nu\rho\sigma}[g_{\mu\nu}]-\frac{1}{\ell^2}\left(g_{\mu\rho}g_{\nu\sigma}-g_{\mu\sigma}g_{\nu\rho}\right) \, ,
\end{equation}
where we subtract from the Riemann tensor a term that is the Riemann tensor of a maximally symmetric spacetime. Then, by definition, $\hat R_{\mu\nu\rho\sigma}$ vanishes when the metric is isomorphic to $\bar g_{\mu\nu}$. 

Next, we consider this subtracted tensor at the linearised level. Note that this is invariant under an arbitrary linearised diffeomorphism $\xi^\mu$, 
\begin{equation}
    \hat {R}_{\mu\nu\rho\sigma}\left[\bar g_{\mu\nu}+\varepsilon h_{\mu\nu}+\varepsilon \mathcal{L}_{\xi}\bar g_{\mu\nu}\right]=\hat {R}_{\mu\nu\rho\sigma}\left[\bar g_{\mu\nu}+\varepsilon h_{\mu\nu}\right] + \mathcal{O}(\varepsilon^2) \,,
\end{equation}
where $\mathcal{L}_{\xi}$ is the Lie derivative with respect to the vector field $\xi^\mu$. This allows us to use the linearised $\hat R_{\mu\nu\rho\sigma}$ as a measure of the bulk gravitational perturbations.\footnote{In fact, $\hat R_{\mu\nu\rho\sigma}$ admits the following physical interpretation. The geodesic deviation equation between two nearby geodesics with a tangent vector $T^\mu$ is given by
\begin{equation}
    a^\mu = \frac{1}{\ell^2}X^\mu + \hat R^\mu{}_{\nu\rho\sigma}T^\nu T^\rho X^\sigma \, ,
\end{equation}
where $X^\mu$ is an infinitesimal displacement between the geodesic and $a^\mu\equiv T^\nu\nabla_\nu(T^\rho \nabla_\rho X^\mu)$ denotes its acceleration along $T^\mu$. The first term on the right hand side describes the gravitational force due to the expansion of spacetime, while the second term, proportional to $\hat R_{\mu\nu\rho\sigma}$, describes the tidal force caused by the perturbations.}

Using the ansatz \eqref{eqn: spherical l>2 ansatz}, it is straightforward to show, at the linearised level,
\begin{equation}
    \hat R_{rtti}= \varepsilon \, \frac{l(l+1)-2}{4r}\left(-2\ell\partial_t\Phi^{(V)}\mathbb{V}_i+\left(1-\frac{r^2}{\ell^2}\right)\ell \partial_r\Phi^{(S)}\tilde\nabla_i \mathbb{S}\right)\,.
\end{equation}
Similarly, another component of this subtracted Riemann tensor,
\begin{equation}
    \hat R_{trtr}= -\varepsilon \frac{l(l+1)\left(l(l+1)-2\right)\ell}{4r^3} \Phi^{(S)}\mathbb{S}
\end{equation}
was used in \cite{Liu:2025xij} (up to field redefinitions) to characterise gauge-invariant scalar perturbations.

\textbf{Physical diffeomorphisms.} The Kodama-Ishibashi formalism is a method to obtain gauge-invariant bulk modes. The description presented is valid for $l\geq 2$. For any $l\geq0$, in the presence of a boundary, in principle, additional boundary modes can exist that are locally diffeomorphisms, but do not satisfy the boundary conditions for allowed diffeomorphisms. 

It can be verified, however, that there are no such physical diffeomorphisms for conformal boundary conditions for $l\geq 1$. This result easily generalises from the analysis done for physically diffeomorphic perturbations around AdS$_4$ in \cite{Anninos:2024xhc}. 

The only physical diffeomorphism is the spherically symmetric one. In fact, one can obtain a non-linear equation for this $l=0$ mode \cite{Liu:2024ymn,Galante:2025tnt}, that in dS$_4$ can be written in terms of its conformal energy $E_\text{conf}$ as
\begin{equation}
\label{eqn: brane eqn}
    \partial_t^2\bomega =  - \frac{1}{2}(\partial_t \bomega)^2-\frac{1}{2\r^2}\left(1-\frac{\r^2}{\ell^2}\right) - \frac{3 G_N^2 E_\text{conf}^2}{2\r^4} e^{-4\bomega} + \left(1-\frac{\r^2}{\ell^2}\right) \left(K^2 + \frac{9}{\ell^2}\right)\frac{e^{2\bomega}}{6} \,,
\end{equation}
where, following \eqref{eqn: def Econf}, 
\begin{equation}\label{Econfzero}
    E_\text{conf}= \frac{\r\sqrt{1-\frac{\r^2}{\ell^2}}}{3G_N}  \left( K \r  e^{3 \bomega}-3 \sigma e^{2 \bomega} \sqrt{1  - \frac{\r^2}{\ell^2} e^{2\bomega}+ \frac{\left(\r \partial_t\bomega\right)^2}{1-\frac{\r^2}{\ell^2}} }\right)\,,
\end{equation}
is constant in time. In   (\ref{Econfzero}) we have considered the conformal energy of hypersurfaces locally embedded in the static patch. (More generally, there may also be other hypersurfaces, such as dS$_3$ hypersurfaces of dS$_4$, for the same value of $K\ell$. Being umbilic, these would have $E_\text{conf}=0$). Indeed, computing the conformal stress-tensor for this mode, we obtain 
\begin{equation}
    T_{mn}dx^m dx^n = \frac{E_\text{conf}}{8\pi \r^2\sqrt{1-\frac{\r^2}{\ell^2}}} \left(2\left(1-\frac{\r^2}{\ell^2}\right)dt^2 +\r^2 d\Omega_2^2\right)  \,,
\end{equation}
such that \eqref{eqn: brane eqn} is compatible with \eqref{eq: 2.12}.
Linearising around the $\bomega = 0$ solution gives, 
\begin{equation}
   \r^2 \partial_t^2 \delta \bomega (t) = \left( 2 - \frac{\r^2}{\ell^2} \right) \delta \bomega (t) \,,
   \label{eq: lin phys diff eom}
\end{equation}
with solution $\delta \bomega (t) = e^{- i \omega^{(0)} t }$, for
\begin{equation}
    \omega^{(0)} \r = \pm i \sqrt{ 2 - \frac{\r^2}{\ell^2} } \,.
\end{equation}
The linearised conformal stress tensor $\delta T_{mn}$ for this solution vanishes identically.

Using the relation \eqref{eq:KdS} between $K$ and $\mathfrak{r}$, we can expand the frequency of the physical diffeomorphism in the stretched horizon limit, obtaining 
\begin{equation}
   \tilde{\omega}^{(0)}  = \pm \frac{2\pi i}{\tilde\beta} \left[ 1+ \frac{1}{(K \ell)^2}-\frac{9}{2 (K\ell)^4}+\mathcal{O} \left(\left(K\ell\right)^{-6}\right) \right] \,, \quad\quad \tilde{\omega}^{(0)} \equiv \frac{\omega^{(0)} \r}{\sqrt{1-\frac{\r^2}{\ell^2}}}~.
    \label{eq: frequency l=0}
\end{equation}
Here we have written the solution in terms of the cosmic horizon inverse conformal dimensionless temperature $\tilde \beta \equiv \tfrac{\beta_{\text{dS}}}{\r}\sqrt{1-\tfrac{\r^2}{\ell^2}}$ as measured by the boundary clock. We have also defined  $\beta_{\text{dS}} \equiv 2\pi \ell$. The dimensionless frequency $\tilde{\omega}^{(0)}$ is that measured by the conformal boundary clock. The leading form of $\tilde{\omega}^{(0)}$ is reminiscent of complex frequencies appearing in the quantum chaos black hole literature \cite{Maldacena:2016upp,Knysh:2024asf}. Note that the leading correction away from $\omega^{(0)} \ell = +i$ is positive and all terms in the expansion are analytic.\footnote{We note that in Euclidean signature, with periodically identified Euclidean time, at least at the linearised level and for the pole patch, there will be a corresponding  thermal solution at the (conformal) temperatures $\tilde{\beta}_n \equiv \tfrac{2\pi n}{|\tilde{\omega}^{(0)}|}$ with constant $K$ and $n\in \mathbb{Z}^+$. This is due to the specific functional form  $\delta \bomega (t) = e^{- i \omega^{(0)} t}$ which allows periodic behaviour for Euclidean time. These configurations appear to spontaneously break the $U(1)$ thermal isometry to some discrete subgroup. It will be interesting to understand what thermodynamic consequences these modes might have on the pole patch. A similar, but more restrictive, situation holds for the cosmic patch, see Eq. $(2.10)$ of \cite{Galante:2025tnt} for a related discussion.}

As this solution is a physical diffeomorphism, the analysis of this spherically symmetric mode is the same both for the cosmic and the pole patch (up to the sign $\sigma$ in the conformal energy).

\subsection{Pole and cosmic patch} \label{ssec:polecosmicpatch}

To proceed with a detailed description of the linearised dynamics, we must solve the master field equation of motion \eqref{eq: master equation}. We will assume the master fields admit a Fourier decomposition, $\Phi^{(S/V)}(t,r)=\Re \, e^{-i\omega t} \, \phi^{(S/V)}(r)$. We further demand that the master fields  obey the linearised conformal boundary conditions; imposing \eqref{eq: vector bdy condition} and  \eqref{eq: deltaK scalar} for the vector and scalar sectors, respectively. This will lead to a set of allowed frequencies $\omega$. Further, since solutions to (\ref{eq: master equation}) take a different form for the pole and the cosmic patch, we consider each one separately.

\subsubsection{Pole patch}\label{subsectionPolePatch}

\begin{figure}[h!]
  \centering
  \begin{minipage}[b]{0.48\textwidth}
    \begin{subfigure}[b]{\textwidth}
      \includegraphics[width=\linewidth]{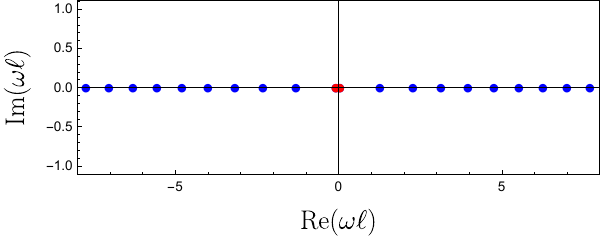}
      \caption{Real scalar modes.}
    \label{subfig:realscalmodes}\end{subfigure}

    \vspace{5mm} 

    \begin{subfigure}[b]{\textwidth}
      \includegraphics[width=\linewidth]{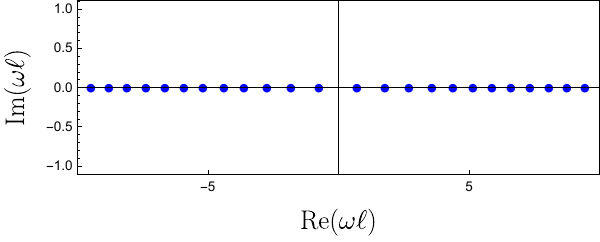}
      \caption{Real vector modes.}
    \end{subfigure}
  \end{minipage}
  \hfill
  \begin{minipage}[b]{0.48\textwidth}
    \begin{subfigure}[b]{\textwidth}
     \begin{picture}(0,0)
  \put(133,196){\makebox(0,0){\scalebox{0.65}{\rotatebox{90}{$K\ell\to-\infty$}}}}
  \put(180,196){\makebox(0,0){\scalebox{0.65}{\rotatebox{38}{$K\ell\to\infty$}}}}
   \put(133,55){\makebox(0,0){\scalebox{0.65}{\rotatebox{-90}{$K\ell\to-\infty$}}}}
  \put(180,55){\makebox(0,0){\scalebox{0.65}{\rotatebox{-35}{$K\ell\to\infty$}}}}
\includegraphics[height=\linewidth,width=\linewidth]{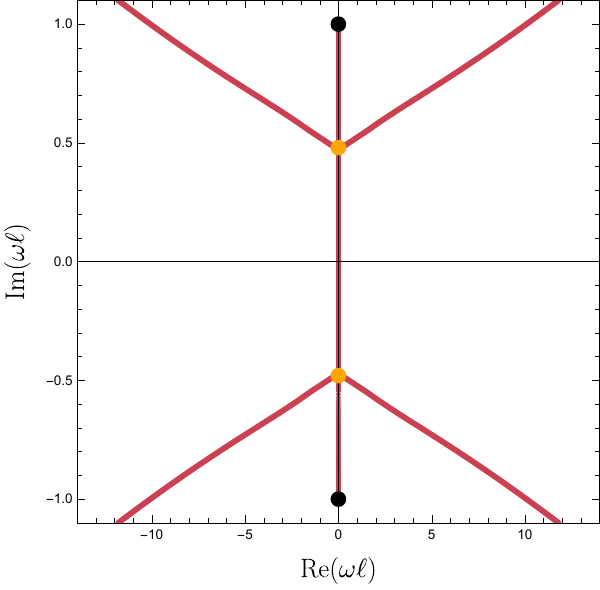}
\end{picture}
      \caption{Complex scalar modes.}
    \end{subfigure}
  \end{minipage}
  \caption{Modes in scalar and vector sector of metric perturbations in the pole patch. Here angular momentum is fixed to be $l=10$. In (a) and (b) we show real modes for fixed $K\ell\approx -70$ $(\mathfrak{r}=.9999 \ell)$. Further, in (a) we highlight (in red) a pair of real modes lying near the origin, $\omega\ell\approx\pm0.075$.
 There are also complex modes that we display in (c). In the pole patch, allowed frequencies appear in complex conjugates pairs. The solid lines represent how modes evolve as we change $K\ell$ (with fixed $l=10$). As $K\ell \to -\infty$ (the stretched horizon limit), frequencies with positive (negative) imaginary part converge to $\omega \ell=+i$ ($\omega \ell=-i$), shown as a large black dot. Away from that limit, $|\text{Im}(\omega \ell)|$ decreases until reaching a critical value $(|K\ell|)_{\text{cr}}\approx 20.75$ (orange dot) where $\omega \ell$ begins to develop a real component. The purely imaginary modes eventually cross the real line near the origin, i.e., the red dots in (a).}
  \label{fig:PPmodes}
\end{figure}

For the pole patch, we select the solution to the master field equation (\ref{eq: master equation}) that is regular at the origin. This yields, for each allowed frequency,
\begin{equation} \label{polepatchgeneral}
    \Phi^{(S/V)} = 
    \Re \, e^{-i\omega t}\left(1-\frac{r^2}{\ell^2}\right)^{-{i\omega\ell/2}}\left(\frac{r^2}{\ell^2}\right)^{\frac{l+1}{2}} {}_2F_1\left(\frac{l - i \omega \ell +2}{2},\frac{l - i \omega \ell +1}{2}, \frac{2l+3}{2},\frac{r^2}{\ell^2}\right) \, .
\end{equation}
Imposing boundary conditions \eqref{eq: vector bdy condition} and \eqref{eq: deltaK scalar} selects the allowed frequencies in the vector and scalar sector, respectively. In figure \ref{fig:PPmodes} we plot the imaginary and real parts of the mode frequencies for both sectors. We observe that only real frequencies are allowed in the vector sector of metric perturbations. Meanwhile, in the scalar sector, in addition to the normal real frequencies, we find complex frequencies, some of which have positive imaginary part. In the limit where the boundary approaches the cosmological horizon $K\ell\to-\infty$, frequencies with positive (negative) imaginary part coalesce at $\omega \ell=+i$ ($\omega \ell=-i$). As the boundary moves away from the horizon, $|\text{Im}(\omega \ell)|$ decreases until it reaches a critical value $(K\ell)_{\text{cr}}$, where $\omega \ell$ begins to develop a real component. 
Note that by plugging in \eqref{polepatchgeneral} into \eqref{eq: deltaK scalar} it is easy to verify that if $\omega \ell$ is a solution, minus its complex conjugate is also a solution, implying a symmetry about the imaginary axis. Further, the pole patch master field is invariant under $\omega \ell \rightarrow -(\omega\ell)$.
Together, the two remarks imply a symmetry about the real axis.

\subsubsection{Cosmic patch}

\begin{figure}[h!]
  \centering

  \begin{subfigure}[b]{0.48\textwidth}
    \begin{picture}(230,230)
      \put(130,168){\makebox(0,0){\scalebox{0.55}{\rotatebox{90}{$K\ell\to\infty$}}}}
      \put(180,138){\makebox(0,0){\scalebox{0.55}{\rotatebox{-20}{$K\ell\to-\infty$}}}}
      \put(142,48){\makebox(0,0){\scalebox{0.55}{\rotatebox{-57}{$K\ell\to-\infty$}}}}
      \includegraphics[width=\linewidth]{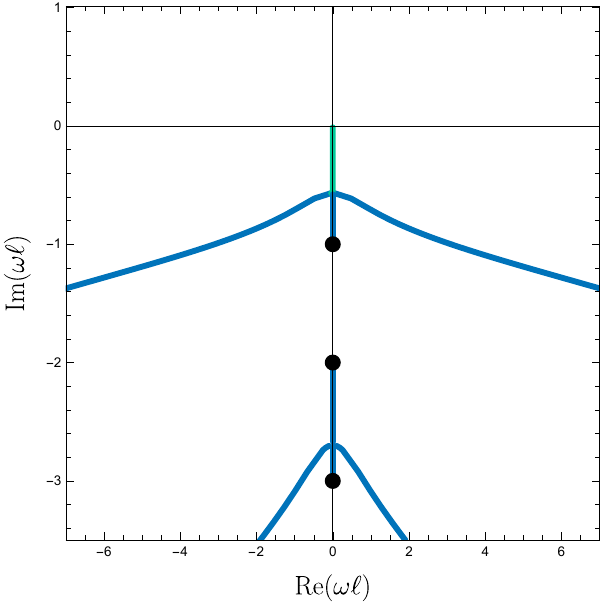}
    \end{picture}
    \caption{Complex vector modes.}
    \label{fig:complex-vector}
  \end{subfigure}
  \hfill
  \begin{subfigure}[b]{0.48\textwidth}
    \begin{picture}(230,230)
      \put(130,207){\makebox(0,0){\scalebox{0.65}{\rotatebox{90}{$ K\ell\to\infty$}}}}
      \put(180,186){\makebox(0,0){\scalebox{0.65}{\rotatebox{7}{$K\ell\to-\infty$}}}}
      \put(132,150){\makebox(0,0){\scalebox{0.65}{\rotatebox{-90}{$K\ell\to\infty$}}}}
      \put(181,60){\makebox(0,0){\scalebox{0.65}{\rotatebox{-41}{$K\ell\to-\infty$}}}}
      \includegraphics[width=\linewidth]{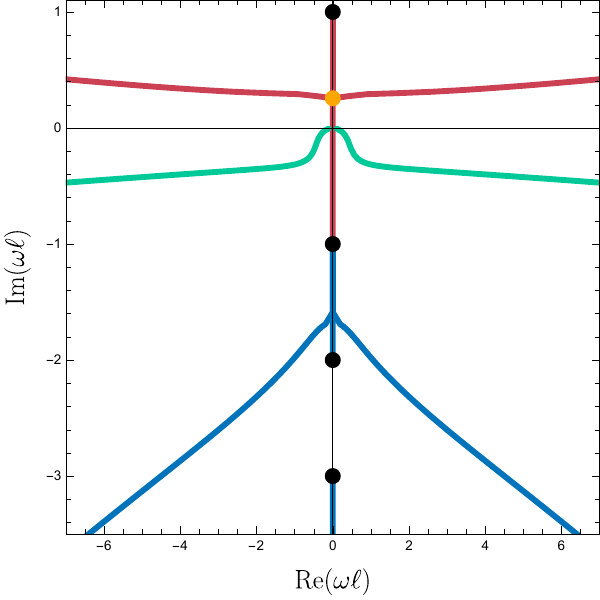}
    \end{picture}
    \caption{Complex scalar modes.}
    \label{fig:complex-scalar}
  \end{subfigure}

  \caption{Modes in the scalar and vector sector of metric perturbations in the cosmic patch. Here angular momentum is fixed to be $l=10$. \emph{Left.} Complex vector modes all have negative imaginary part. In the cosmological horizon limit the frequencies become purely imaginary; fluid shear modes (green), and `new' quasinormal modes (blue). Not shown are the standard dS QNMs which are recovered in the strict worldline limit. \emph{Right.} Complex scalar modes can have positive imaginary part. Modes with positive imaginary part coalesce at $\omega=+i$ in cosmological horizon limit (red curve). Away from the strict cosmological horizon limit, the $\omega=\pm i$ modes develop a real component at a critical value of $K\ell$ (center orange dot). Fluid sound modes (green) develop negative imaginary component as the boundary approaches worldline, and disappear in the strict $K\ell\to\infty$ limit. New (purely imaginary) QNMs arise as the boundary approaches cosmological horizon; real components develop as the boundary moves away from the cosmological horizon  (lower blue).}
  \label{fig:CPmodes}
\end{figure}

For the cosmic patch, we select the solution to (\ref{eq: master equation}) that is purely outgoing at the horizon. For each frequency,
\begin{equation}\label{PhiS}
    \Phi^{(S/V)} = \Re\, e^{-i\omega t}\left(1-\frac{r^2}{\ell^2}\right)^{-i\omega\ell/2} \left(\frac{r^2}{\ell^2}\right)^{i\omega\ell/2} {}_2F_1\left(-l-i\omega \ell,1+l-i\omega\ell;1-i\omega\ell;\frac{1}{2}-\frac{\ell}{2r}\right) \, .
\end{equation}
Imposing boundary conditions \eqref{eq: vector bdy condition} and \eqref{eq: deltaK scalar} on $\Phi^{(S/V)}$ selects the allowed frequencies in the vector and scalar sector, respectively. This problem was studied in detail in \cite{Anninos:2024wpy}, which we now briefly summarize; see figure \ref{fig:CPmodes} for an illustration.

\textbf{Vector modes.} Numerically scanning \eqref{eq: vector bdy condition} in the plane of complex $\omega$, there exist only complex vector modes with negative imaginary part, see figure \ref{fig:CPmodes}(\subref{fig:complex-vector}). In the cosmological horizon limit, $K\ell\to+\infty$ there are purely imaginary modes, while real deviations from the purely imaginary modes develop as one moves toward the worldline, $K\ell\to-\infty$.  
Displayed in  figure \ref{fig:CPmodes}(\subref{fig:complex-vector}) are two sets of modes: purely imaginary shear modes (green) and a tower of new purely imaginary quasinormal modes and deviations thereof (bottom, blue). The shear modes were identified in \cite{Anninos:2011zn} and, together with the sound modes in the scalar sector (see below), reflect the emergence of linearised conformal fluid dynamical behaviour near the horizon \cite{Anninos:2024wpy}. Not displayed (recall that the figure has $l=10$), but also allowed, are the familiar quasinormal modes of empty dS$_{4}$ \cite{Lopez-Ortega:2006aal}, 
\begin{equation} \label{qnmdeSitter}
    \omega^{\text{qnm}}\ell=-i(l+n+1) \, , \quad\quad  n \in \mathbb{N} \,,
\end{equation}
which arise in the strict worldline limit. 

\textbf{Scalar modes.} Scalar modes are found by scanning the boundary condition \eqref{eq: deltaK scalar}, in which, notably, there also exist complex frequencies with positive imaginary part, see Figure \ref{fig:CPmodes}(\subref{fig:complex-scalar}). Firstly, there are modes with purely (positive and negative) imaginary parts.
As the boundary moves away from the horizon, a real component to the frequency develops (at a critical value of $K\ell$) with positive imaginary part. In the strict cosmological horizon limit, all modes with positive imaginary part coalesce at either $\omega\ell=\pm i$. Additionally, there are (relativistic) conformal fluid-like sound modes, with a dominant real component in the large-$K\ell$ limit, and negative imaginary deviations. In the strict cosmological horizon limit, the sound modes disappear. 
Approaching the worldline, one also has real component deviations from the standard dissipative quasinormal modes of dS$_4$ (which are recovered in the strict worldline limit).

\subsection{Characterising the radial profile} \label{sec: radial profile}

In the following sections, we will characterise the allowed frequencies (both for the pole and the cosmic patch) in two regimes. First, in section \ref{sec: large l} we study the large angular momentum-$l$ limit. Second, in sections \ref{sec: large K pole} and \ref{sec: large K cosmic} we study the stretched horizon limit, where the trace of the extrinsic curvature is large. In these limits, some of the allowed frequencies have a radial profile that localises very close to the boundary $\Gamma$. Moreover, in some cases, the master field at the boundary will diverge. This can be clearly seen, for instance, from the pole patch master field \eqref{polepatchgeneral}: when $r\to\ell$, the hypergeometric function diverges. In order to compare if and how the different modes localise close to the boundary, we define the following quantity,
\begin{equation}
    \tilde{\phi}^{(S/V)} (s) \equiv  \frac{\phi^{(S/V)}(s)}{\left. \phi^{(S/V)} \right|_{\Gamma}} \,, \quad s \equiv \left| \int_\r^r \frac{dr}{f(r)} \right| =  \ell |\arcsin(\mathfrak{r}/\ell)-\arcsin(r/\ell)| \,.
    \label{eq: normalised phi}
\end{equation}
Here $s$ is the proper distance from the boundary ($s=0$), along a line of constant $(t, \theta, \varphi)$. Note that by definition $\tilde \phi^{(S/V)}$ at the boundary is equal to one. From the worldline to the horizon the proper distance corresponds to $s=\tfrac{\pi\ell}{2}$. For small displacements $\delta s$, the proper distance close to the boundary can be approximated as
\begin{equation} \label{eq: short proper distance}
    \delta s=\frac{|\mathfrak{r}-r|}{\sqrt{1-\frac{\mathfrak{r}^2}{\ell^2}}}+\mathcal{O}((\mathfrak{r}-r)^2) \,.
\end{equation}
Localised modes have a profile $\tilde\phi^{(S/V)}$ that rapidly decays to a very small value in the interior of the spacetime. To characterise that value, we define a radial width $D$ as the proper distance from the boundary at which the radial profile decays to a particular fraction $0<\alpha<1$, 
\begin{equation}
    \tilde\phi^{(S/V)} (s=D) = \alpha \,.
\end{equation}
This radial width certainly depends on $\alpha$. For a given frequency whose radial profile localises, we will numerically compute $D$ for various $\a$. To compare between different allowed frequencies, we fix $\alpha$ and look at how the radial width changes with the value of $K\ell$ at the boundary.

\section{The large angular momentum limit} \label{sec: large l}

We start by considering the large angular momentum $l$ limit of the metric perturbations around pure dS$_4$ in both pole and cosmic patches, keeping $K\ell$ fixed.  
Our strategy is to first solve the master equation \eqref{eq: master equation} in the large angular momentum limit, and then impose the conformal boundary conditions. 

\textbf{Master field dynamics as an effective 1D description.} Recall both the vector and the scalar master fields obey the bulk equation of motion (\ref{eq: master equation}), repeated here for convenience,
\begin{equation}
\left(1-\frac{r^{2}}{\ell^{2}}\right)^{2}\partial_{r}^{2} \Phi^{(S/V)} -\frac{2r}{\ell^{2}}\left(1-\frac{r^{2}}{\ell^{2}}\right)\partial_{r} \Phi^{(S/V)}-\left[\frac{l(l+1)}{r^{2}}\left(1-\frac{r^{2}}{\ell^{2}}\right)+ \partial_t^2 \right] \Phi^{(S/V)} =  0~.
\label{eq: master field large l again}
\end{equation}
To analyse the large-$l$ limit, it is convenient to define the tortoise coordinate $r^*$,
\begin{equation}\label{eqn: def of R}
    r^*(r) \equiv \int_0^r \frac{dr'}{1-\frac{r'^2}{\ell^2}} = \ell \tanh^{-1}\frac{r}{\ell} \,.
\end{equation}
In this coordinate, the worldline ($r=0$) and the cosmological horizon ($r=\ell$) correspond to $r^*=0$ and $r^*\to\infty$, respectively. It will prove useful to rewrite the angular momentum $l$ in terms of a new parameter $\L$,
\begin{equation}
    \L \equiv \sqrt{l(l+1)} \, , \qquad \L = l + \frac{1}{2} - \frac{1}{8l}+\mathcal{O}(l^{-2}) \, ,
\end{equation}
such that, at leading order, the large-$\L$ limit is equivalent to the large-$l$ limit. Using these variables and plugging in $\Phi^{(S/V)}=\Re e^{-i\omega t}\phi^{(S/V)}(r^*)$, the master field equation of motion \eqref{eq: master field large l again} can be rewritten as
\begin{equation}\label{eqn: Meqn dS large l}
    \left(-\frac{1}{\L^2}\partial_{r^*}^2 + V_{\text{eff}}(r^*)\right)\phi^{(S/V)} (r^*)= \frac{\omega^2}{\L^2}\phi^{(S/V)} (r^*) \, , \qquad V_{\text{eff}}(r^*) \equiv \frac{1-\frac{r(r^*)^2}{\ell^2}}{r(r^*)^2} = \frac{1}{\ell^2 \sinh^2\frac{r^*}{\ell}} \, .
\end{equation}
This equation takes the form of a one-dimensional Schr\"{o}dinger equation for a single particle with energy $\tfrac{\omega^2}{\L^2}$ subject to the potential $V_{\text{eff}}(r^*)$. The potential is a monotonically decreasing function over the range $r^{*}\in[0,\infty]$, diverging at the worldline ($r^*=0$) and vanishing at the cosmological horizon ($r^*\to\infty$). 

We also rewrite the linearised boundary conditions in terms of the new variables $(r^{*},\L)$.
The boundary condition for vector perturbations \eqref{eq: vector bdy condition} becomes,
\begin{equation}\label{eqn: bdry cond vec dS large l}
    \left. \left(\tanh\frac{r^*}{\ell}V_\text{eff} (r^*)+\partial_{r^*} \right)\phi^{(V)} (r^*) \right|_{r^*=\r^*} = 0 \, ,
\end{equation}
 where the boundary is at $r^* = \r^*$. 
The scalar boundary condition \eqref{eq: deltaK scalar} is given by,
\begin{equation}\label{eqn: bdry cond dS large l}
   \left.\left(a_1 +a_2\right) \phi^{(S)} (r^*)+\left(a_3+a_4\right) \ell\partial_{r^*} \phi^{(S)} (r^*)\right|_{r^*=\r^*} = 0 \, ,
 \end{equation}
where 
\begin{equation}\label{eqn: bdry cond dS large l coeffs}
    \begin{cases}
        a_1=-2\ell^4\left(\omega^2-\mathfrak{L}^2V_\text{eff} (r^*)\right)^2 \, , \\
        a_2 = -2 \ell^2\left(\omega^2-\mathfrak{L}^2 V_\text{eff} (r^*)\right)\left(1+2\text{csch}^2\frac{r^*}{\ell}\right)-\ell^2\mathfrak{L}^2V_\text{eff} (r^*)\left(1+\text{csch}^2\frac{r^*}{\ell}\right) \, , \\
        a_3 = 2\ell^2\left(\omega^2-\mathfrak{L}^2V_\text{eff}(r^*)\right)\text{csch}\frac{r^*}{\ell}\text{sech}\frac{r^*}{\ell}-\ell^2\mathfrak{L}^2V_\text{eff}(r^*)\coth\frac{r^*}{\ell}\, , \\
        a_4 = \left(3+\cosh\frac{2r^*}{\ell}\right)\text{csch}^3\frac{r^*}{\ell}\text{sech}\frac{r^*}{\ell} \, .
    \end{cases}
\end{equation}

So far we have just re-written the problem in a suggestive way. In the following, we consider taking the large-$\L$ limit of this system, while keeping $K\ell$ (or $\r^*$) fixed. Using the Schr\"{o}dinger equation analogy, we observe that $\L$ plays the role of $\hbar^{-1}$ in \eqref{eqn: Meqn dS large l}. Therefore, the large-$\L$ limit can be analysed using a standard WKB approximation, which we carry out in appendix \ref{app: WKB}.

\textbf{General solutions to the bulk equation.} Given $\omega$, we define a turning point $r^*_\text{t}$ as the location at which the potential is equal to the energy, 
\begin{equation}\label{eqn: def R_t}
    \omega^2= \L^2V_\text{eff}(r^*_\text{t}) \, .
\end{equation}
Since $V_\text{eff}(r^*)$ is a monotonic function, there exists a unique real solution of $r^*_\text{t}$. The large-$\L$ solution to \eqref{eqn: Meqn dS large l} can be obtained in two different regimes, depending on whether $r^*$ is near or away from $r^*_\text{t}$. 

In the region away from $r^*_\text{t}$, the general solution is well-approximated by
\begin{equation}\label{eqn: WKB sol 1}
    \phi^{(S/V)}(r^*) = \tilde c_1 \frac{\exp\left(i \mathfrak{L} \int^{r^*} dx\, \sigma_0^{+}(x)\right)}{\left|V_\text{eff}(r^*)-V_\text{eff}(r^*_\text{t})\right|^{1/4}}+\tilde c_2 \frac{\exp\left(i \mathfrak{L} \int^{r^*} dx\, \sigma_0^{-}(x)\right)}{\left|V_\text{eff}(r^*)-V_\text{eff}(r^*_\text{t})\right|^{1/4}} + \mathcal{O}(\mathfrak{L}^{-1}) \, ,
\end{equation}
for constants of integration $\tilde c_{1,2}$, and $\sigma^{\pm}_0(r^*)$ are two branches of solutions to
\begin{equation}
    (\sigma^{\pm}_0(r^*))^2  = V_\text{eff}(r^*_\text{t})-V_\text{eff}(r^*) \, .
\end{equation}
The solution \eqref{eqn: WKB sol 1} is only valid for $|r^*-r^*_\text{t}|\gg \L^{-2/3}$, i.e., far away from the turning point. See appendix \ref{app: WKB} for a derivation. The precise power $\L^{-2/3}$ will play a predominant role in the following discussion.

On the other hand, in the near turning point region, where $|r^*-r^*_\text{t}|\sim \L^{-2/3}$, the general solution is instead given by a linear combination of Airy functions Ai$(x)$ and Bi$(x)$,
\begin{equation}\label{eqn: WKB sol 2}
    \phi^{(S/V)}(r^*) = \tilde c_3 \text{Ai}\left(\left(-\mathfrak{L}^2 \partial_{r^*}V_\text{eff}(r^*_\text{t})\right)^{1/3}\left(r^*_\text{t}-r^*\right)\right)+\tilde c_4  \text{Bi}\left(\left(-\mathfrak{L}^2 \partial_{r^*}V_\text{eff}(r^*_\text{t})\right)^{1/3}\left(r^*_\text{t}-r^*\right)\right) + \mathcal{O}(\mathfrak{L}^{-1}) \, ,
\end{equation}
where $\tilde c_{3,4}$ are constants of    integration, and the cubic root is chosen to be real. 

\begin{figure}[t!]
\centering
 \includegraphics[width=.5\textwidth]{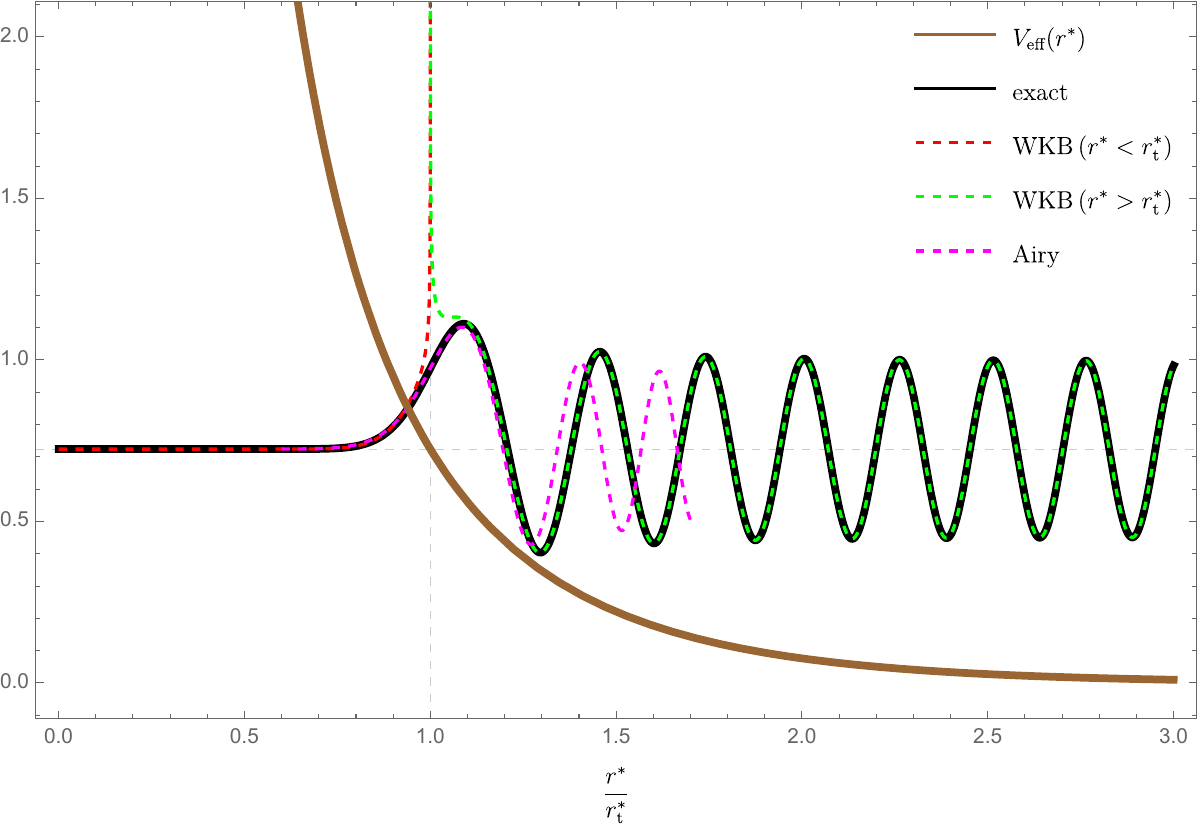}
\caption{Potential $V_{\text{eff}}(r^{*})$ in one-dimensional Schr\"{o}dinger equation effectively characterising the master field dynamics. Here we display $\omega/\L = (\sinh 1)^{-1}$ and $\mathfrak{L}=30$.}
\label{fig:effpotlarL}
\end{figure}

Requiring the solution to be smooth across $r^*_\text{t}$ imposes connecting conditions between \eqref{eqn: WKB sol 1} and \eqref{eqn: WKB sol 2}. In particular, we match \eqref{eqn: WKB sol 1} in the $|r^*-r^*_\text{t}|\to0$ limit to the asymptotic expansion of the Airy functions \eqref{eqn: WKB sol 2}. This results in relations among $\tilde{c}_i$. We perform this analysis in appendix \ref{app: WKB}, and show the full solution with the different approximations in figure \ref{fig:effpotlarL}.

\textbf{Boundary conditions when the boundary is far.} The large-$\mathfrak{L}$ limit also simplifies the boundary conditions \eqref{eqn: bdry cond vec dS large l} and \eqref{eqn: bdry cond dS large l}. As with the bulk solutions, it is useful to analyse these boundary conditions when the boundary is far or close to the turning point,  $|r^*_\text{t}-\r^*|\gg \mathfrak{L}^{-2/3}$ and $|r^*_\text{t}-\r^*|\sim \mathfrak{L}^{-2/3}$.

We first look at the boundary condition for the vector perturbations. To obtain the leading boundary conditions in this case with $|r^*_\text{t}-\r^*|\gg \mathfrak{L}^{-2/3}$, we plug in \eqref{eqn: WKB sol 1} to \eqref{eqn: bdry cond vec dS large l}. To leading order, this leads to a Neumann-like boundary condition,
\begin{equation}\label{eqn: lead large l vec bdry cond 1}
    \left.\partial_{r^*}\phi^{(V)} (r^*) \right|_{r^*=\r^*}= 0 \, ,
\end{equation}
where we have suppressed sub-leading contributions. 

Next, consider scalar perturbations. From \eqref{eqn: bdry cond dS large l coeffs} and assuming $\tfrac{\omega}{\mathfrak{L}}$ is finite, we see that $a_2$ and $a_4$ scale as $\mathfrak{L}^2$, while $a_1$ and $a_3$ scale as $\mathfrak{L}^4$. Hence, the large $\mathfrak{L}$ limit suppresses the $a_2$ and $a_4$ terms. Using \eqref{eqn: def R_t}, the boundary condition \eqref{eqn: bdry cond dS large l} becomes (to leading order), 
\begin{equation}\label{eqn: bdry cond dS large l 2}
    \left.2\mathfrak{L}^4\left(V_\text{eff}(r^*_\text{t})-V_\text{eff}(\r^*)\right)^2 \phi^{(S)} -\mathfrak{L}^2\left(4\left(V_\text{eff}(r^*_\text{t})-V_\text{eff}(\r^*)\right)\text{csch}\frac{2r^*}{\ell}-V_\text{eff}(\r^*)\coth\frac{\r^*}{\ell}\right)\partial_{r^*} \phi^{(S)}\right|_{r^*=\r^*}=0 \,.
\end{equation}
For the far away solution, we substitute \eqref{eqn: WKB sol 1} into \eqref{eqn: bdry cond dS large l 2}. Acting $\partial_{r^*}$ on \eqref{eqn: WKB sol 1} brings an additional factor of $\mathfrak{L}$, such that the second term scales as $\mathfrak{L}^3$. Then in the large-$\L$ limit, the boundary condition \eqref{eqn: bdry cond dS large l 2} is dominated by the first term, effectively making the boundary condition (to leading order) Dirichlet-like,
\begin{equation}\label{eqn: lead large l bdry cond 1}
    \left.\phi^{(S)} (r^*)\right|_{r^*=\r^*}=0\,.
\end{equation}

\textbf{Boundary conditions when the boundary is near the turning point.} A more involved class of solutions occur when the turning point is located near the boundary, $|r^*_\text{t}-\r^*| \sim \mathfrak{L}^{-2/3}$. This requires inserting \eqref{eqn: WKB sol 2} into \eqref{eqn: bdry cond vec dS large l} for vector perturbations and \eqref{eqn: bdry cond dS large l} for scalar perturbations. To analyse these solutions, we express both the solutions and frequencies in terms of the potential evaluated at the boundary, $V_\text{eff}(\r^*)$. Define a near-boundary radial coordinate
\begin{equation}\label{eqn: near bdry xi}
    \xi \equiv \left(-
    \mathfrak{L}^2 \partial_{r^*}V_\text{eff}(\r^*)\right)^{1/3}\left(\r^*-r^*\right)\,,
\end{equation}
where the cubic root is chosen such that $\xi \in \mathbb{R}$, and we note $\xi=0$ at the boundary. Inserting \eqref{eqn: near bdry xi} into \eqref{eqn: Meqn dS large l}, the large $\mathfrak{L}$ expansion of the left-hand-side of \eqref{eqn: Meqn dS large l} requires the frequencies to have the expansion of the form 
\begin{equation}\label{eqn: ansatz near Rt}
    \omega = \mathfrak{L}\sqrt{V_\text{eff}(\r^*)}+\mathfrak{L} \sum_{n=1}^\infty \omega_n \mathfrak{L}^{-2n/3}\, ,
\end{equation}
where $\omega_n$ is undetermined but independent of $r^*$ and $\mathfrak{L}$.
Using \eqref{eqn: near bdry xi} and \eqref{eqn: ansatz near Rt}, the near boundary solution \eqref{eqn: WKB sol 2} can be simply written as
\begin{equation}\label{eqn: WKB sol 3}
    \phi^{(S/V)}(\xi)  = c_1 \text{Ai}(\xi-\nu) + c_2 \text{Bi}(\xi-\nu) + \mathcal{O}\left(\mathfrak{L}^{-2/3}\right) \, , \qquad \nu \equiv \frac{2 \sqrt{V_\text{eff}(\r^*)} }{\left(-\partial_{r^*}V_\text{eff}(\r^*)\right)^{2/3}}\omega_1 \, .
\end{equation}
The derivation of this solution and its sub-leading corrections is detailed in appendix \ref{sec: derive Ai(xi-nu)}. Now we use \eqref{eqn: WKB sol 3} to analyse the boundary conditions. 

For the vector perturbations, plugging \eqref{eqn: WKB sol 3} into \eqref{eqn: bdry cond vec dS large l} yields again a Neumann-like boundary condition to leading order,
\begin{equation}\label{eqn: lead large l vec bdry cond 2}
    \left.\partial_\xi \phi^{(V)} (\xi)\right|_{\xi=0}=0 \,.
\end{equation}

For the scalar perturbations, by plugging in \eqref{eqn: WKB sol 3} to \eqref{eqn: bdry cond dS large l 2}, we now find that the first term is of the same order as the last term. The resulting boundary condition, to leading order, is
\begin{equation}\label{eqn: lead large l bdry cond 2}
    \left.4 \nu^2 \phi^{(S)} (\xi) - \partial_\xi \phi^{(S)} (\xi)\right|_{\xi=0}=0 \, .
\end{equation}

Together, equations (\ref{eqn: lead large l vec bdry cond 2}) and (\ref{eqn: lead large l bdry cond 2}) are the general boundary conditions in the near-turning point region. To complete the analysis, we need to further select  regularity conditions, depending on whether the solution corresponds to a pole or a cosmic patch solution. We do this below, treating each patch separately.

\subsection{Pole patch}\label{sec: pole large L}

We start by considering the pole patch solution in the large-$\mathfrak{L}$ limit, for the case in which the turning point is near the boundary, $|r^*_\text{t}-\r^*| \ll \mathfrak{L}^{-2/3}$. 

The pole patch solution is regular at $r^*=0$,  so
\eqref{eqn: WKB sol 3} must be exponentially decaying as $\xi \to +\infty$. As a result, the solution for the master field (including 
leading and first correction) is given by
\begin{equation}\label{eqn: large L pol sol}
    \phi^{(S/V)}(\xi) = \phi_0(\xi)+\left(\frac{\mathfrak{a}_1}{5}\xi \phi_0(\xi)-\left(\mathfrak{a}_1+\mathfrak{a}_2\xi^2+\frac{4\mathfrak{a}_2}{15}\left(\nu-\xi\right)\left(2\nu+3\xi\right)\right)\partial_\xi \phi_0(\xi)\right)\mathfrak{L}^{-2/3}+\mathcal{O}(\mathfrak{L}^{-4/3})\, ,
\end{equation}
where the function $\phi_0(\xi)$ and $\xi$-independent parameters $\mathfrak{a}_1$ and $\mathfrak{a}_2$ are 
\begin{equation}
    \phi_0(\xi)=c_1 \text{Ai}(\xi-\nu)\, , \qquad\mathfrak{a}_1 = \frac{\nu^2}{2^{4/3}f(\r)^{1/3}}+\frac{2^{1/3}\r \omega_2}{f(\r)^{1/6}} \, , \qquad \mathfrak{a}_2 = -\frac{2+f(\r)}{2^{4/3}f(\r)^{1/3}} \, .
\end{equation}
We recall that the parameter $\nu$ is related to $\omega_1$ via \eqref{eqn: WKB sol 3}, so that the solution to this order is completely specified by $\omega_1$ and $\omega_2$. The derivation of this solution is shown in appendix \ref{sec: derive Ai(xi-nu)}.

\textbf{Allowed frequencies.} Let us now impose the  boundary conditions on the solution \eqref{eqn: large L pol sol} to determine the form of allowed frequencies. We first analyse the vector sector of perturbations. To leading order, the boundary condition \eqref{eqn: lead large l vec bdry cond 2} with \eqref{eqn: large L pol sol} yields a simple condition on the allowed values of $\nu$,
\begin{equation}\label{eqn: Airy prime bdry cond}
   -\partial_\nu \text{Ai}(-\nu)=0 \, .
\end{equation}
The zeros of the derivative of the Airy function all lie in the negative real axis. For instance, the numerical value of the first five lowest solutions are 
\begin{equation}\label{eqn: real nus vec}
    \nu^{(1)}=1.0188  \, , \quad \nu^{(2)}=3.2482 \, , \quad \nu^{(3)}=4.8201 \, , \quad \nu^{(4)}=6.1633 \, , \quad \nu^{(5)}=7.3722 \, .
\end{equation}
The first sub-leading correction to the boundary condition fixes $\omega_2$ in terms of $\nu$.
Inserting $\omega_1$ and $\omega_2$ back to \eqref{eqn: ansatz near Rt}, we obtain the vector perturbation frequency in the large angular momentum limit,
\begin{equation}\label{eqn: large l disper pole vec}
    \frac{\omega \r}{\sqrt{f(\r)}} = \mathfrak{L} + \frac{\nu}{f(\r)^{1/3}} \left(\frac{\mathfrak{L}}{2}\right)^{1/3} + \frac{6+\nu^3+\left(33+8 \nu^3\right)f(\r)}{60 \nu f(\r)^{2/3}} \left(\frac{\mathfrak{L}}{2}\right)^{-1/3} + \mathcal{O}\left(\mathfrak{L}^{-1}\right) \, .
\end{equation}

Let us now turn to the scalar perturbations. Inserting \eqref{eqn: large L pol sol} into the boundary condition \eqref{eqn: lead large l bdry cond 2}, yields to leading order, 
\begin{equation}\label{eqn: Airy bdry eqn}
    -\partial_\nu\text{Ai}(-\nu)=4\nu^2 \text{Ai}(-\nu) \, .
\end{equation}
This equation was first uncovered in \cite{Liu:2024ymn}, and cannot be solved analytically. Upon numerical scanning for solutions, we find there are two complex solutions \cite{Liu:2024ymn},
\begin{equation}\label{eqn: complex nu}
    \nu^{(\pm)} =0.0674\pm0.4279i \, ,
\end{equation}
and a tower of positive real solutions. For example, the five lowest real solutions have
\begin{equation}\label{eqn: real nus}
    \nu^{(1)}=2.2905  \, , \quad \nu^{(2)}=4.0728 \, , \quad \nu^{(3)}=5.5123 \, , \quad \nu^{(4)}=6.7812 \, , \quad \nu^{(5)}=7.9402 \, .
\end{equation}
Similar to the vector perturbation, the first sub-leading order correction to the boundary condition fixes $\omega_2$ in terms of $\nu$.
Inserting $\omega_1$ and $\omega_2$ back to \eqref{eqn: ansatz near Rt}, we obtain 
\begin{equation}\label{eqn: large l disper pole}
    \frac{\omega \r}{\sqrt{f(\r)}} = \mathfrak{L} + \frac{\nu}{f(\r)^{1/3}} \left(\frac{\mathfrak{L}}{2}\right)^{1/3}  + \frac{36+89\nu^3+16\nu^6+\left(3+472\nu^3+128\nu^6\right)f(\r)}{60\nu(-7+16\nu^3)f(\r)^{2/3}} \left(\frac{\mathfrak{L}}{2}\right)^{-1/3} + \mathcal{O}\left(\mathfrak{L}^{-1}\right) \, .
\end{equation}
The higher order corrections to the frequency can also be written as a function of $\nu$. 

In all, we found that in both vector and scalar perturbations, \eqref{eqn: large l disper pole vec} and \eqref{eqn: large l disper pole}, the parameter $\nu$ uniquely characterises the set of allowed frequencies. For the vector perturbations, $\nu$ is a solution to \eqref{eqn: Airy prime bdry cond}, while for scalar perturbations $\nu$ is a solution to \eqref{eqn: Airy bdry eqn}. Both equations contain a tower of solutions with real frequencies. The frequency difference between two nearby solutions scale as $\mathfrak{L}^{1/3}$. A distinguishing feature of the scalar perturbation is the existence of complex frequency solutions due to \eqref{eqn: complex nu}. In particular, the leading imaginary part of these modes scales as $\mathfrak{L}^{1/3}$. Given the discussion in section \ref{subsectionPolePatch}, we note that given an allowed frequency in the pole patch, minus its complex conjugate is also allowed.

Let us analyse the worldline and stretched horizon limits. In the worldline limit, $\tfrac{\r}{\ell}\to 0$ (or $K\ell \to +\infty$), we observe the vector modes have 
\begin{equation}\label{eqn: omega L vec world pole}
    \omega\r = \mathfrak{L} + \nu \left(\frac{\mathfrak{L}}{2}\right)^{1/3} + \frac{13+3\nu^3}{20\nu}\left(\frac{\mathfrak{L}}{2}\right)^{-1/3}+\mathcal{O}(\mathfrak{L}^{-1}) \, ,
\end{equation}
while the scalar modes have
\begin{equation}\label{eqn: omega L sca world pole}
    \omega\r = \mathfrak{L} + \nu \left(\frac{\mathfrak{L}}{2}\right)^{1/3} + \frac{13 + 187 \nu^3 + 48 \nu^6}{20\nu(-7+16\nu^3)}\left(\frac{\mathfrak{L}}{2}\right)^{-1/3}+\mathcal{O}(\mathfrak{L}^{-1}) \, .
\end{equation}
 Recall from (\ref{eq:limitsbackK}) that to leading order $\r=2/K$. We see then our results agree with the frequency of the Minkowskian modes with large angular momentum analysed in \cite{Anninos:2023epi,Liu:2024ymn,Liu:2025xij}. 

On the other hand, the stretched horizon limit, $\r \to \ell$ (or $K \ell \to -\infty$), for vector perturbations yields 
\begin{equation}\label{eqn: omega L vec strch pole}
    \omega \ell = \frac{\mathfrak{L}}{|K\ell|} + \frac{\nu}{2^{1/3}} \left(\frac{\mathfrak{L}}{|K\ell|}\right)^{1/3}+ 2^{1/3}\frac{6+\nu^3}{60\nu} \left(\frac{\mathfrak{L}}{|K\ell|}\right)^{-1/3} + \mathcal{O}(\mathfrak{L}^{-1})\, ,
\end{equation}
and 
\begin{equation}\label{eqn: omega L sca strch pole}
    \omega \ell = \frac{\mathfrak{L}}{|K\ell|} + \frac{\nu}{2^{1/3}} \left(\frac{\mathfrak{L}}{|K\ell| }\right)^{1/3}+ 2^{1/3}\frac{36+89\nu^3+16\nu^6}{60\nu(-7+16\nu^3)} \left(\frac{\mathfrak{L}}{|K\ell|}\right)^{-1/3} + \mathcal{O}(\mathfrak{L}^{-1})\, ,
\end{equation}
for scalar perturbations. Notably, in the stretched horizon limit, the positive imaginary part of the allowed frequencies is suppressed by a factor of $|K\ell|^{-1/3}$. In fact, it is easy to see that this large-$\L$ expansion is only valid for
\begin{equation}
    1 \ll |K\ell| \ll \mathfrak{L} \,.
\end{equation}
Thus, in the strict $|K\ell| \to \infty$ limit, the allowed frequencies with positive imaginary part are not present. 

\textbf{Conformal stress-tensor.} Let us now consider the conformal stress tensor for solutions \eqref{eqn: large L pol sol} with (scalar mode) frequencies \eqref{eqn: large l disper pole}. 
To do so, it is useful to note that, the leading order linearised Weyl factor is given by
\begin{equation}
    \delta \bomega = \Re e^{-i\omega t} \left(-\frac{\ell\mathfrak{L}^2(2f(\r)-1)c_1}{4\r}\right)\text{Ai}(-\nu)\mathbb{S}+\mathcal{O}(\mathfrak{L}^{-4/3})\,.
\end{equation}
Consequently, via (\ref{eq:compsdeltaTweyl}), the linearised conformal stress tensor (for scalar perturbations) has components
\begin{equation}\label{eqn: Tmn large l pole}
    \begin{cases}
        8\pi G_N \delta T_{tt} =&  \left(\frac{2f(\r)^{3/2}}{\r}\nu  \left(\frac{4\mathfrak{L}^4}{ f(\r)}\right)^{1/3}+\mathcal{O}(\mathfrak{L}^{0})\right)\delta \bomega \, , \\
        8\pi G_N \delta T_{ti}=& -\frac{1}{\mathfrak{L}^2}\left(2 \r \sqrt{f(\r)}\nu  \left(\frac{4\mathfrak{L}^4}{ f(\r)}\right)^{1/3}+\mathcal{O}(\mathfrak{L}^{0})\right)\partial_t\tilde{\nabla}_i\delta \bomega \, , \\
        8 \pi G_N \delta T_{ij} =& -\frac{1}{\mathfrak{L}^2}\left(2 \r \sqrt{f(\r)}\nu  \left(\frac{4\mathfrak{L}^4}{ f(\r)}\right)^{1/3}+\mathcal{O}(\mathfrak{L}^{0})\right)\left(\tilde\nabla_i\tilde\nabla_j-\frac{1}{2}\tilde g_{ij} \tilde \nabla^2 \right)\delta \bomega \\ &+ \left(\r \sqrt{f(\r)}\nu  \left(\frac{4\mathfrak{L}^4}{ f(\r)}\right)^{1/3}+\mathcal{O}(\mathfrak{L}^{0})\right)\tilde g_{ij} \delta \bomega \, ,
    \end{cases}
\end{equation}
which depends on the crucial coefficient $\nu$. It is direct to check that $\delta T_{mn}$ is traceless and transverse with respect to the conformal representative $\bar{g}_{mn}$ in \eqref{eq: induced}.\footnote{To see $\delta T_{mn}$ is transverse, note $\tilde{\nabla}^{2}\delta\bomega=-\L^{2}\delta\bomega$, such that $
    \tilde\nabla^i \left(\tilde \nabla_i \tilde \nabla_j - \frac{1}{2}\tilde g_{ij} \tilde \nabla^2\right)\delta \bomega = \left(-\frac{\mathfrak{L}^2}{2}+1\right)\tilde \nabla_j \delta \bomega \, .$}
Further,
\begin{align} \label{TdTlargeL}
    64\pi^{2}G_{N}^{2}\bar{T}^{mn}\delta T_{mn} &= \bigg( \frac{2}{\r^2}\nu  \left(\frac{4\mathfrak{L}^4}{ f(\r)}\right)^{1/3} + \mathcal{O}(\mathfrak{L}^0)\bigg)\delta\bomega \,.
\end{align}

Given this, let us consider the linearised Weyl mode equation \eqref{linearized Weyl eq} in the large $\mathfrak{L}$ limit,
\begin{align}
    -\frac{2 \mathbb{S}}{\mathfrak{r}^2}\delta\bomega(t) \mathfrak{L}^2+\frac{2\mathbb{S}}{\mathfrak{r}^2 f(\mathfrak{r})}\bigg(\mathfrak{r}^2 \partial_t^2\delta\bomega(t)-(f(\mathfrak{r})+1)\delta\bomega(t)\bigg)=64\pi^{2}G_{N}^{2}\bar{T}^{mn}\delta T_{mn}.
\end{align}
It follows from \eqref{TdTlargeL} that, to leading order, the right hand side is subleading such that
\begin{equation}
    -\frac{2 \mathbb{S}}{\mathfrak{r}^2}\delta\bomega(t) \mathfrak{L}^2+\frac{2\mathbb{S}}{f(\mathfrak{r})} \partial_t^2\delta\bomega(t)= \mathcal{O}(\mathfrak{L}^{4/3}) \,.
\end{equation}
 Assuming $\delta \bomega(t) = N_\bomega e^{-i\omega t}$, this gives, to leading order, $\omega = \pm \sqrt{f(\r)} \L/\r$. To obtain the next order correction, we add the contribution from $\bar T_{mn} \delta T^{mn}$ in \eqref{TdTlargeL} to self-consistently obtain
 \begin{equation} \label{eq: freq repeat}
     \frac{\omega \r}{\sqrt{f(\r)}}= \mathfrak{L} + \frac{\nu}{f(\r)^{1/3}}\left(\frac{\mathfrak{L}}{2}\right)^{1/3} +\mathcal{O}(\mathfrak{L}^{-1/3})  \,.
 \end{equation}

For completeness, we report on the components of the conformal stress-energy tensor in the vector sector. The non-vanishing components in the large-$\L$ limit are given by
\beq
\begin{cases}\label{eqn: Tmn large l pole vec}
 8\pi G_{N}\delta T_{ti}=\left( -\frac{\sqrt{f(\mathfrak{r})}}{2\mathfrak{r}} \mathfrak{L}^2
 +\mathcal{O}(\mathfrak{L}^{0})
 \right) \Phi^{(V)}(t,\mathfrak{r})\mathbb{V}_{i}\;,\\
 8\pi G_{N}\delta T_{ij}=\frac{\mathfrak{r}}{2\sqrt{f(\mathfrak{r})}}\partial_t \Phi^{(V)}(t,\mathfrak{r})(\tilde{\nabla}_{i}\mathbb{V}_{j}+\tilde{\nabla}_{j}\mathbb{V}_{i})\;,
\end{cases}
\eeq
where, using \eqref{eqn: large L pol sol}, 
   $ \Phi^{(V)}(t,\mathfrak{r}) = \Re e^{-i\omega t} \text{Ai}(-\nu)+\mathcal{O}(\mathfrak{L}^{-2/3})\;,$
with frequencies $\omega$ given by \eqref{eq: freq repeat} and $\nu$ solutions to \eqref{eqn: Airy prime bdry cond}.

\textbf{The bulk radial profile.}  Using the asymptotic behaviour of the Airy functions, the master field away from the boundary (and to leading order) is proportional to
\begin{equation}\label{eqn: width L^1/3 dS pole}
    \Phi^{(S/V)}(t,r) \propto \Re e^{-i\omega t} \left(\frac{\delta s}{D_{(\L)}}-\nu\right)^{-1/4}\text{exp}\left(-\frac{2}{3}\left(\frac{\delta s}{D_{(\L)}}-\nu\right)^{3/2}\right)\, , \qquad \frac{|\delta s|}{D_{(\L)}}\gg1\, , 
\end{equation}
where $\delta s$ and $D_{(\L)}$ are defined as
\begin{equation}\label{eqn: width L^1/3 dS pole 2}
    \delta s \equiv \frac{\r-r}{\sqrt{1-\frac{\r^2}{\ell^2}}} \, , \qquad D_{(\L)} \equiv \left(\frac{1}{2\mathfrak{L}^2}\sqrt{1-\frac{\r^2}{\ell^2}}\right)^{1/3}\r \, .
\end{equation}
In this limit, $\delta s$ is the proper displacement away from the boundary along a line of constant $(t,\theta,\varphi)$, where $r$ is taken to be near $\mathfrak{r}$, see \eqref{eq: short proper distance}.
The exponential decay in \eqref{eqn: width L^1/3 dS pole} implies that the perturbation is localised near the boundary, with a width given by $D_{(\L)}$ in \eqref{eqn: width L^1/3 dS pole 2}. In the worldline limit, the width becomes
\begin{equation}
    D_{(\L)} = \frac{2^{2/3} \ell}{|K\ell| \mathfrak{L}^{2/3}} + \mathcal{O}(|K\ell|^{-2/3}) \, .
\end{equation}
On the other hand, in the stretched horizon limit, the width is given by 
\begin{equation}\label{widthpole}
    D_{(\L)} = \frac{\ell}{2^{1/3}|K\ell|^{1/3} \mathfrak{L}^{2/3}} + \mathcal{O}(|K\ell|^{-1}) \, .
\end{equation}

\subsubsection*{Comparison with global AdS$_4$}

It is instructive to compare the behaviour of the large angular momentum modes in the pole patch with the case of global AdS. In the following, we focus on the four-dimensional case, but a generalisation to arbitrary dimension is straightforward. The background metric is given by \eqref{eq:dSmetric} with $f(r)=1+\frac{r^2}{\ell_\text{AdS}^2}$, where $\ell_\text{AdS}$ is the AdS$_4$ radius. We focus on a ``pole patch" region between the origin and a boundary $\Gamma$ at $r=\mathfrak{r}$. The trace of the extrinsic curvature at the boundary is
\begin{align}
    K\ell_\text{AdS}|_{\Gamma}&=\frac{2\ell_\text{AdS}^2+3\mathfrak{r}^2}{\mathfrak{r}\sqrt{\ell^2_\text{AdS}+\mathfrak{r}^2}} =3+\frac{\ell^2_\text{AdS}}{2\mathfrak{r}^2}+\mathcal{O}(\mathfrak{r}^{-4}) \,.
\end{align}
The analysis of linearised perturbations around  global AdS$_4$ using the Kodama-Ishibashi formalism was performed in \cite{Anninos:2024xhc}. One can solve the master field equation in the large angular momentum limit using an analogous WKB approximation to the one shown for de Sitter. In a region of width $\sim \L^{-2/3}$ away from the boundary, the solutions to the master field equation are given by Airy functions. In both scalar and vector sectors, the allowed frequencies take the following form,
\begin{equation}\label{global AdS frequencies}
    \frac{\omega \r}{\sqrt{f(\r)}} = \mathfrak{L} + \frac{\nu}{f(\r)^{1/3}} \left(\frac{\mathfrak{L}}{2}\right)^{1/3} +\mathcal{O}(\mathfrak{L}^{-1/3})\, .
\end{equation}
In the vector sector, $\nu$ is real for all the modes. In the scalar sector, there exists complex solutions $\nu$ with positive imaginary part. Considering further the limit $K\ell_\text{AdS} \to 3$, we obtain
\begin{equation}\label{global AdS frequencies asymptotic}
    \omega \ell_\text{AdS} = \mathfrak{L} + \nu (K\ell_\text{AdS}-3)^{1/3}\mathfrak{L}^{1/3} +\mathcal{O}(\mathfrak{L}^{-1/3})\, .
\end{equation}
The expansion is valid for $\mathfrak{L}\gg \frac{1}{K\ell_\text{AdS}-3}$. Outside the regime of validity, one needs to consider first the limit $K\ell_\text{AdS}\rightarrow 3$, in which no complex frequencies appear \cite{Anninos:2024xhc}.

We note the similarity between the $K\ell_\text{AdS}\to3$ limit of AdS$_4$ and the large-$|K\ell|$ limit for the de Sitter pole patch. In the de Sitter case, there still exist complex modes approaching $\omega \ell =\pm i$, however, due to the proximity to the cosmological horizon.

\subsubsection{Summary of key results}

To compare with the next sections, we summarise here the key results obtained in the large angular momentum limit. Considering the {stretched horizon limit}, $|K\ell|\gg 1$, we find:

\begin{center}
\begin{tcolorbox}[tab2,
  title=\textbf{Large angular momentum modes for pole patch in the stretched horizon limit},
  boxrule=2pt,
  width=16.7cm,
  fontupper=\normalsize,
  fonttitle=\normalsize,
]
\begin{itemize}
    \item For $\mathfrak{L} \gg |K\ell| \gg1$, the allowed frequencies take the form 
    \begin{equation}
         \omega \ell = \frac{\mathfrak{L}}{|K\ell|} + \frac{\nu}{2^{1/3}} \left(\frac{\mathfrak{L}}{|K\ell|}\right)^{1/3}
        + \mathcal{O}(\mathfrak{L}^{-1/3}) \, ,
        \label{eq: summary 1}
    \end{equation}
    where $\nu$ solves \eqref{eqn: Airy prime bdry cond} for vector or
    \eqref{eqn: Airy bdry eqn} for scalar perturbations. The scalar sector contains a
    pair of complex modes with positive imaginary part, whose linearised Weyl
    factor grows exponentially in time.

    \item The leading coefficient in \eqref{eq: summary 1} follows from the linearised
    boundary–mode equation \eqref{linearized Weyl eq}, assuming the perturbed conformal stress tensor
    contributes only subleading corrections. These corrections are needed to fix 
    the subleading frequency shifts.

    \item All these modes localise very close to the boundary. Their radial
    profile decays exponentially with the proper distance, and their width $D_{(\L)}$ scales as
    \begin{equation}
        \frac{D_{(\L)}}{\ell} \sim \mathfrak{L}^{-2/3}|K\ell|^{-1/3}.
    \end{equation}
\end{itemize}
\end{tcolorbox}
\captionof{table}{Summary for pole patch large angular momentum modes in the stretched horizon limit. \label{tab:large-l}}
\end{center}

\subsection{Cosmic patch}

Following an analogous procedure as in the pole patch, we consider the large $\mathfrak{L}$ solution for the case in which the turning point and the boundary are close, $|r^*_\text{t}-\r^*|\ll \mathfrak{L}^{-2/3}$. This time we look for purely outgoing solutions as $r^*\to \infty$, or equivalently \eqref{eqn: WKB sol 3}  that is purely-outgoing as $\xi \to -\infty$. The solution (including its first correction) takes the same form as the pole patch solution \eqref{eqn: large L pol sol},
\begin{equation}\label{eqn: large L cosmic sol}
    \phi^{(S/V)}(\xi) = \phi_0(\xi)+\left(\frac{\mathfrak{a}_1}{5}\xi \phi_0(\xi)-\left(\mathfrak{a}_1+\mathfrak{a}_2\xi^2+\frac{4\mathfrak{a}_2}{15}\left(\nu-\xi\right)\left(2\nu+3\xi\right)\right)\partial_\xi \phi_0(\xi)\right)\mathfrak{L}^{-2/3}+\mathcal{O}(\mathfrak{L}^{-4/3})\, ,
\end{equation}
with the same $\mathfrak{a}_1$ and $\mathfrak{a}_2$ but with a different function $\phi_0(\xi)$,
\begin{equation}
    \phi_0(\xi)=c_1 \text{Ai}(e^{\frac{2\pi i}{3}}(\xi-\nu))\, , \qquad\mathfrak{a}_1 = \frac{\nu^2}{2^{4/3}f(\r)^{1/3}}+\frac{2^{1/3}\r \omega_2}{f(\r)^{1/6}} \, , \qquad \mathfrak{a}_2 = -\frac{2+f(\r)}{2^{4/3}f(\r)^{1/3}} \, .
\end{equation}
The solution up to this order is completely specified by $\nu$ and $\omega_2$.

\textbf{Allowed frequencies.} Now we impose the conformal boundary conditions on the solution \eqref{eqn: large L cosmic sol}. Let us first look at the vector perturbations. We plug in the solution \eqref{eqn: large L cosmic sol} to \eqref{eqn: lead large l vec bdry cond 2}, yielding
\begin{equation}\label{eqn: Airy prime bdry cond 2}
    - \partial_\nu\text{Ai}(-e^{\frac{2\pi i}{3}}\nu) = 0 \, .
\end{equation}
We observe that this equation is identical to the pole patch case, \eqref{eqn: Airy prime bdry cond}, upon identifying
\begin{equation}\label{eqn: nu cos and pole}
    \nu^{(\text{cosmic})}=e^{-\frac{2\pi i}{3}}\nu^{(\text{pole})} \, .
\end{equation}
It then follows that the solution to \eqref{eqn: Airy prime bdry cond 2} is given by a tower of real numbers, \eqref{eqn: real nus vec}, rotated by a phase $-\tfrac{2\pi}{3}$, rendering them complex with a negative-definite imaginary part. For the scalar perturbations, we plug the solution \eqref{eqn: large L cosmic sol} into \eqref{eqn: lead large l bdry cond 2}, leading to
\begin{equation}\label{eqn: Airy bdry eqn 2}
    -\partial_\nu\text{Ai}(-e^{-\frac{2\pi i}{3}}\nu)=4\nu^2 \text{Ai}(-e^{-\frac{2\pi i}{3}}\nu) \, .
\end{equation}
Similarly, this is identical to the pole patch case, \eqref{eqn: Airy bdry eqn}, upon using \eqref{eqn: nu cos and pole}. The solution to \eqref{eqn: Airy bdry eqn 2} is then given by rotating the pair of complex solutions \eqref{eqn: complex nu} to,
\begin{equation}\label{eqn: complex nu 2}
    \nu^{(+)}=0.3369 - 0.2723 i \, , \qquad \nu^{(-)}=-0.4043 + 0.1556i \, ,
\end{equation}
and a tower of real solutions \eqref{eqn: real nus}, also rotated by a phase $-\tfrac{2\pi}{3}$. In summary, the allowed frequencies expanded in terms of $\nu$ take the same form as in the pole patch, \eqref{eqn: large l disper pole vec} for the vector perturbations and \eqref{eqn: large l disper pole} for the scalar perturbations.  
Concretely, to leading order in the stretched horizon limit of the cosmic patch, the allowed frequencies are given by,
\begin{equation}
     \omega \ell = \frac{\mathfrak{L}}{|K\ell|} + \frac{\nu}{2^{1/3}} \left(\frac{\mathfrak{L}}{|K\ell| }\right)^{1/3}+ \mathcal{O}(\L^{-1/3}) \,,
     \label{eq: frequencies cosmic large l large K}
\end{equation}
with $\nu$ now a solution to either \eqref{eqn: Airy prime bdry cond 2} or \eqref{eqn: Airy bdry eqn 2}. Similarly to the pole patch solutions, in the cosmic patch if $\omega$ is an allowed solution, the same frequency but with minus its real part, is also a solution. Note that remarkably, these frequencies can be generalised for generic spherically symmetric backgrounds with non-degenerate horizons, see appendix \ref{app: large l generic}.

\textbf{The conformal stress-tensor.} Now we look at the conformal stress-tensor for solutions \eqref{eqn: large L cosmic sol} in the scalar sector. In particular, by expressing in terms of the linearised Weyl factor, this tensor takes the same form as \eqref{eqn: Tmn large l pole} with $\nu$ now given by \eqref{eqn: Airy bdry eqn 2}. The discussion regarding its relation to the linearised Weyl mode equation \eqref{linearized Weyl eq} follows what presented in the pole patch section. 

In the vector sector, the conformal stress-tensor is similarly given by \eqref{eqn: Tmn large l pole vec}, with $\Phi^{(V)}(t,r)$ replaced by the cosmic patch solution \eqref{eqn: large L cosmic sol} and $\nu$ a solution to \eqref{eqn: Airy prime bdry cond 2}.

\textbf{The bulk radial profile.} By repeating the same process as in the pole patch, one can examine the behaviour of master field away from the boundary. In this case we attain
\begin{equation}\label{eqn: width L^1/3 dS cosmic}
    \Phi^{(S/V)}(t,r) \propto \Re e^{-i\omega t}e^{\frac{i \pi}{12}} \left(-\frac{\delta s}{D_{(\L)}}-\nu\right)^{-1/4}\text{exp}\left(-i\frac{2}{3}\left(\frac{\delta s}{D_{(\L)}}+\nu\right)^{3/2}\right)\, , \qquad \frac{|\delta s|}{D_{(\L)}}\gg1\, , 
\end{equation}
where $\delta s$ and $D_{(\L)}$ are given in \eqref{eqn: width L^1/3 dS pole 2}. The bulk master field is rapidly oscillating with a decaying amplitude and a characteristic wavelength given by $D_{(\L)}$. Given in the large-$K\ell$ limit, the boundary approaches the horizon, care must be taken in order to define how localised these modes are. We postpone a proper treatment to section \ref{sec: rindler}, where we work in a Rindler approximation.

\section{The large mean curvature limit: pole patch} \label{sec: large K pole}

In this section, we analyse the stretched horizon limit of the pole patch. That is, we take the large-$|K\ell|$ limit of the linearised pole patch dynamics, characterising the behaviour of modes frequencies illustrated in figure \ref{fig:PPmodes}, as well as the conformal stress-tensor, Weyl factor, and the master fields.

Assuming $|K\ell|$ is the largest scale in the problem, it is straightforward to expand the hypergeometric and linearised boundary condition \eqref{eq: deltaK scalar} close to $\r = \ell$ to find the allowed frequencies. A large-$|K\ell|$ expansion of \eqref{eq: deltaK scalar} gives to leading order,
\begin{equation}\label{eq largeKl pole}
     0= (\omega\ell)^2((\omega\ell)^2+1)
\bigg[
\frac{ 2^{-i\omega\ell} \, \Gamma(-i\omega\ell) }
{ \Gamma(1 + l - i\omega\ell) } \, |K \ell|^{-i\omega\ell}
+ 
\frac{ 2^{i\omega\ell} \, \Gamma(i\omega\ell) }
{ \Gamma(1 + l + i\omega\ell) } |K \ell|^{i\omega\ell}\bigg] + \mathcal{O}((K \ell)^{-2}) \,,
\end{equation}
where we have dropped a non-vanishing overall term. The zeros of this equation are found to be at $\omega \ell = 0, \pm i$ and at a discrete set of real frequencies\footnote{Care should be taken because the term in the bracket diverges for $\omega\ell=\pm i,0$. Analyzing the divergence of the master field at the boundary and the leading order correction to the $\omega\ell=\pm i,0$ modes, however, reveals the bracket diverges slower than the rate at which the pre-factors vanish.} coming from the cancellation of the two terms in between squared brackets.\footnote{These terms come from 
the master field evaluated at the boundary $\Phi^{(S)}|_\Gamma$, in the large-$|K\ell|$ limit. Indeed, one can explicitly check that the term proportional to $\partial_r \Phi^{(S)}$ in \eqref{eq: deltaK scalar} will be always subleading in the large-$|K\ell|$ expansion. This is true both for the cosmic and the pole patch, with the difference that in the cosmic patch this will lead to complex (instead of real) frequencies. See section \ref{sec: large K cosmic}.} In what follows, we separately study each set of modes.

Further, using the boundary condition (\ref{eq: deltaK scalar}) to eliminate a radial derivative in (\ref{eq: weyl scalar}), we can express the linearised Weyl factor as 
\beq \delta\bomega =- \Re \frac{e^{-i\omega t}\ell}{4\mathfrak{r}}\frac{l(l+1)(l-1)(l+2)}{2(l^{2}+l-1)\frac{\mathfrak{r}^{2}}{\ell^{2}}+(4-3l(l+1)+2\mathfrak{r}^{2}\omega^{2})}\mathbb{S}\;\phi^{(S)}(r)\;.\label{eq:weylfactSphi}\eeq
Below we will evaluate this expression for each set of frequencies and determine the components of the stress-tensor  (\ref{eq:compsdeltaTweyl}).

\subsection{Soft modes}
We start by analysing those modes that in the strictly infinite $|K\ell|$ limit have $\omega \ell = \pm i$. We will first determine the allowed frequencies in a large-$|K\ell|$ expansion, incorporating deviations away from the strict limit. We will then compute the conformal stress tensor evaluated for such modes. Finally, we will analyse the bulk radial profile for these modes.

\subsubsection{Allowed frequencies}

\begin{figure}[t!]
\centering
 \includegraphics[width=.5\textwidth]{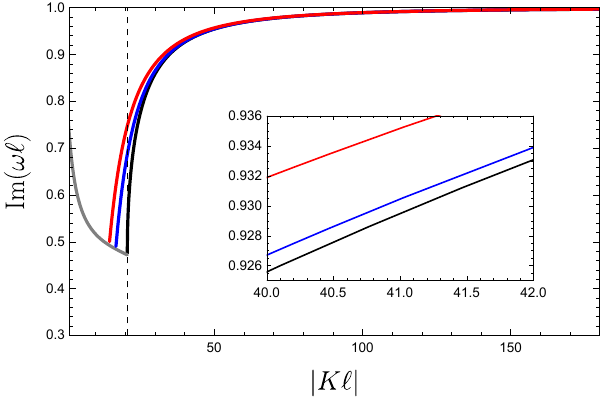}
\caption{Large-$|K\ell|$ behaviour of $\omega\ell$. Lower black curve is exact numerical result for $l=10$ 
; middle blue (top red) curve corresponds to the analytic expansion (\ref{eq: frequency expansion}) truncated at $\omega_{2}$ ($\omega_{1}$). The dashed vertical line denotes $(|K\ell|)_{\text{cr}}\approx 20.75$, below which $\text{Im}(\omega\ell)$ increases and the scalar modes develop a real component.}
\label{fig:fitomdatlargeKpole}
\end{figure}

Expanding the linearised boundary condition \eqref{eq: deltaK scalar}, we find the first correction to the $\omega\ell=\pm i$ frequencies enter at order $\omega \ell = \pm i \pm i \omega_1 (K\ell)^{-2}$, where  
$\omega_{1}$ should satisfy
\begin{equation}
    0 = (\omega_1 + (l(l+1) -1)) \frac{i 2^{l+4}\Gamma(l+\tfrac{3}{2})}{\sqrt{\pi}\Gamma(l+2)} + \mathcal{O}((K\ell)^{-2}\log(|K\ell|)) \,.
\end{equation}
Solving determines the allowed value for the correction $\omega_1$. 
Continuing this process order-by-order, we found that the set of allowed frequencies in the large-$K\ell$ limit is given by 
\begin{equation}
    \omega\ell = \pm i \pm i \omega_1 (K\ell)^{-2} \pm i \omega_2 (K\ell)^{-4} \log |2K\ell| \pm i \omega_3 (K \ell)^{-4} + \mathcal{O}((K\ell)^{-6}(\log|K\ell|)^2) \,,
    \label{eq: frequency expansion}
\end{equation}
with 
\begin{equation}
\begin{cases}
    \omega_1 = -(l(l+1) -1) \,, \\ \omega_2=-\frac{1}{4}l^2(l+1)^2 \,,  \\ \omega_3 =\frac{1}{4}l^2(l+1)^2 H_{l-1}-\frac{1}{4}\left(3l^4+5l^3-22l^2-24l+18\right) \,,
    \end{cases}
    \label{eq: subleading frequencies pole patch}
\end{equation}
where $H_l=\sum_{k=1}^{l}  \frac{1}{k}$ is the $l^{\text{th}}$ harmonic number. Note that all corrections are purely imaginary and they lower the absolute value of the allowed frequencies, in accordance with the general expectation from figure \ref{fig:PPmodes}. Furthermore, we see precise agreement between the analytic large-$|K\ell|$ expansion (\ref{eq: frequency expansion}) and the numerical solutions to \eqref{eq: deltaK scalar} in figure~\ref{fig:fitomdatlargeKpole}.

Even though this analysis is in principle valid only for $l\geq 2$, we observe that the extrapolation to $l=0$ coincides with the analysis for the spherically symmetric mode in \eqref{eq: frequency l=0}. In particular the logarithmic contributions go to zero when $l=0$.\footnote{Note this extrapolation does not hold for $l=1$, where there are no physical metric perturbations \cite{Anninos:2024wpy}.} Finally, it is easy to see this expansion breaks down when $l$ is of order of $|K\ell|$.

To leading order, we observe that the Weyl factor (\ref{eq:weylfactSphi}) goes like
\begin{equation}
    \delta \bomega =e^{\pm t/\ell} \frac{2^{l-1} \left(l^2+l-2\right)  \Gamma \left(l+\frac{3}{2}\right) \mathbb{S}}{\sqrt{\pi } \Gamma (l+2)} |K\ell|+ \mathcal{O}\left(|K\ell|\right)^{-1} \,,
    \label{eq: weyl w=i}
\end{equation}
where the $\pm$ corresponds to $\omega\ell =  \pm i$, respectively. The plus sign exhibits exponential growth of the Weyl factor to linearised order in the perturbation. 

Interestingly, the leading frequencies $\omega\ell = \pm i$ are independent of the angular momentum $l$. Thus, their spatial distribution on the horizon $S^2$ is frozen at the linearised level, to leading order in $K\ell$. They are somewhat reminiscent of soft modes discussed in the context of asymptotically Minkowksi space \cite{Hawking:2016msc, Raclariu:2021zjz}, which can similarly have a non-trivial spatial dependence on the celestial sphere whilst having vanishing frequency. As we shall see momentarily, the stress-tensor of these modes is also `soft'.

\subsubsection{The conformal stress-tensor} \label{sssec:confstpmiPP}

Here we compute the conformal stress-tensor \eqref{eq: tmn} perturbatively in the large-$|K\ell|$ expansion. As in the previous subsection we focus on the conformal stress tensor for the (perturbed) $\omega \ell = \pm i$ modes. One immediate feature of evaluating the conformal stress tensor perturbatively in $|K\ell|$ is that different components will scale differently with $|K\ell|$. This is due to the fact we are measuring time with respect to the de Sitter inertial observer, i.e., the conformal representative chosen for this calculation has $[g]_{tt} = f(\r) dt^2 \approx (K\ell)^{-2} dt^2$, in the large $|K\ell|$ limit. Transversability of the stress tensor with respect to $\bar{g}_{mn}$ in \eqref{eq: induced}, directly implies, for instance, that the $|\delta T_{tt}/\delta T_{ij}| = \mathcal{O}((K\ell)^{-4})$.

To see this explicitly, it is useful to express the components of the conformal stress tensor in terms of the Weyl factor at the boundary. In the stretched horizon limit we find,
\begin{equation}\label{deltaTcomponents}
    \begin{cases}
         8\pi G_N \delta T_{tt} = \left( \frac{2l(l+1)}{K^3\ell^4} + \mathcal{O}(|K\ell|^{-5}\log(K\ell)) \right) \delta \bomega\,, \\
         8 \pi G_N \delta T_{ti} = \left( -\frac{2}{K\ell}\tilde{\nabla}_i + \mathcal{O}(|K\ell|^{-3}\log(K\ell)) \right) \delta \bomega \, , \\
        8\pi G_N \delta T_{ij}= \left( \frac{4K\ell^2}{l(l+1)-2}\left(\tilde{\nabla}_i \tilde{\nabla}_j - \frac{1}{2}\tilde{g}_{ij} \tilde{\nabla}^2\right) + \mathcal{O}(|K\ell|^{-1}\log(K\ell)) \right) \delta \bomega\, ,        
    \end{cases}
\end{equation}
where we recall $(i,j)=(\theta,\varphi)$ denote angular indices on the two-sphere, and $\tilde{g}_{ij}$ and $\tilde{\nabla}_i$ are the metric and covariant derivative of the unit two-sphere. To leading order, $\delta \bomega$ is given by \eqref{eq: weyl w=i},  it will receive subleading corrections in the large-$|K\ell|$ expansion.

The traceless property of $T_{mn}$ is manifest to this order in a curious way. The $\delta T_{ij}$ components are by themselves traceless with respect to the two-dimensional metric $\tilde g_{ij}$, while $g^{tt} \delta T_{tt}$ is subleading to first order.

We proceed by explicitly computing the stress tensor order-by-order in the stretched horizon limit. The first observation is, even though the Weyl factor at the boundary is non-trivial to leading order, a slightly off-shell\footnote{By slightly off-shell we mean considering a mode with frequency $\omega \ell = \omega_0 i + \mathcal{O} (K\ell)^{-2}$. In that case, the stress tensor will have a non-vanishing contribution at leading order for all values of $\omega_0 \in \mathbb{R}$, except for the case when $\omega_0 = \pm1$, when this leading contribution will vanish.}  computation of the conformal stress tensor shows that for the exactly $\omega\ell = \pm i$, the stress tensor exhibits a leading cancellation. The subleading piece that persists goes as
\begin{equation}
  \delta T_{tt} = \mathcal{O} (|K\ell|^{-2}) \quad , \quad \delta  T_{ij} = \mathcal{O}(|K\ell|^2)\, \quad , \quad    \delta T_{ti} = \mathcal{O} (1)  \,.
   \label{eq: leading tmn w=i}
\end{equation}
As a reminder, the background $\bar{T}_{tt}$ scales like $|K\ell|$ in this limit. 

To next order, we use $\delta\bomega$ in \eqref{eq: weyl w=i} and \eqref{deltaTcomponents} to find the expressions for the perturbed stress tensor. Note that to this order only the $tt$-component is subleading but both $\delta T_{ii}$ will contribute. An interesting exercise is to consider the stress tensor to this order but with an off-shell frequency. In that case, we observe a curious cancellation in the $tt$-component that only happens when we select the frequency to be on the allowed value. See appendix \ref{app: cancellation} for details.

Next, we analyse the contribution from this mode from the point of view of \eqref{eq: linearised 2.12}, in the large-$|K\ell|$ limit. For this, the relevant quantity to consider is $\bar{T}^{mn} \delta T_{mn}$. Given the background stress tensor is traceless, one can easily write,
\begin{equation}
     64\pi^{2}G_{N}^{2}\bar{T}^{mn}\delta T_{mn} = -\frac{1}{\mathfrak{r}^3\sqrt{1-\frac{\mathfrak{r}^2}{\ell^2}}} \tilde{g}^{ij}\delta T_{ij} = \frac{1}{\ell^3}  \left(- |K\ell| + \mathcal{O}((K\ell)^{-1})) \right) \tilde{g}^{ij}\delta T_{ij} =  \mathcal{O} (|K\ell|) \,.
\label{eq:TdeltaTleaddSomi}\end{equation}
Hence, the leading contribution $\delta T_{ij}$ in \eqref{deltaTcomponents}, even though non-zero, will not contribute to $\bar T_{mn}\delta T^{mn}$ to leading order, because $\delta T_{ij}$ is traceless with respect to $\tilde{g}_{ij}$. 
We will compute the finite contribution momentarily. 

Writing $\delta\bomega = \delta \bomega(t) \mathbb{S}$ such that we use the linearised Weyl equation (\ref{linearized Weyl eq}), we find, in the large-$|K\ell|$ limit, 
\begin{equation} \label{weylpmi1}
    K^2 \ell^2 \left( \partial_t^2 \delta \bomega(t) + \ell^{-2} \delta \bomega(t) \right) = \mathcal{O}(|K\ell|) \,,
\end{equation}
where we assumed $|K\ell| \gg l$, such that the spatial part of the covariant derivative is subleading. Note, to obtain that the $\delta T_{mn}$ contribution is subleading we used information about $\omega_{1}$ in the large-$|K\ell|$ expansion \eqref{eq: frequency expansion}. 

Solving \eqref{weylpmi1} self-consistently gives $\delta \bomega(t) = N_\bomega e^{\pm t/\ell} + \cdots$, where the ellipsis indicates subleading contributions in the large-$|K\ell|$ expansion and $N_\bomega$ is some (time-independent) constant that will become important at next order.

To obtain a non-trivial contribution to $\bar{T}^{mn} \delta T_{mn}$, we need to include all corrections to the frequency up to order $(K\ell)^{-4}$. In that case a direct calculation reveals that,
\begin{equation}
    64\pi^2 G_N^2 \bar{T}^{mn} \delta T_{mn} = \frac{
    2^{l} \, e^{\pm t/\ell} \, (2 + l) \, \Gamma\!\left(\tfrac{3}{2} + l\right) \, \mathbb{S}
}{
    \ell^2\sqrt{\pi} \, \, \Gamma(l - 1)
} |K\ell| + \mathcal{O} (|K\ell|^{-1}\log |K\ell|)\,.
    \label{eq: finite tmn}
\end{equation}

We can continue to expand \eqref{eq: linearised 2.12} to next order in the large-$|K\ell|$ limit. Now this contribution from $\bar T^{mn} \delta T_{mn}$ will become relevant. In particular, to have a consistent equation implies $\delta \bomega$ should have an overall normalisation that goes like $|K\ell|$, which agrees with the normalisation chosen in \eqref{eq: weyl scalar}. Using \eqref{eq: weyl w=i}, we can identify $N_\bomega$ and rewrite $\bar{T}^{mn} \delta T_{mn}$ as
\begin{align}
     64\pi^2 G_N^2 \bar{T}^{mn} \delta T_{mn} = \frac{2l(l+1)
}{
    \ell^2 } N_{\bomega} e^{\pm t/\ell} + \mathcal{O} (|K\ell|^{-1}\log |K\ell|) \,.
\end{align}
Inserting $\delta\bomega(t)= N_{\bomega} \exp \left(\tfrac{t}{\ell} (\pm 1 \pm \omega_1 |K\ell|^{-2} + \cdots ) \right)$ into the linearised Weyl equation \eqref{linearized Weyl eq} at large-$|K\ell|$, we obtain
\begin{align}
    \pm \frac{4 e^{\pm t/\ell}}{\ell^2}\bigg( l(l+1)-1 +\omega_1 \bigg) =\mathcal{O}(|K\ell|^{-2}) \,,
\end{align}
which self-consistently sets $\omega_1$ to the correct value \eqref{eq: subleading frequencies pole patch}.

\subsubsection{The bulk radial profile} \label{subsectionBulkRadialProfilePmI}
Given that to leading order the $\omega \ell = \pm i$ can be obtained directly from the boundary dynamics equation \eqref{weylpmi1}, it is interesting to analyse the radial profile for these modes in the bulk. As we will see, these modes are localised close to the boundary. 

In the strict large-$|K\ell|$ limit, the pole patch master field \eqref{polepatchgeneral} for the modes with frequencies $\omega \ell= \pm i$ diverges  as $|K\ell|$ as we approach the boundary. For this reason, and in order to compare with other modes, we adopt the criterion specified in \ref{sec: radial profile}. Namely, we observe the width of the normalised master field $\tilde\phi (s)$, see \eqref{eq: normalised phi}, as a function of the proper distance away from the boundary.

\begin{figure}[h!]
    \centering
    
    \begin{subfigure}[t]{0.48\textwidth}
        \centering
        \includegraphics[width=\textwidth]{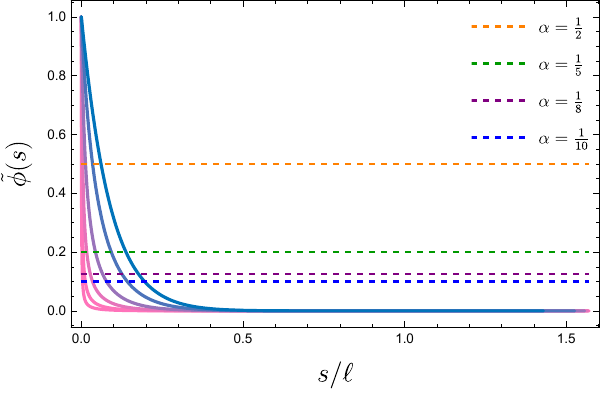}
        \caption{Radial profiles.}
        \label{fig:omiprofsPP}
    \end{subfigure}
    \hfill
    \begin{subfigure}[t]{0.48\textwidth}
        \centering
        \begin{picture}(0,0)
        \put(110,120){\small $\sim |K\ell|^{-1}$} 
    \end{picture}%
        \includegraphics[width=1.04\textwidth]{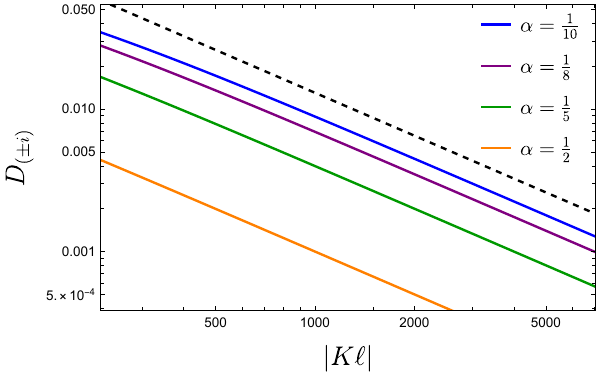}
        \caption{Width.}
        \label{fig:disomiPP}
    \end{subfigure}
    \caption{Normalised radial profile $\tilde{\phi}(s)$ and width $D_{(\pm i)}$ for $\omega_{(\pm i)}\ell$ modes and $l=10$. \emph{Left.} Bulk profiles (solid curves)  at various $|K\ell|$: from right to left $|K\ell|=7,22,71,224,707,2236$. As $|K\ell|$ increases, the profiles become more localised. Horizontal dashed lines denote different values of $\alpha$. \emph{Right.}  Width $D_{(\pm i)}$ for the various $\alpha$'s: top to bottom $\alpha=\tfrac{1}{10},\tfrac{1}{8},\tfrac{1}{5},\tfrac{1}{2}$. As a guide, the black dashed line shows a scaling of $D_{(\pm i)} \sim |K\ell|^{-1}$.
    }
    \label{fig:bulkprofsandDomi}
\end{figure}

In figure \ref{fig:omiprofsPP}, we plot this radial profile for a variety of $|K\ell|$, increasing in magnitude. Definitely the radial profile localises as $|K\ell| \to \infty$. To characterise the localisation, we consider the proper distance for which the profile decayed to a fraction $\alpha$ of its original value. We plot this width $D_{(\pm i)}$ for different values of $\alpha$ in figure \ref{fig:disomiPP}. We observe that, when $\alpha$ is any order one fraction, the width decays as $D_{(\pm i)}  \sim |K\ell|^{-1}$, in the large $|K\ell|$ limit. An analytic characterisation of the master field profile is left to appendix \ref{app: radial profile}.

\subsection{Gapless modes}\label{subsec:om0modes}

Recall figure \ref{subfig:realscalmodes}, where we highlighted the existence of a pair of real scalar modes near the origin. Below we analyse their behaviour, proceeding in the same way as with the $\omega \ell = \pm i$ modes: we find perturbatively the values for the allowed frequencies, compute their stress-energy tensor and characterise their radial profile.

\subsubsection{Allowed frequencies}
Expanding \eqref{eq: deltaK scalar} in the large-$K\ell$ limit, we find that the first correction to the $\omega=0$ modes enters at order $|K\ell|^{-1}$, while the next corrections involve logarithms\footnote{More generally, choosing an ansatz $\omega\ell=\beta_1 |K\ell|^{-1}+\beta_2|K\ell|^{-2}+\beta_3 |K\ell|^{-3}+...$, one can solve for the coefficients $\beta_i$, which are functions of $\log|K\ell|$. The even coefficients $\beta_{2n}$ vanish. 
For instance, to leading order this gives, 
\begin{align}
    \omega \ell&=\pm \frac{
\sqrt{\, l(1+l)\,\left( \log\!\left(2 |K\ell|\right) - H_l -1\right) +2\,}
}{
\sqrt{\,2 (\log\!\left(2 |K\ell|\right) - H_l)
\,}
}|K\ell|^{-1}+ \mathcal{O}(|K\ell|^{-3}) \,,
\label{eq:gaplessfreqsPP}\end{align}
where $H_l=\sum_{k=1}^l \frac{1}{k}$ is the $l$-th harmonic number. Using this expansion for $l=10$ and $|K\ell|\approx 70.68$, we recover $\omega\ell\approx\pm .075$, matching the numerics highlighted in figure \ref{fig:PPmodes}.
}
\begin{align}
    \omega\ell&= \pm \sqrt{\frac{l(l+1)}{2}} \frac{1}{|K\ell|} \mp \frac{l(l+1)-2}{2\sqrt{2l(l+1)}}  \frac{1}{|K\ell|\log|K\ell|} +\mathcal{O}\big(|K\ell|^{-1}\log^{-2}|K\ell|\big) \,.
\label{eq:omega0expan}
\end{align}

To leading order, the Weyl factor (\ref{eq:weylfactSphi}) corresponding to these modes is given by  
\begin{align}
    \delta\bomega&= \frac{2^{l-1}l(l+1)\Gamma(l+\frac{3}{2})\mathbb{S}}{\sqrt{\pi}\Gamma(l+1)} \log|K\ell|+\mathcal{O}(1) \,. \label{eq: deltaw w=0}
\end{align}
As expected, the leading order contribution to the Weyl factor is time-independent.

\subsubsection{The conformal stress-tensor}

Next we evaluate the conformal stress tensor for these gapless modes. Note that since the leading frequency is zero, the conformal stress tensor becomes time-independent to leading order. Moreover, it can be verified that it does not depend on the leading correction to the frequency. For the first non-trivial contribution in this case, we obtain, 
\begin{equation}
\begin{cases}
    8\pi G_N\delta T_{tt} 
    = \left( -\frac{2}{\ell|K\ell|} +\mathcal{O}(|K\ell|^{-3})  \right) \delta\bomega \,, \\
    8\pi G_N\delta T_{ti} = \left( \frac{2|K\ell|}{l(l+1)}\tilde{\nabla}_i+ \mathcal{O}(|K\ell|^{-1})\right)\partial_t\delta\bomega \,,\\
     8\pi G_N\delta T_{ij} 
     = \left( -\ell |K\ell| \tilde{g}_{ij}+ \mathcal{O}(|K\ell|^{-1}) \right) \delta\bomega  \,.
    \end{cases}
\end{equation}
Using the expansion for $\delta \bomega$ given in \eqref{eq: deltaw w=0}, it follows that $\delta T_{ti}$ vanishes at leading order in the large-$|K\ell|$ limit. However, $\delta T_{ti}$ will receive non-zero subleading corrections, by virtue of the time-dependence in the full linearised Weyl factor (\ref{eq:weylfactSphi}).

It is also straightforward to compute
\begin{equation} \label{TdT pole omega0}
     64\pi^2 G_N^2 \bar{T}^{mn} \delta T_{mn} = \frac{2^{l}l(l+1)\Gamma(l+\frac{3}{2})\mathbb{S}}{\ell^2\sqrt{\pi}\Gamma(l+1)} |K\ell|^2\log|K\ell| + \mathcal{O}(|K\ell|^{-2}) = \left( \frac{2}{\ell^2}|K\ell|^2  + \mathcal{O}(1) \right) \delta\bomega\;.
\end{equation}
The subdominant contribution follows from substituting in the first correction in the $\omega = 0$ expansion (\ref{eq:omega0expan}) on-shell. At leading order, the linearised Weyl equation \eqref{linearized Weyl eq} gives
\begin{align}\label{leadingorderWeyleq}
    2|K\ell|^2 \big(\partial_t^2\delta\bomega(t) + \ell^{-2} \delta\bomega(t) \big) =64\pi^{2}G_{N}^{2}\bar{T}^{mn}\delta T_{mn}\;.
\end{align}
Observe that the second term on the left hand side equals \eqref{TdT pole omega0}. Thence, requiring \eqref{linearized Weyl eq} be satisfied, we deduce the Weyl factor is time-independent at leading order, consistent with the explicit computation \eqref{eq: deltaw w=0}. 

Importantly, and in contrast to the $\pm i$ modes, we  need a non-vanishing (but time-independent) contribution from the conformal stress-tensor in order to obtain the correct leading order $\omega\ell=0$ frequency. This implies these modes require more information about the bulk than the $\omega\ell=\pm i$ modes. We will confirm this below with the bulk radial profile which, though localised, decays slower than the profile for the soft modes.

\subsubsection{The bulk radial profile}
In the strict large-$|K\ell|$ limit, the pole patch master field for the modes with $\omega\ell=0$ also diverges at the boundary. The difference with respect to the $\omega\ell=\pm i$ modes is that in this case the divergence is logarithmic in $|K\ell|$. 

\begin{figure}[t!]
    \centering
    
    \begin{subfigure}[t]{0.48\textwidth}
        \centering
        \includegraphics[width=\textwidth]{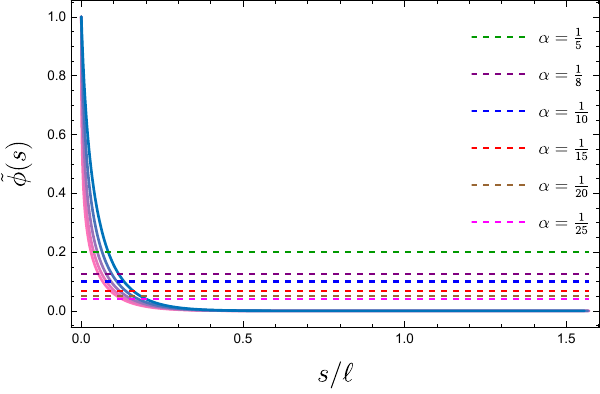}
        \caption{Radial profiles.}
        \label{fig:om0profsPP}
    \end{subfigure}
    \hfill
    \begin{subfigure}[t]{0.48\textwidth}
        \centering
         \begin{picture}(0,0)
        \put(108,33){\tiny $\alpha$} 
    \end{picture}%
        \includegraphics[width=1.03\textwidth]{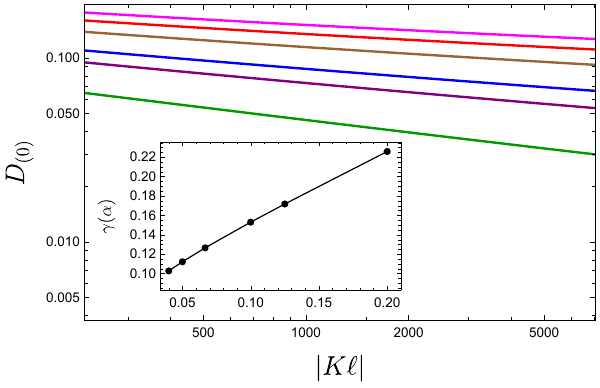}
        \caption{Width.}
        \label{fig:disom0PP}
    \end{subfigure}
    \caption{Normalised radial profile $\tilde{\phi}(s)$ and width $D_{(0)}$ for gapless  modes (here $l=10$). \emph{Left.} Bulk profiles (solid curves)  at various $|K\ell|$: from right to left $|K\ell|=71,224,707,2236,7071$. As $|K\ell|$ increases, the profiles become more localised. Horizontal dashed lines denote magnitudes of $\alpha\tilde{\phi}$. \emph{Right.} Width $D_{(0)}$ at various $\alpha$: top to bottom $\alpha=\tfrac{1}{25},\tfrac{1}{20},\tfrac{1}{15},\tfrac{1}{10},\tfrac{1}{8},\tfrac{1}{5}$. Inset displays $\gamma(\alpha)$ for the values of $\alpha$ analysed.
    }
    \label{fig:bulkprofsandDom0}
\end{figure}

As with the previous modes, we plot in figure \ref{fig:om0profsPP} the normalised master field \eqref{eq: normalised phi} as a function of the proper distance from the boundary. In figure \ref{fig:disom0PP}, we show the characteristic width for different values of $\alpha$. For all $\alpha$'s analysed, the width goes to zero at large-$|K\ell|$, with a rate that is slower to the one for the soft modes. As opposed to those modes, in this case, the rate of decay depends on $\alpha$, and for $\alpha$ sufficiently small, we observe that $D_{(0)} \sim |K\ell|^{-\gamma(\alpha)}$, with $\gamma$ positive and smaller than one. It would be interesting to have an analytical understanding of this scaling. 

\subsection{Scalar normal modes} \label{subsec:normaltowerPP}

We now turn to the final set of solutions to \eqref{eq largeKl pole}, corresponding to the vanishing of the square bracket. It is straightforward to verify, at leading order in a large-$|K\ell|$ expansion, these solutions have a vanishing master field evaluated at the boundary. Consequently, as we will show, these modes have vanishing (linearised) Weyl factor.

Setting the bracket in the expansion \eqref{eq largeKl pole} to zero yields an infinite tower of solutions parametrised by a non-zero integer $n$,
\begin{align}\label{pole tower eq}
    \omega_n \ell&=i\frac{
\log\!\left(
\frac{-2^{\,i\ell\omega_n}\,\Gamma(1 + l - i\ell\omega_n)\,\Gamma( i\ell\omega_n)}
{\Gamma(1 + l + i\ell\omega_n)\,\Gamma(- i\ell\omega_n)}
\right) - 2\pi i n
}{
\log(2|K\ell|^2)
},\hspace{1cm}n\in\mathbb{Z}/\{0\}\;.
\end{align}

Solutions to the tower (\ref{pole tower eq}) can be numerically determined and are real. Note $n=0$ is not a solution to \eqref{eq largeKl pole}.

In the following, we find analytic approximations of these solutions in the large-$|K\ell|$ limit. Since the solutions depend on a new parameter, $n$, in order to fully specify the large-$|K\ell|$ limit, we need to state how $n$ scales with $|K\ell|$. This will distinguish between two sets of modes.

First, there are low frequency modes, which behave as $\omega_n\ell\rightarrow 0$ in the large-$|K\ell|$ limit. There is another set of modes whose frequency does not vanish in the stretched horizon limit. Its frequencies satisfy $\omega_n\ell\gg l$. In the following, we will study each set separately, characterising the range of validity of each approximation. 

\subsubsection{Real scalar modes with low frequencies} 
We first study the set of real modes that have vanishing frequency in the strict large-$|K\ell|$ limit.

\textbf{Allowed frequencies.} We consider an expansion ansatz of the form
\begin{align}\label{ansatz pole tower}
    \omega_n\ell&= \frac{\pi n }{\log |K\ell|}\bigg(a_0+ \frac{a_1}{\log|K\ell|}+ \frac{a_2}{\log^2|K\ell|}+\frac{a_3}{\log^3|K\ell|}+\mathcal{O}(\log^{-4}|K\ell|)\bigg)\;.
\end{align}
Again, this expansion is valid for small $\omega_n\ell$, which implies $n\ll \log|K\ell|$.
Demanding consistency between the left and right sides of the tower (\ref{pole tower eq}), the coefficients are determined to be
\begin{align}
\begin{cases}
    a_0=1,\\a_1=H_l-\log 2,\\a_2=(H_l-\log 2)^2, \\ a_3=\bigl(H_l - \log 2\bigr)^3
- \frac{n^2 \pi^2}{6}
\left( \psi^{(2)}(1 + l) + 2 \zeta(3) \right) \,,
\end{cases}
\end{align}
where $\zeta(3)$ is the Riemann zeta function and $\psi^{(2)}$ is the polygamma function of order two. Note that the expansion (\ref{ansatz pole tower}) actually breaks down when $n^2$ is of order $\log|K\ell|$. In this case, $\omega_n\ell$ is large enough such that ansatz (\ref{ansatz pole tower}) is insufficient. We analyse those modes in the next subsection.

In the large-$|K\ell|$ limit, the space between neighbouring modes,
\begin{align}
    \Delta \omega \equiv \omega_{n+1}-\omega_n =\frac{\pi/\ell}{\log|K\ell|}\;,
\label{eq:spacingscalnormmodes}\end{align}
vanishes and we obtain a continuum of real modes.

As noted above, the Weyl factor corresponding to the $n$th mode in the tower, with $\omega_n\ell$ given in \eqref{ansatz pole tower}, vanishes at leading order in the large-$|K\ell|$ expansion. To find the first nontrivial correction to $\delta\bomega$ requires going beyond \eqref{pole tower eq}. Expanding the $\delta K=0$ boundary condition at next order, we find that the first correction to the tower of real modes \eqref{pole tower eq} is given by
\begin{align} \label{correction tower}
\omega_n\ell&=\frac{i}{\log\!\left(2 |K \ell|^2\right)}
\bigg[
\log\!\left(
-\frac{
    2^{i \omega_n\ell} \, 
    \Gamma(1 + l - i \omega_n\ell) \, 
    \Gamma(i \omega_n\ell)
}{
    \Gamma(-i \omega_n\ell) \, 
    \Gamma(1 + l + i \omega_n\ell)
}\right)
\nonumber\\
&+
\frac{1}{2 |K \ell|^2}
\left(
\frac{
    i \left(
        4(-2 + l + l^2)
        + (-9 + 2 l (1 + l)) (\omega_n\ell)^2
        - (\omega_n\ell)^4
    \right)
}{
    2 \omega_n\ell (1 + (\omega_n\ell)^2)
}
\bigg)
- 2 \pi i n
\right]\;.
\end{align}
Considering an ansatz of the form $\omega_n=\omega_{n0}+\omega_{n1} |K\ell|^{-2}$, where $\omega_{n0}$ solves the leading contribution to the tower \eqref{pole tower eq}, we find, for each $n$, that $ \omega_{1n}$ may be expressed in terms of $ \omega_{0n}$. Furthermore, in the large-$|K\ell|$ limit, $\omega_{1n}$ can be further expanded in an infinite series of logarithmic terms.

To leading order, the linearised Weyl factor is  given by
\begin{align}
    \delta\bomega&=\frac{(-1)^{n + 1} \, 2^{l-2} \, l (l + 1) (l - 1) (l + 2) \, \Gamma\!\left(\tfrac{3}{2} + l\right) \, \mathbb{S} \,}
{n^2 \, \pi^{5/2} \, \Gamma(1 + l) \,} \frac{\log^2|K\ell|}{|K\ell|^2} + \mathcal{O}\bigg(\frac{\log|K\ell|}{|K\ell|^2}\bigg)\;.
\end{align}

\textbf{The conformal stress tensor.} The conformal stress tensor associated to \eqref{ansatz pole tower} is given by 
\begin{equation}\label{stresstensor tower}
\begin{cases}
    8\pi G_N\delta T_{tt}
    = \left( -\frac{2}{\ell|K\ell|} +\mathcal{O}(|K\ell|^{-1}\log^{-2}|K\ell|)  \right) \delta\bomega \,, \\
   8\pi G_N\delta T_{ti} = \left( \frac{2|K\ell|}{l(l+1)} \tilde{\nabla}_i+ \mathcal{O}(|K\ell| \log^{-2}|K\ell|)\right)\partial_t\delta\bomega \,,\\
    8\pi G_N \delta T_{ij} =\left( \frac{4\pi^2 n^2\ell}{l(l+1)(l-1)(l+2)}\frac{|K\ell|^3}{\log^2|K\ell|}\left(\tilde{\nabla}_i \tilde{\nabla}_j - \frac{1}{2}\tilde{g}_{ij} \tilde{\nabla}^2\right) + \mathcal{O}(|K\ell|^{3}\log^{-3}|K\ell|) \right) \delta \bomega\, ,
    \end{cases}
\end{equation}
while, 
\beq 64\pi^{2}G_{N}^{2}\bar{T}^{mn}\delta T_{mn}=\left(\frac{2|K\ell|^{2}}{\ell^{2}}+\mathcal{O}\left(|K\ell|^{2}\log^{-2}|K\ell|\right)\right)\delta\bomega\;.\eeq
Using it in the Weyl mode equation \eqref{linearized Weyl eq}, we consistently obtain that the Weyl mode is time-independent at leading order, hence the frequency is 0.

Thus, akin to the $\omega\ell=0$ case, and different from the $\omega\ell=\pm i$ modes, we need to consider the contribution from the conformal stress-tensor in the Weyl mode equation (\ref{eq: linearised 2.12}) to maintain consistency. 

\textbf{The bulk radial profile.} 
In contrast to the other sets of modes studied so far, the pole patch master field for the tower of normal modes does not diverge at the boundary, but instead vanishes. For a given $n$, the radial profile oscillates close to the boundary, and then decays to zero, so it is still reasonable to compute the radial width of the master field at large $|K\ell|$. 

\begin{figure}[t!]
    \centering
    
    \begin{subfigure}[t]{0.48\textwidth}
        \centering
        \includegraphics[width=\textwidth]{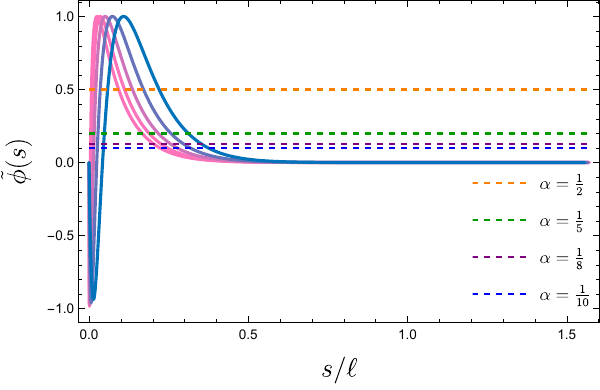}
        \caption{Radial profiles.}
        \label{fig:omrealprofsPPn2}
    \end{subfigure}
    \hfill
    \begin{subfigure}[t]{0.48\textwidth}
        \centering
        \begin{picture}(0,0)
        \put(80,45){\rotatebox{-12}{\small $\sim \log|K\ell|^{-1}$}} 
    \end{picture}%
        \includegraphics[width=1.03\textwidth]{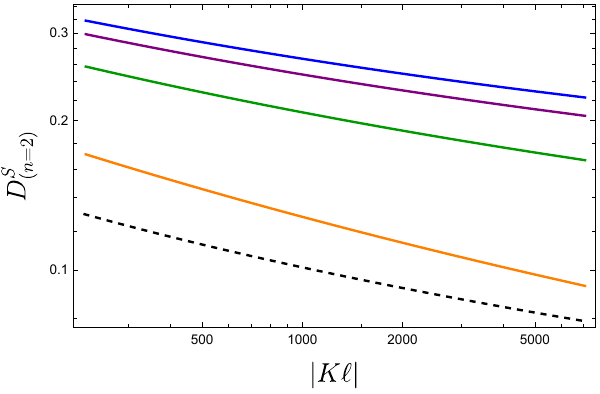}
        \caption{Width.}
        \label{fig:disomrealPPn2}
    \end{subfigure}
    \caption{Normalised radial profile $\tilde{\phi}(s)$ and width $D_{(n=2)}^S$ for tower of real scalar modes with low-frequency (here $l=10$, $n=2$). \emph{Left.} Bulk profiles (solid curves)  at various $|K\ell|$; from right to left $|K\ell|=71,224,707,2236,7071$. As $|K\ell|$ increases, the width shrinks. Horizontal dashed lines show different values of $\alpha$, where we compute the width. \emph{Right.}  Width $D_{(n=2)}^S$ at various $\alpha$: top to bottom $\alpha=\tfrac{1}{10},\tfrac{1}{8},\tfrac{1}{5},\tfrac{1}{2}$. The width still decays at large $|K\ell|$, but much slower, at a logarithmic rate (indicated by black dashed line).}
    \label{fig:bulkprofsandDomrealn2}
\end{figure}

We observe that the maximum value of the master field for these modes grows logarithmically with $|K\ell|$. As such, to characterise the radial profile we define the normalised field $\tilde\phi(s)$, by dividing the master field by its maximum value. Note this slightly differs from the definition \eqref{eq: normalised phi}. In this case, the normalised field $\tilde\phi (s)$ will be zero at the boundary, then oscillate and reach a maximum of one near the boundary, before finally decaying to zero. This is shown in figure \ref{fig:omrealprofsPPn2} for $n=2$. We also compute the width at different heights $\alpha$ of the normalised field. These are shown in figure \ref{fig:disomrealPPn2}, where we still observe a decay at large $|K\ell|$, but it is of a much slower nature. For the values of $\alpha$ analysed in the plot, we observe that $D_{(n=2)}^S \sim 2 \log^{-1} |K\ell|$. Similar analysis for other values of $n$ reveal $D_{(n)}^S \sim n \log^{-1} |K\ell|$, at large $|K\ell|$. 

\subsubsection{Real scalar modes with high frequencies}
Next we proceed with a similar analysis for the high frequency modes which do not vanish in the large-$|K\ell|$ limit. As we shall see in a moment, they correspond to $n \gg \log |K\ell|$.

\textbf{Allowed frequencies.} In order to obtain high frequencies, we assume that they have $\omega_n\ell\gg l$, so that we can ignore the $l$-dependence in \eqref{pole tower eq}. In this case, we obtain the same leading order result $\omega_n\ell=\frac{\pi n}{\log|K\ell|}$, but which is valid now for $n\gg\log|K\ell|$. A better estimate for the high frequencies is obtained using Stirling's approximation. We find
\begin{align}\label{solutionlargen}
    \omega_n\ell&=\frac{(2n+l)\pi }{2\log(2|K\ell|)}- \frac{l(l+1)}{(2n+l)\pi}  +\mathcal{O}\bigg(\frac{\log^2|K\ell|}{n^3}\bigg),
\end{align}
Clearly, the space between neighbouring modes  vanishes in the large-$|K\ell|$ limit and we obtain a continuum of real modes with high frequencies. 

\textbf{The conformal stress tensor.} 
Using \eqref{eq:TdeltaTweyl}, we find\footnote{To be more precise about the limit we are considering, we expand first in large-$|K\ell|$ and then in large $n$. The same result is obtained if we expand first in large $\omega$ and then in large $|K\ell|$.}
\begin{align}
     64\pi^{2}G_{N}^{2}\bar{T}^{mn}\delta T_{mn}=\left(\frac{2 n^2 \pi^2 |K\ell|^{2}}{\ell^{2} \log^2|K\ell|}+\mathcal{O}\left(n^2|K\ell|^{2}\log^{-3}|K\ell|\right)\right)\delta\bomega\;.
\end{align}
Keeping only the leading order terms in the large-$|K\ell|$ limit, on both sides of the Weyl mode equation \eqref{linearized Weyl eq}, we find
\begin{align}
    2|K\ell|^2 \big(\partial_t^2\delta\bomega(t) + \ell^{-2} \delta\bomega(t) \big) = \frac{2 n^2 \pi^2 |K\ell|^{2}}{\ell^{2} \log^2|K\ell|}\delta \bomega(t)\;.
\end{align}
Assuming an ansatz $\delta\bomega(t)=N_{\bomega}e^{-i\omega t}$, we obtain the correct leading order frequency $\omega_n\ell=\frac{\pi n}{\log|K\ell|}$. Note the $\delta \bomega(t)$ term is subleading compared to the stress tensor contribution. Just like in the case of the low frequency modes, it is crucial to consider the contribution from the conformal stress-tensor in the Weyl mode equation \eqref{linearized Weyl eq} for consistency. The difference between the low and high frequency modes consists in which term, the Weyl mode or its double time derivative dominates in \eqref{linearized Weyl eq} in the large-$|K\ell|$ limit. %

\textbf{The bulk radial profile.} For the large $\omega_n\ell$ modes, for which \eqref{solutionlargen} is a good approximation, the radial profile is not localised near the boundary. Instead, it exhibits oscillatory behaviour through the whole static patch. 

\begin{figure}[t!]
    \centering
    
    \begin{subfigure}[t]{0.47\textwidth}
        \centering
        \includegraphics[width=\textwidth]{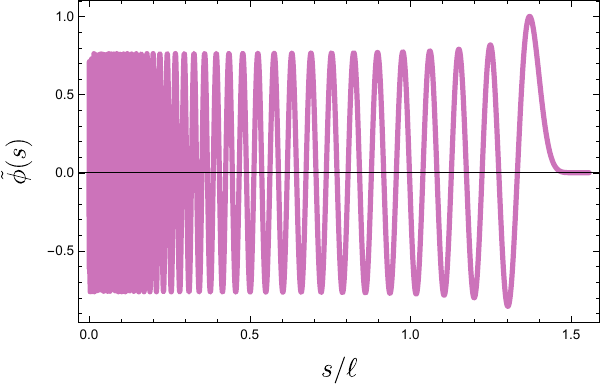}
        \caption{Scalar normal mode.}
        \label{fig:scaltower-all}
    \end{subfigure}
    \hspace{5mm}
    \begin{subfigure}[t]{0.47\textwidth}
        \centering
        \includegraphics[width=\textwidth]{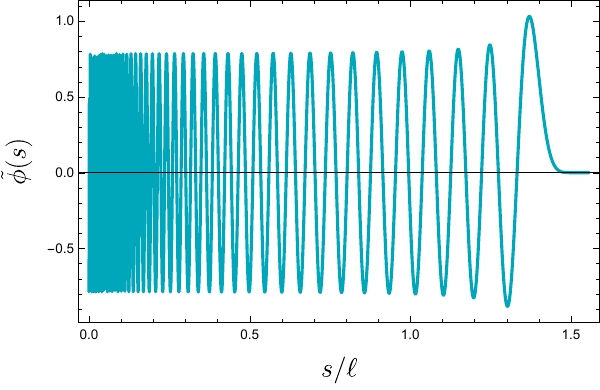}
        \caption{Vector normal mode.}
        \label{fig:vectower-all}
    \end{subfigure}
    \caption{Normalised $\tilde{\phi}(s)$ for normal mode with $n=100$ of (a) scalar, (b) vector sector
    ($l=10, |K\ell|\approx 71$). }
    \label{fig: tower vector and scalar}
\end{figure}

We plot the normalised radial profile for these high frequency modes in figure \ref{fig:scaltower-all}, which  exhibits qualitatively different behaviour compared to the low frequency modes in figure \ref{fig:bulkprofsandDomrealn2}. As with the $n=2$ tower modes, we normalise with respect to the maximum value of the profile. 

\subsection{Vector normal modes}

So far we have only focused on the large-$|K\ell|$ behaviour of modes characterizing the scalar sector of metric perturbations in the pole patch. Let us now turn to the vector modes. These are found by expanding the boundary condition \eqref{eq: vector bdy condition} in the large-$|K\ell|$ limit. At leading order we find
\begin{align} \label{large Kl vector}
    0&=\frac{2^{-i \omega\ell} \, \Gamma(1 - i \omega\ell)}{\Gamma(1 + l - i \omega\ell)\,} 
\;|K\ell|^{- i \omega\ell}+\;
\frac{2^{i \omega\ell} \Gamma(1+i \omega\ell)}{\, \Gamma(1 + l + i \omega\ell)}|K\ell|^{i \omega\ell}\;,
\end{align}
up to an overall non-vanishing factor. We can rewrite the equation as 
\begin{align}\label{vectorpoleeqq}
    \omega_n \ell&=i\frac{
\log\!\left(
\frac{2^{\,i\ell\omega_n}\,\Gamma(1 + l - i\ell\omega_n)\,\Gamma( i\ell\omega_n)}
{\Gamma(1 + l + i\ell\omega_n)\,\Gamma(- i\ell\omega_n)}
\right) - 2\pi i n
}{
\log(2|K\ell|^2)
}\;,\hspace{1cm}n\in\mathbb{Z}/\{0\}\;.
\end{align}
All solutions are real and, as mentioned in section \ref{section 2.2:linearised dynamics}, the corresponding Weyl factor vanishes at the boundary. 
In the following, we obtain analytic approximations of the solutions, depending on how $n$ scales with $|K\ell|$. The structure of equation \eqref{vectorpoleeqq} is the same as of \eqref{pole tower eq}. The only difference is a minus sign in the argument of the logarithm. Given this, our discussion will be brief. 

\subsubsection{Vector modes with low frequencies}

\textbf{Allowed frequencies.} We consider an expansion ansatz of the form
\begin{align}\label{ansatz pole tower vector}
    \omega_n\ell&= \frac{(2n-1)\pi  }{2\log |K\ell|}\bigg(a_0+ \frac{a_1}{\log|K\ell|}+ \frac{a_2}{\log^2|K\ell|}+\frac{a_3}{\log^3|K\ell|}+\mathcal{O}(\log^{-4}|K\ell|)\bigg)\;,
\end{align}
for $n\ll \log|K\ell|$. The coefficients are determined to be
\begin{equation}
   a_0=1, \quad a_1=H_l-\log 2, \quad a_2=(a_1)^2, \quad a_3=(a_1)^3
- \frac{(2n-1)^2 \pi^2}{24}
\left( \psi^{(2)}(1 + l) + 2 \zeta(3) \right) \,.
\end{equation}
As in the case of the tower of scalar normal modes, the large-$|K\ell|$ expansion breaks down for $n^2\sim \log |K\ell|$. Meanwhile, at leading order in the expansion, the spacing between neighbouring modes takes the form
$
    \Delta \omega \ell =\pi \log^{-1} |K\ell|\;,
$
and goes to zero in the large-$|K\ell|$ limit.

\textbf{The conformal stress tensor.} We analyse the conformal stress tensor for the modes \eqref{ansatz pole tower vector}. While the Weyl factor vanishes at the boundary,  the master field at the boundary is nonzero. Specifically,
\begin{align}\label{vectorfieldbd}
    \Phi^{(V)}|_{\Gamma}&=\frac{(-1)^n 2^{l+2}\Gamma(l+\frac{3}{2})}{(2n-1)\pi^{3/2}\Gamma(l+1)} \log|K\ell| + \mathcal{O}(1)\;.
\end{align}
We observe the master field at the boundary diverges in the strict large-$|K\ell|$ limit.

Consequently, the nonzero components of the stress tensor (\ref{eq:deltaTmnvecgen}) in a large-$|K\ell|$ expansion are, 
\beq  \label{componentsvectortower}
\begin{cases}
 8\pi G_{N}\delta T_{ti}=\left(-\frac{(l-1)(l+2)}{2|K\ell|}+\mathcal{O}((|K\ell|\log|K\ell|)^{-1})\right)\Phi^{(V)}\mathbb{V}_{i}\;,\\
 8\pi G_{N}\delta T_{ij}=\left(\frac{|K\ell|}{2}+\mathcal{O}(|K\ell|^{-1})\right)\partial_t\Phi^{(V)}(\tilde{\nabla}_{i}\mathbb{V}_{j}+\tilde{\nabla}_{j}\mathbb{V}_{i})\;.
\end{cases}
\eeq
where $\Phi^{(V)}$ is evaluated at the boundary. At leading order in the large-$|K\ell|$ expansion, it is given in \eqref{vectorfieldbd} and it is time-independent, which implies that to leading order $\delta T_{ij}$ is zero. 

\begin{figure}[t!]
    \centering
    
    \begin{subfigure}[t]{0.48\textwidth}
        \centering
        \includegraphics[width=\textwidth]{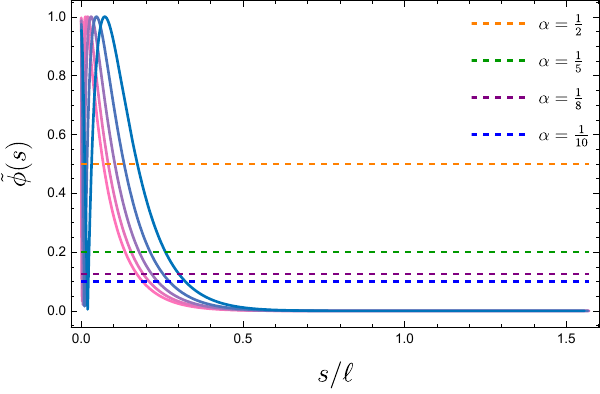}
        \caption{Radial profiles.}
        \label{fig:omrealvecprofsPPn2}
    \end{subfigure}
    \hfill
    \begin{subfigure}[t]{0.48\textwidth}
        \centering
        \begin{picture}(0,0)
        \put(85,42){\rotatebox{-12}{\small $\sim \log|K\ell|^{-1}$}} 
    \end{picture}%
        \includegraphics[width=1.03\textwidth]{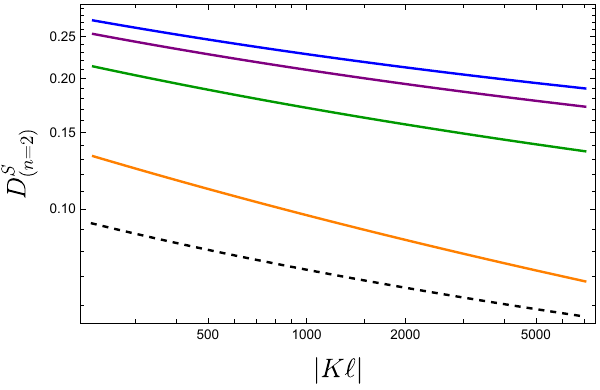}
        \caption{Width.}
        \label{fig:disomrealvecPPn2}
    \end{subfigure}
    \caption{Normalised radial profile $\tilde{\phi}(s)$ and width $D_{(n=2)}^V$ for tower of real vector modes with low-frequency (here $l=10$, $n=2$). \emph{Left.} Bulk profiles (solid curves)  at various $|K\ell|$; from right to left $|K\ell|=71,224,707,2236,7071$. As $|K\ell|$ increases, the width shrinks. Horizontal dashed lines show different values of $\alpha$, where we compute the width. \emph{Right.}  Width $D_{(n=2)}^V$ at various $\alpha$: top to bottom $\alpha=\tfrac{1}{10},\tfrac{1}{8},\tfrac{1}{5},\tfrac{1}{2}$. The width still decays at large $|K\ell|$, but much slower, at a logarithmic rate (indicated by black dashed line).}
    \label{fig:bulkprofsandDomrealvecn2}
\end{figure}

\textbf{The bulk radial profile.} The bulk radial profile of the vector modes is very similar to the one of the real scalar modes. 

To characterise the radial profile, we again  normalise the master field by dividing by its maximum value; the normalised field $\tilde\phi (s)$ starts at a nonzero value at the boundary, reachs a maximum of unity near the boundary and then decays to zero. This is shown in figure \ref{fig:omrealvecprofsPPn2} for $n=2$. Similar to the other modes analysed, we compute the width at different heights $\alpha$ of the normalised field. These are shown in figure \ref{fig:disomrealvecPPn2}, where we observe a logarithmic decay at large $|K\ell|$. More precisely, for the values of $\alpha$ shown in the plot, we observe that $D_{(n=2)}^V \sim \frac{3}{2} \log^{-1} |K\ell|$. Similar analysis for other values of $n$ reveals $D_{(n)}^V \sim \frac{2n-1}{2} \log^{-1} |K\ell|$, at large $|K\ell|$.

\subsubsection{Vector modes with high frequencies}

\textbf{Allowed frequencies.} 
As before, we use Stirling's approximation to obtain 
\begin{align}\label{solutionlargenVector}
    \omega_n\ell&=\frac{(2n+l+1)\pi }{2\log(2|K\ell|)}- \frac{l(l+1)}{(2n+l+1)\pi}  +\mathcal{O}\bigg(\frac{\log^2|K\ell|}{n^3}\bigg),
\end{align}
which is the vector sector analogue of \eqref{solutionlargen}. The space between neighbouring modes still vanishes in the large-$|K\ell|$ limit. 

\textbf{The conformal stress tensor.} Since \eqref{eq:deltaTmnvecgen} only depends on the frequency through the master field evaluated at the boundary, the nonvanishing components of the conformal stress-tensor are still given by \eqref{componentsvectortower}, but for $\Phi^{(V)}$ replaced by 
\begin{align}
    \Phi^{(V)}|_{\Gamma}&=\mathfrak{R} e^{-i\omega_n t} \frac{2^{l+1}\Gamma(l+\frac{3}{2})}{\sqrt{\pi}}\frac{1}{(\omega_n\ell)^{l+1}}\cos\bigg(\omega_n\ell \log|K\ell|-\frac{\pi}{2}(l+1)\bigg), 
\end{align}
at leading order in the large-$|K\ell|$ and large $\omega$ expansion. As we can check numerically, increasing $n$ for $l,\mathfrak{r}$ fixed, the absolute value of the master at the boundary decreases.

\textbf{The bulk radial profile.} For the regime $n\gg\log|K\ell|$, in which the frequencies do not vanish in the large-$|K\ell|$ limit, the radial profile is not localised near the boundary.
Just like for the high frequency modes in the scalar sector, the master field exhibits oscillatory behaviour through the whole static patch. We plot the radial profile for these modes in figure \ref{fig:vectower-all}.

\subsection{Connection to quasinormal modes}\label{sectionQNM}

In the strict $|K\ell|\rightarrow\infty$ limit, the boundary coincides with the cosmological horizon. Therefore, it is natural to inquire whether the de Sitter quasinormal modes \eqref{qnmdeSitter} can be recovered from the two towers of the pole patch scalar and vector normal modes. A key observation supporting this possibility is that, in both cases, the leading-order term of the linearised Weyl factor vanishes at the boundary,\footnote{In the vector sector, the linearised Weyl factor vanishes identically, while for the scalar sector tower of normal modes it only vanishes at leading order in the large $|K\ell|$ expansion} so one recovers essentially the Dirichlet `brickwall' problem \cite{tHooft:1984kcu}.

The encoding of quasinormal modes in a set-up with a brickwall near a horizon can be formulated in terms of a scattering matrix \cite{Law:2022zdq}. Specifically, the Dirichlet boundary condition at the wall gives rise to an infinite set of real normal modes which accumulate in the limit the wall approaches the horizon. In the limit, the density of real modes can be expressed in terms of an S-matrix
\begin{align}\label{scatteringphase}
    S(\omega) =\frac{A^{\text{out}}(\omega)}{A^{\text{in}}(\omega)}\;.
\end{align}
Here $A^{\text{out(in)}}$ denote the outgoing (ingoing) components of the solutions to the equation of motion for the field in question, e.g., a scalar field propagating on a black hole background, near the horizon.
Poles of the S-matrix occur for frequencies when the ratio (\ref{scatteringphase}) has poles. The S-matrix determines a density of states $\varrho(\omega)$ that diverges when the Dirichlet cutoff is removed. To ameliorate this divergence, it is common to subtract a reference background and analyze a regulated density of states, $\Delta\varrho(\omega)$. Notably, this regulated density of states defines a renormalized bulk thermal canonical partition function, which, for a specific choice of reference background,  equals the one-loop Euclidean gravitational path integral \cite{Anninos:2020hfj,Law:2022zdq}. A direct consequence of this observation is that the poles of $\Delta\varrho$ precisely coincide with the quasinormal modes of both the original and reference backgrounds.\footnote{In particular, the specific choice of reference background is flat Rindler, such that the poles $\Delta\varrho$ also include the Rindler quasinormal modes.  We will return to this point in section \ref{sec: rindler}.}

In our context, we impose regularity of the pole patch solutions to the master field equation at the origin, which implies the solutions are a linear combination of purely ingoing and outgoing components at the boundary. As explained above, our boundary condition, essentially Dirichlet at leading order in the large-$|K\ell|$ expansion, gives rise to an infinite tower of real normal modes, whose spacing \eqref{eq:spacingscalnormmodes} vanishes in the strict $|K\ell|\rightarrow \infty$ limit. Explicitly, the scattering phase \eqref{scatteringphase} in our case is given by\footnote{These phases can be obtained in a straightforward manner by taking the ratios of the outgoing and ingoing components in (\ref{eq largeKl pole}) for the scalar and for the vector (\ref{large Kl vector}), respectively.}
 \begin{align}\label{QNMssmatrix}
        \bigg(\frac{A^{\text{out}}}{A^{\text{in}}}\bigg)_{\hspace{-1mm}S}&=  \frac{2^{i\omega\ell} \, \Gamma( i\omega \ell) \, \Gamma(1 + l - i\omega\ell)}
{\Gamma(1 + l + i\omega\ell) \, \Gamma(- i\omega\ell)}\;,\quad  \bigg( \frac{A^{\text{out}}}{A^{\text{in}}}\bigg)_{\hspace{-1mm} V}= \frac{2^{i\omega\ell} \, \Gamma(1 + i\omega \ell) \, \Gamma(1 + l - i\omega\ell)}
{\Gamma(1 + l + i\omega\ell) \, \Gamma(1 - i\omega\ell)}\;.
    \end{align}
In either case, among the poles of the scattering phase are the de Sitter quasinormal modes (\ref{qnmdeSitter}), $\omega_n\ell =-i(n+l+1)$.

The poles of the scattering phase correspond to the zeros of $A^{\text{in}}$, the coefficient of the purely ingoing component. Thus, the encoding of the quasinormal modes in our analysis is natural: instead of solving for $\omega$ from the linearised boundary condition imposed on solutions that are regular at the origin, here we solve for the modes that come from zeros of the purely ingoing component of the solution at the boundary.

The return of the ringing quasinormal modes in systems with discrete real normal modes that accumulate in the limit the brickwall responsible for the discreteness is pushed towards a horizon can be formulated in terms of retarded Green's functions (see, e.g., \cite{Anninos:2011af,Son:2002sd,Law:2022zdq,Giusto:2023awo,Banerjee:2024dpl,Banerjee:2024ivh,Das:2024fwg}). 
As poles accumulate, the Green's function, which is directly related to the scattering phase \eqref{scatteringphase}, develops a branch cut on the real axis with the discontinuity given by the density of states, 
\begin{align}
    G(\omega+i\varepsilon) - G(\omega-i\varepsilon)=2\pi i \varrho(\omega) =2 i \text{Im} G_R(\omega)\;.
\end{align}
This discontinuity is nothing but the imaginary part of the retarded Green's function, whose poles
correspond to the quasinormal modes.

\subsection{Summary}

A careful analysis of the linearised gravitational dynamics in the stretched horizon limit of the pole patch reveals an interesting fellowship of modes that can be divided into three categories: purely boundary, near-boundary, and bulk modes.

\textbf{Purely boundary modes.} These modes are defined as those that can be obtained directly from \eqref{eq: linearised 2.12}, without needing any contribution from the linearised stress tensor $\delta T_{mn}$. In the large-$|K\ell|$, we found two distinct set of modes with this characteristic, that differ on how we take the large-$|K\ell|$ limit compared to the large angular momentum $\L$ limit. 

\begin{itemize}
    \item \textbf{Large angular momentum modes} have $\L/|K\ell| \gg 1$. These were analysed in section \ref{sec: pole large L}. They are the most localised modes, with a radial profile that decays exponentially away from the boundary.  
    See table \ref{tab:large-l} for summary.
    \item \textbf{Soft modes} appear in the opposite limit, $\L/|K\ell| \ll 1$. To leading order their frequency is $\omega \ell = \pm i$, independent of $\L$. Their radial profile localises with a width that decays as $|K\ell|^{-1}$ in the large-$|K\ell|$ limit.
\end{itemize}
Both types of modes have frequencies with positive imaginary part, so the linearised Weyl factor grows exponentially with time.

\textbf{Near-boundary modes.} These modes appear as a non-trivial solution to \eqref{eq: linearised 2.12}, where the contribution of $\delta T_{mn}$ is needed even at leading order at large-$|K\ell|$. We call them near-boundary modes, because nevertheless, their radial profile localises close to the boundary. We summarise their properties below.

\begin{itemize}
    \item \textbf{Gapless scalar modes} have vanishing frequency and constant-in-time stress tensor to leading order in the large-$|K\ell|$ limit. The first correction scales as $|K\ell|^{-2}$.  The radial profile for these modes  decays away from the boundary with a power-law behaviour, but slower than the soft modes.
    \item \textbf{Real vector and scalar modes.} Both sectors admit an infinite number of normal modes, labeled by integer $n$. In the limit $n \ll \log |K\ell|$, their frequencies decay to zero logarithmically, while their radial profile decays logarithmically with the proper distance away from the boundary.
\end{itemize}

\textbf{Bulk modes.} In the opposite limit, $n \gg \log |K\ell|$, there exist an infinite set of normal modes both in the vector and scalar sectors that do not localise close the boundary. Instead, their radial profiles have support over the whole static patch.

\begin{figure}[t!]
\centering
 \includegraphics[width=.8\textwidth]{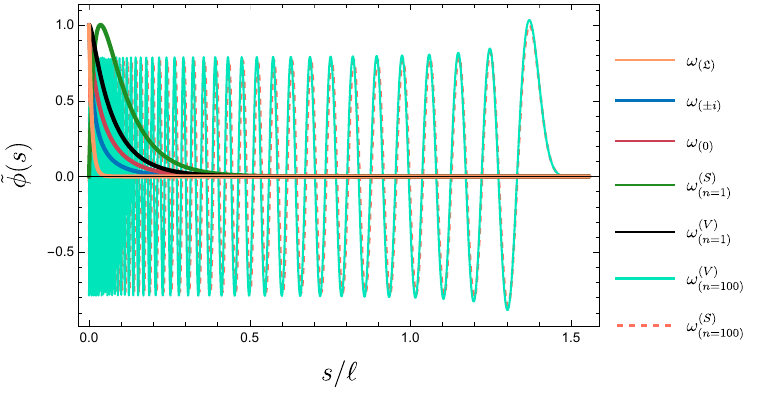}
\caption{Exact pole patch radial profiles for $|K\ell|\approx 70.7$. The large angular momentum mode (orange) has $l=100$. All other modes have $l=10$. The radial profiles of the normalised master field  $\tilde{\phi}(s)$ are shown from most to least localised.
}
\label{fig:PPradprofiles}
\end{figure}

We summarise these findings in table \ref{tab:modes}, while we provide a graphical representation of this hierarchy of stretched horizon modes in figure \ref{fig:PPradprofiles}, where we show the radial profiles for the normalised master fields for each type of mode.

\begin{center}
\begin{tcolorbox}[tab2,tabularx={c||c||c||c},title= Pole patch in the stretched horizon limit,boxrule=2pt, width=12.8cm, code={\setstretch{1.7}}]
   $\textbf{Modes}$ & Frequencies $\omega\ell$ & $D/\ell$ & $\bar{T}^{mn}\delta T_{mn}$ \\ 
   \hline \hline
   $\omega_{(\L)}$ & $\frac{\L}{|K\ell|}+\frac{\nu}{2^{1/3}}\big(\frac{\L}{|K\ell|}\big)^{1/3} + \mathcal{O}(\L^{-1/3})$ & $\L^{-2/3} |K\ell|^{-1/3}$ & $\times$ \\
   \hline \hline
 $\omega_{(\pm i)}$ & $\pm i + \mathcal{O}(|K\ell|^{-2})$ & $|K\ell|^{-1}$ & $\times$ \\
   \hline \hline
   $\omega_{(0)}$ & $\pm \tfrac{1}{\sqrt{2}}\frac{\L}{|K\ell|}+\mathcal{O}(|K\ell|^{-1}\log^{-1}|K\ell|)$ & $|K\ell|^{-\gamma}$ \,, $\gamma<1.$ & \checkmark \\
   \hline \hline
   $\omega_{n\neq0}^{(S)}$ & $\frac{\pi n}{\log|K\ell|}+\mathcal{O}(\log^{-2}|K\ell|)$ & $\log^{-1}|K\ell|$ & \checkmark \\
    \hline \hline
     $\omega_{n\neq0}^{(V)}$ & $\frac{(2n-1)\pi}{2\log|K\ell|}+\mathcal{O}(\log^{-2}|K\ell|)$ & $\log^{-1}|K\ell|$ & \checkmark 
\end{tcolorbox}
\end{center}
\begin{center}
\begin{minipage}[c]{\textwidth}
\captionof{table}{Pole patch at large-$|K\ell|$. Listed is the essential behaviour of each mode that localises in the pole patch. Note there are also scalar and vector normal modes whose radial profiles do not localise near the boundary. \label{tab:modes}}
\end{minipage}
\end{center}

\section{The large mean curvature limit: cosmic patch} \label{sec: large K cosmic}

Here we analyse the behaviour of the linearised dynamics of the cosmic patch in the stretched horizon limit. The modes of the scalar and vector sectors of the metric perturbations were categorized in \cite{Anninos:2024wpy}, see figure \ref{fig:CPmodes} for a summary. For the scalar sector, expanding the $\delta K|_{\Gamma}=0$ boundary condition (\ref{eq: deltaK scalar}) at large-$K\ell$ gives, up to an overall nonvanishing multiplicative factor,
\begin{align}\label{cosmicpatchdeltaK}
    0&= (\omega\ell)^2(1+(\omega\ell)^2)\phi^{(S)}(\mathfrak{r}) +\mathcal{O}((K\ell)^{-2})\;.
\end{align}
The leading-order contribution clearly vanishes for frequencies $\omega\ell=0, \pm i$, just like in the pole patch. There also exists another set of solutions corresponding to the master field vanishing at the boundary, analogous with the pole patch. For the vector sector, we expand in the large-$K\ell$ limit \eqref{eq: vector bdy condition}. Below we evaluate the Weyl factor and conformal stress-tensor components for these frequencies and their $\tfrac{1}{K\ell}$ corrections. Here, however, we do not analyse the associated bulk radial profiles. This is because, in the strict large-$K\ell$ limit, the cosmic patch has no bulk region. In principle one could zoom into the region between the boundary and the cosmological horizon with an appropriate choice of coordinates. We reserve such analysis to section \ref{sec: rindler}, where we work with a Rindler approximation.

\subsection{Soft modes}

\noindent \textbf{Allowed frequencies.} In contrast with the pole patch, the stretched horizon limit of $\omega\ell=\pm i$ modes are not complex conjugates of each other, and their large-$(K\ell)$ corrections differ.

We begin with the $\omega\ell=+i$ modes. Solving order by order the $\delta K|_{\Gamma}=0$ condition (\ref{eq: deltaK scalar}) in a large-$K\ell$ expansion, we find 
\begin{equation}
    \omega\ell =  i + i \omega_1 (K\ell)^{-2} + i \omega_2 (K\ell)^{-4}  +i \omega_3 (K \ell)^{-6} + \mathcal{O}((K\ell)^{-8}) \,,
    \label{eq:cosmicplusIexp}
\end{equation}
with
\begin{equation}
\begin{cases}
    \omega_1 =-\frac{l^2+l-2}{2} \,, \\ \omega_2=-\frac{3(l-1)(l+2)(l-3)(l+4)}{16} \,,  \\ \omega_3 = -\frac{(l - 1)(2 + l)\left(49l^{4} + 98l^{3} - 677l^{2} - 726l + 4704\right)}{384}
 \,.
    \end{cases}
    \label{eq: subleading frequencies cosmic patch}
\end{equation}

The $\omega\ell=-i$ frequencies admit a similar expansion,
\begin{equation}
    \omega\ell =  -i + i \omega_1 (K\ell)^{-2} + \mathcal{O}((K\ell)^{-4}) \,,
    \label{cosmicminusIexp}
\end{equation}
however, the first correction follows from solving a quadratic equation yielding
\begin{equation}
    \omega_1 = \frac{1}{8} \Biggl( -4 + 3 \, l (1 + l) \pm \sqrt{16 + l (1 + l) \bigl(-40 + 17 \, l (1 + l)\bigr)} \Biggr)\;.
    \label{eq: subleading frequencies cosmic patch minus i}
\end{equation}
The next correction is fully determined by $\omega_1$. Hence, there are two sets of modes, due to the `$\pm$' sign, as depicted in figure \ref{fig:CPmodes}.
Note that, just like for the pole patch, the naive continuation to $l=0$ gives the correct result. This is not the case for $l=1$, however.

\vspace{2mm}

\noindent \textbf{Weyl factor.} Substituting the frequency expansion for the $+i$ modes (\ref{eq:cosmicplusIexp}) into the Weyl factor (\ref{eq:weylfactSphi}) gives
\beq 
\delta\bomega=\frac{e^{t/\ell}}{4K\ell}(l-1)(l+2)\mathbb{S}+\mathcal{O}((K\ell)^{-3})\;.
\eeq
For the $-i$ modes, the leading order term depends on the first correction $\omega_1$ in \eqref{eq: subleading frequencies cosmic patch minus i},
\begin{align}
    \delta\bomega&=\frac{e^{-t/\ell}}{4}(l-1)(l+2) \frac{2\omega_1+l(l+1)}{2\omega_1}\mathbb{S} (K\ell) + \mathcal{O}((K\ell)^{-1}\log(K\ell))\;.
\end{align}

\noindent \textbf{Conformal stress-tensor.} The components of the stress-tensor $\delta T_{mn}$ (\ref{eq:compsdeltaTweyl}) evaluated at leading order for the $+i$ expansion are
\beq
\begin{cases}
    8\pi G_N\delta T_{tt}
    = \bigg(-\frac{l(l-1)(l+2)(l+1)}{4\ell(K\ell)^{5}}+\mathcal{O}((K\ell)^{-7}\bigg) \delta\bomega\,, \\
   8\pi G_N\delta T_{ti} = \bigg(\frac{(l-1)(l+2)}{4(K\ell)^{3}}+\mathcal{O}((K\ell)^{-5}) \bigg)\tilde{\nabla}_{i}\delta\bomega \,,\\
    8\pi G_N \delta T_{ij} =\bigg(-\frac{\ell}{2(K\ell)}\left(\tilde{\nabla}_{i}\tilde{\nabla}_{j}-\frac{1}{2}\tilde{g}_{ij}\tilde{\nabla}^{2}\right)+ \mathcal{O}((K\ell)^{-3} )\bigg)\delta\bomega\, .
    \end{cases}
\label{eq:stresstenpimodesCP}\eeq
Despite appearances, the linearised stress-tensor is traceless, $\bar{g}^{mn}\delta T_{mn}$. This follows from a cancellation between $\bar{g}^{tt}\delta T_{tt}$ and a term at $\mathcal{O}((K\ell)^{-3})$ in $\bar{g}^{ij}\delta T_{ij}$. We note that, just like for the pole patch, the two-dimensional conformal stress tensor $T_{ij}$ is traceless at leading order in the large-$K\ell$ limit with respect to the metric on the unit two-sphere.

Further, from (\ref{eq:TdeltaTweyl}) we have 
\beq 64\pi^{2}G_{N}^{2}\bar{T}^{mn}\delta T_{mn}=\bigg(-\frac{(l(l+1)-2)}{4(K\ell)^{2}\ell^{2}}+\mathcal{O}((K\ell)^{-4})\bigg)\tilde{\nabla}^{2}\delta\bomega\;.
\eeq

Meanwhile, for the $-i$ modes, the conformal stress-tensor components are
\beq
\begin{cases}
    8\pi G_N\delta T_{tt}
    = \bigg( \frac{2(l^2+l-2-\omega_1)}{\ell(K\ell)^3} + \mathcal{O}((K\ell)^{-5})\bigg)\delta\bomega\,, \\
   8\pi G_N\delta T_{ti} = \bigg( \frac{2(l^2+l-2-\omega_1)}{l(l+1)(K\ell)} + \mathcal{O}((K\ell)^{-3})\bigg) \tilde{\nabla}_i \delta\bomega \,,\\
    8\pi G_N \delta T_{ij} = \bigg(\frac{4\ell(l^2+l-2-\omega_1)}{l(l+1)(l-1)(l+2)}(K\ell) \left(\tilde{\nabla}_{i}\tilde{\nabla}_{j}-\frac{1}{2}\tilde{g}_{ij}\tilde{\nabla}^{2}\right) + \mathcal{O}((K\ell)^{-1})\bigg)\delta\bomega\, ,
    \end{cases}
\eeq
and
\beq 64\pi^{2}G_{N}^{2}\bar{T}^{mn}\delta T_{mn}=
\bigg(-\frac{2}{\ell^2}(l(l+1)-2-\omega_1)+\mathcal{O}((K\ell)^{-2})\bigg) \delta\bomega
\;,\eeq
where $\omega_1$ is given in \eqref{eq: subleading frequencies cosmic patch minus i}. We note that $T_{ij}$ is traceless with respect to the metric on the unit two-sphere for the $-i$ modes also.

\vspace{2mm}

Let us now comment on the Weyl mode equation \eqref{linearized Weyl eq}. For the $+i$ modes, we obtain at leading order in the large-$K\ell$ limit,
\begin{equation}
    K^2 \ell^2 \left( \partial_t^2 \delta \bomega(t) + \ell^{-2} \delta \bomega(t) \right) = \mathcal{O}((K\ell)^{-2}) \,,
\end{equation}
where the only difference with respect to \eqref{weylpmi1} is the scaling of the right hand side. The leading order solution takes the same form as in the pole patch, $\delta \bomega(t) = N_\bomega e^{\pm t/\ell} + \cdots$, for some time-independent constant $N_\bomega$. At next order, we can explicitly check the right hand side remains subleading compared to the left hand side, so we do not need the contribution from the conformal stress-tensor to fix the correction to $\delta\bomega$. Indeed, taking  $\delta\bomega(t)= N_{\bomega} \exp \left(\tfrac{t}{\ell} ( 1 + \omega_1 (K\ell)^{-2} + \cdots ) \right)$ we obtain
\begin{align}
    \frac{4 e^{t/\ell}}{\ell^2}\bigg( \frac{l(l+1)-2}{2} +\omega_1 \bigg) =\mathcal{O}((K\ell)^{-2}) \,,
\end{align}
which self-consistently sets $\omega_1$ that agrees with \eqref{eq: subleading frequencies cosmic patch}. 

Thus, different from the pole patch, the first correction to the cosmic patch $+i$ mode can be fixed without additional information from the bulk. For the $-i$ modes, however, due to the different scalings in the large-$K\ell$ limit, the correction cannot be fixed without information about the conformal stress tensor.

\subsection{Quasinormal type modes}
Similar to the pole patch, there exist two towers of modes indexed by a new integer parameter $n$, one belonging to the scalar sector and one to the vector. Because they follow the same structure, we analyse them together.

In the scalar sector, the tower of modes corresponds to the vanishing of the cosmic patch master field at the boundary in \eqref{cosmicpatchdeltaK}. The allowed frequencies are given by\footnote{We analysed $n=1$ in the previous subsection, so here we consider only $n>1$.}
\begin{align}
    \omega\ell=-i n + \frac{i}{(K\ell)^{2n}}\left(\omega_1 +\mathcal{O}((K\ell)^{-2}) \right) \,, \quad  \omega_1= \frac{(-1)^n}{2^{2n} \, \Gamma(n + 1)\, \Gamma(n)} \, , \quad n=2,3,...,l.
\label{eq:QNMlikemodsscal}\end{align}
Hence, for fixed $l$, one obtains a finite set of $(l-1)$ purely imaginary modes, labelled by an integer $n=2,3,...,l$. These modes resemble the tower of scalar normal modes in the pole patch, as they also have vanishing Weyl factor at leading order in the large-$K\ell$ expansion, but different from the pole patch, the tower terminates at $n=l$. Interestingly, a continuation of the leading order term $-i n$ for $n>l$ would give the familiar quasinormal modes of empty dS$_4$ \eqref{qnmdeSitter}, which are nevertheless not present in the large-$K\ell$ limit.

In the vector sector, for each fixed $l$, we find a tower of modes very similar to those in the scalar sector (\ref{eq:QNMlikemodsscal}), 
\begin{align}
    \omega\ell=-i n + \frac{i}{(K\ell)^{2n}}\left(\omega_1 +\mathcal{O}((K\ell)^{-2}) \right)  \,, \quad   \omega_1= \frac{(-1)^{n+1}}{2^{2n} \Gamma(n + 1)\, \Gamma(n)}\frac{\Gamma(l + n + 1)}{\Gamma(l - n + 1)}\; \,, \quad  n=1,2,3,...,l.
\label{eq:QNMlikemodsvec}\end{align}
Note the only difference with respect to the scalar sector is a minus sign in $\omega_1$. As for all perturbations in the vector sector, the linearised Weyl factor identically vanishes for these modes. 

\vspace{2mm}

\noindent \textbf{Conformal stress tensor.} The components of the stress-tensor $\delta T_{mn}$ (\ref{eq:compsdeltaTweyl}) for the quasinormal mode scalar sector in the cosmic patch are given by
\beq
\begin{cases}
    8\pi G_N\delta T_{tt}
    = \left(\frac{2(n^{2}-1)}{\ell K\ell}+\mathcal{O}((K\ell)^{-3})\right) \delta\bomega\,, \\
   8\pi G_N\delta T_{ti} =\left(\frac{2n(n^{2}-1)}{l(l+1)}K\ell\tilde{\nabla}_{i}+\mathcal{O}((K\ell)^{-1})\right)\delta\bomega \,,\\
    8\pi G_N \delta T_{ij} =\left(\frac{4n^{2}(n^{2}-1)\ell}{l(l+1)(l+2)(l-1)}(K\ell)^{3}\left(\tilde{\nabla}_{i}\tilde{\nabla}_{j}-\frac{1}{2}\tilde{g}_{ij}\tilde{\nabla}^{2}\right)+\mathcal{O}((K\ell))\right)\delta\bomega\, .
    \end{cases}
\label{eq:stresstenpimodesQNM}\eeq
Further, we have
\begin{align}
    64\pi^{2}G_{N}^{2}\bar{T}^{mn}\delta T_{mn}=\bigg( \frac{2(1-n^2)}{\ell^2}(K\ell)^2  +\mathcal{O}(1)\bigg)\delta\bomega\;.
\end{align}

In the scalar sector, the Weyl mode is nontrivial. We comment on its linearised equation \eqref{linearized Weyl eq}, which in the large-$(K\ell)$ limit gives \eqref{leadingorderWeyleq}. Using (\ref{eq:stresstenpimodesCP}) on the right hand side of \eqref{leadingorderWeyleq}, it follows
\begin{align}
    \partial_t^2\delta\bomega(t)+\frac{n^2}{\ell^2} \delta\bomega(t)=0\;,
\end{align}
which sets the frequency to the correct value at leading order. Hence, for this set of modes also, a contribution from the conformal stress-tensor is essential in order to recover the correct results.

\subsection{Fluid dynamical modes} \label{sec: lin fluid}

\subsubsection{Shear modes}

In the vector sector, solving the linearised boundary condition order by order (\ref{eq: vector bdy condition}) in the stretched horizon limit about $\omega\ell=0$
yields a set of modes with purely imaginary frequencies \cite{Anninos:2011zn, Anninos:2024wpy}
\begin{align}
    \omega_{\text{shear}}\ell&=-\frac{i (l(l+1) - 2)}{2(K\ell)^2}
- \frac{3 i (l - 1)(l + 2)(l - 3)(l + 4)}{16(K\ell)^4}+\mathcal{O}((K\ell)^{-6})\;.
\label{eq:shearmodes}\end{align}
Interestingly, at least up to this order in the expansion, the above corrections are precisely the same corrections to the $+i$ modes of the scalar sector (\ref{eq:cosmicplusIexp}). At the next order $(K\ell)^{-6}$ and beyond, this is no longer the case. 

The master field evaluated at the boundary for the shear modes is
\begin{align}
    \Phi^{(V)}|_{\Gamma}&=1 +\frac{l (l+1) - 2}{2 (K\ell)^2}\log(K\ell) +\mathcal{O}((K\ell)^{-2})\;.
\end{align}
In terms of the master field, the non-vanishing components of the conformal stress tensor (\ref{eq:deltaTmnvecgen}) are
\beq 
\begin{split}
\begin{cases}
 8\pi G_{N}\delta T_{ti}=\left( -\frac{(l-1)(l+2)}{2(K\ell)} +\mathcal{O}((K\ell)^{-3})\right)\Phi^{(V)}|_{\Gamma}\mathbb{V}_{i}\;,\\
 8\pi G_{N}\delta T_{ij}=\left(-\frac{\ell(l-1)(l+2)}{4(K\ell)}  +\mathcal{O}((K\ell)^{-3})\right)\Phi^{(V)}|_{\Gamma}(\tilde{\nabla}_{i}\mathbb{V}_{j}+\tilde{\nabla}_{j}\mathbb{V}_{i})\;.
\end{cases}
\end{split}
\eeq

These shear modes were first noted in \cite{Bredberg:2011xw, Anninos:2011zn}. Supplemented with the sound modes that we describe next, they form a conformal fluid. The non-linear generalisation of this conformal fluid can be found in section \ref{sec: fluid result}, where we also compute transport coefficients associated to this fluid.

\subsubsection{Sound modes}

In the scalar sector, solving order by order the boundary condition (\ref{eq: deltaK scalar}) in a large-$K\ell$ expansion, we recover the sound modes \cite{Anninos:2024wpy}
\begin{equation}
\omega_{\text{sound}}\ell=\pm \frac{\sqrt{l(1 + l)}}{\sqrt{2}} \frac{1}{(K\ell)}
- \frac{i\,(l(l+1)-2)}{4} \frac{1}{(K\ell)^2}
+\mathcal{O}((K\ell)^{-3}) \,.
    \label{cosmicsound}
\end{equation}
The dominant contribution is real, while the subleading term has a negative imaginary component. Evidently, the scalar sound modes vanish in the strict stretched  horizon limit. The Weyl factor for the sound modes (\ref{cosmicsound}) is easily worked out to be
\begin{align}
    \delta\bomega&= \frac{l(l+1)}{4}\mathbb{S} + \mathcal{O}((K\ell)^{-1}\log(K\ell))\;.
\end{align}

\noindent \textbf{Conformal stress-tensor.} The stress-tensor components (\ref{eq:compsdeltaTweyl}) for the scalar sound modes are 
\beq
\begin{cases}
    8\pi G_N\delta T_{tt}
    = \bigg( -\frac{2}{(K\ell)\ell}+\mathcal{O}((K\ell)^{-3}) \bigg)\delta\bomega \,, \\
   8\pi G_N\delta T_{ti} =\bigg( \frac{2(K\ell)}{l(l+1)}+\mathcal{O}((K\ell)^{-1}) \bigg) \partial_t\tilde{\nabla}_{i}\delta\bomega\,,\\
    8\pi G_N \delta T_{ij} =\bigg(-\tilde{g}_{ij}\ell (K\ell)+\mathcal{O}(1)\bigg)\tilde{\nabla}^{2}\delta\bomega\, ,\label{eqn: Tmn sound}
    \end{cases}
\eeq
from which it is easy to see $\bar{g}^{mn}\delta T_{mn}=0$. Since the Weyl factor is time-independent at leading order, the leading order term in $\delta T_{ti}$ vanishes. Further, 
\beq 64\pi^{2}G_{N}^{2}\bar{T}^{mn}\delta T_{mn}=\bigg(\frac{2(K\ell)^{2}}{\ell^{2}}+\mathcal{O}(1)\bigg)\delta\bomega \;.\eeq
As with the other sets of modes whose frequencies vanish in the strict large-$K\ell$ limit, the sound modes require a time-independent contribution from the conformal stress-tensor in order to fix their Weyl mode at leading order using \eqref{linearized Weyl eq}.

\section{Non-linear fluid dynamics} \label{sec: non-linear fluid}

In the previous section we obtained linearised gravitational perturbations which, in the stretched horizon limit, have the frequencies analogous to the linearised sound and shear modes of conformal fluids. Here we consider a non-linear treatment of the fluid dynamical sector in the stretched horizon limit.

\subsection{Fluid coordinate system}\label{sec: pure dS fluid}

Let us start by recasting the empty dS$_4$ metric (\ref{eq:dSmetric}) in a $(\tau,\rho,\theta,\phi)$ coordinate system with 
\begin{equation}\label{eqn: fluid coord transf}
     t = \frac{\r \tau}{\sqrt{\ell^2-\r^2}} + \ell\tanh^{-1}\frac{r}{\ell}\, , \qquad r=\r-\left(\frac{\ell^2-\r^2}{4\ell^2-6\r^2}\right)\rho\,,
\end{equation}
where $\r$ is the location of the boundary in the static patch coordinate. We will call $\tau$, the retarded time. In this coordinate system, the metric satisfies
\begin{equation}
    g_{\rho\rho}=g_{\rho i}=0\, , \qquad g_{\rho \tau} = -\frac{1}{2 K \ell}\,.
\end{equation}
The boundary is located at $\rho=0$ for any value of $K\ell$, while the horizon is now at $\rho=\ell+\mathcal{O}((K\ell)^{-2})$. In the large-$K\ell$ limit, the pure de Sitter metric becomes,
\begin{equation}\label{eqn: fluid 0th sol}
    ds^2 = \left[ -\left(1-\frac{\rho}{\ell}\right)d\tau^2-\frac{1}{K\ell} d\rho d\tau \right]+\ell^2d\Omega_2^2 + \mathcal{O}\left((K\ell)^{-2}\right) \, .
\end{equation}
This takes the form of a two-dimensional Rindler space times a round two-sphere of radius $\ell$. At the boundary, the induced metric is given by
\begin{equation}\label{eqn: conf rep}
    \left.ds^2\right|_{\rho=0}=e^{2\bomega}[g]_{mn}dx^m dx^n \equiv e^{2\bomega}\left(-d\tau^2 + \ell^2d\Omega_2^2\right) \, , \qquad e^{2\bomega}= 1 + \mathcal{O}((K\ell)^{-2})\, ,
\end{equation}
where the conformal representative $[g]_{mn}$ is chosen to be time times the round two-sphere of radius $\ell$. A constant $x^m=(\tau,\theta,\varphi)$ line is a null geodesic with $\rho$ being an affine parameter. The corresponding conformal stress-tensor \eqref{eq: tmn} is given by
\begin{equation}\label{eqn: fluid Tmn 0th}
    T_{mn}dx^mdx^n =\frac{K}{24 \pi G_N }\left(2d\tau^2+\ell^2d\Omega_2^2\right)+\mathcal{O} ((K\ell)^{-1}) \, .
\end{equation}

\subsection{The Einstein equations} \label{sec: fluid perturbation}
In this section, we perturbatively construct solutions to the non-linear Einstein field equations \eqref{eqn: Einstein tensor} treating $(K\ell)^{-1}$ as the perturbative parameter and \eqref{eqn: fluid 0th sol} as the zeroth order solution.

First, we fix the gauge to
\begin{equation}\label{eqn: fluid gauge}
    g_{\rho\rho}= 0 \, , \qquad g_{\rho m} = \frac{1}{2 K\ell} u_m \, ,
\end{equation}
where the $x^m$-dependent three-vector field $u_m$ will be defined momentarily. In the following, indices $m,n,...=\tau,\theta,\varphi$ are raised and lowered by $[g]_{mn}$ from \eqref{eqn: conf rep}, while $i,j,...=\theta,\varphi$ are indices on the round two-sphere metric of radius $\ell$.

We consider the expansion of the remaining metric components as
\begin{equation}\label{eqn: fluid metric ansatz}
    g_{mn}= g^{(0)}_{mn} +\sum_{\mathfrak{n}=1}(K\ell)^{-\mathfrak{n}} h_{m n}^{(\mathfrak{n})}  \, ,
\end{equation}
where $h^{(\mathfrak{n})}_{mn}$ are $(\rho,x^m)$-dependent and $g^{(0)}_{mn}$ is chosen to be
\begin{equation}\label{eqn: fluid metric 0th order}
    g^{(0)}_{m n} dx^m dx^n = -\left(1-\frac{\rho}{\ell}\right)d\tau^2 + \ell^2d\Omega_2^2 \,.
\end{equation}
By decomposing the Einstein tensor $G_{\mu\nu}$ \eqref{eqn: Einstein tensor} into components tangent and normal to constant-$\rho$ surfaces, and using \eqref{eqn: fluid gauge}, \eqref{eqn: fluid metric ansatz} and \eqref{eqn: fluid metric 0th order}, we find the Einstein tensor admits the following $(K\ell)^{-1}$ expansion,
\begin{equation}
\begin{cases}

    G_{\perp\perp}&\equiv G_{\mu\nu}n^\mu n^\nu = (K\ell)^{2} \sum_{\mathfrak{n}=1}(K\ell)^{-\mathfrak{n}}  G_{\perp\perp}^{(\mathfrak{n})}  \, , \\ 
     G_{\perp m} &\equiv G_{\mu m}n^\mu = (K\ell)^{2}\sum_{\mathfrak{n}=1}(K\ell)^{-\mathfrak{n}}  G_{\perp m}^{(\mathfrak{n})}\, , \\
     G_{mn} &= (K\ell)^{2}\sum_{\mathfrak{n}=1}(K\ell)^{-\mathfrak{n}}  G_{mn}^{(\mathfrak{n})}\,,
    \end{cases}
\end{equation}
where $n^\mu$ is the outward-pointing, unit-normal vector. Next, we consider the Einstein field equations at order $\mathfrak{n}$.

To solve these, we recall that, according to the Bianchi identity,
\begin{equation}
    \nabla^\mu G_{\mu\nu} =0 \, ,
\end{equation}
the equations $G_{\perp \perp}=G_{\perp m}=0$ are satisfied provided  $G_{mn}=0$ hold everywhere and $G_{\perp \perp}=G_{\perp m}=0$ at the boundary.\footnote{This statement is the radial version of the fact that the Hamiltonian and momentum constraints of the Einstein equations are first-class constraints. Specifically, time evolving Cauchy data obeying these constraints on the Cauchy surface guarantees that the constraints hold everywhere.} Hence, we only have to look at the equations $G^{(\mathfrak{n})}_{\perp\perp}=G^{(\mathfrak{n})}_{\perp m}=0$ at $\rho=0$ and $G^{(\mathfrak{n})}_{mn}=0$.

Let us first  consider $G_{mn}^{(\mathfrak{n})}$. Setting this to zero we obtain
\begin{equation}\label{eqn: fluid Gmn}
    \mathbb{H}\,h_{mn}^{(\mathfrak{n})}= s^{(\mathfrak{n})}_{mn} \, ,
\end{equation}
for tensorial differential operator $\mathbb{H}$ and $s^{(\mathfrak{n})}_{mn}$ denotes the source term, with $s^{(1)}_{mn}=0$. The explicit expression of $\mathbb{H}$ and its zero modes are given in appendix \ref{sec: H}. This operator has the same structure at every order $\mathfrak{n}$ and involves only $\rho$-derivatives.  
The source term $s_{mn}^{(\mathfrak{n})}$ collects contributions from the metric corrections up to order $(\mathfrak{n}-1)$, and thereby depends on $(\rho,x^m)$. All these features render \eqref{eqn: fluid Gmn} a set of linear second-order ordinary differential equations for $h^{(\mathfrak{n})}_{mn}$ with the source $s_{mn}^{(\mathfrak{n})}$.

Setting $G^{(\mathfrak{n})}_{\perp \perp}$ at $\rho=0$ to zero, we find
\begin{equation}\label{eqn: fluid H constraint}
    \left.
    [g]^{ij} \partial_\rho h_{ij}^{(\mathfrak{n})} \right|_{\rho=0}= s^{(\mathfrak{n})}\, ,
\end{equation}
where $s^{(\mathfrak{n})}$ is a function of $x^m$ and only depend on the metric up to order $\mathfrak{n}-1$ evaluated at $\rho=0$, with $s^{(1)}=0$. 

Setting $G^{(\mathfrak{n})}_{\perp m}$ at $\rho=0$ to zero, we obtain the transverse condition for $T_{mn}^{(\mathfrak{n}-1)}$ (recall $K\ell$ is constant), 
\begin{equation}\label{eqn: fluid P constraint}
    \mathcal{D}^m T_{mn}^{(\mathfrak{n}-1)}=0 \, ,
\end{equation}
where $\mathcal{D}_m$ is the covariant derivative compatible with $[g]_{mn}$, and $T^{(\mathfrak{n}-1)}_{mn}$ is the contribution to conformal stress-tensor from the metric up to order $(\mathfrak{n}-1)$. The zeroth order $T_{mn}^{(0)}$ is  given by \eqref{eqn: fluid Tmn 0th}.

\textbf{Conformal boundary conditions.} The conformal boundary conditions can be similarly analyzed at each order. At order $\mathfrak{n}$, the condition that the induced metric is conformally equivalent to $[g]_{mn}$ reads
\begin{equation}\label{eqn: fluid conf bdry 1}
    \left.h_{mn}^{(\mathfrak{n})}-\frac{1}{3}[g]_{mn} [g]^{pq}h_{pq}^{(\mathfrak{n})}\right|_{\rho=0}=0 \, .
\end{equation}
Meanwhile, the condition that the trace of the extrinsic curvature remains unchanged is 
\begin{equation}\label{eqn: fluid conf bdry 2}
     \left.\frac{1}{2}\left(h^{(\mathfrak{n})}_{\tau\tau}-2\ell[g]^{mn}\partial_\rho h_{mn}^{(\mathfrak{n})}\right)\right|_{\rho=0}+ \delta \hat{K}^{(\mathfrak{n})}=0 \, ,
\end{equation}
where $ \delta \hat{K}^{(\mathfrak{n})}$ is the contribution to the trace of the extrinsic curvature at order $\mathfrak{n}$ from the metric up to order $(\mathfrak{n}-1)$.

Taken together, the perturbative in $\tfrac{1}{K\ell}$ strategy for constructing the metric is as follows. We begin by radially integrating the equations \eqref{eqn: fluid Gmn}, which produces solutions involving twelve functions of integration, all depending on the boundary coordinates $x^m$. Nine are fixed algebraically. More specifically, one is fixed by \eqref{eqn: fluid H constraint}. Imposing purely outgoing condition fixes two more.\footnote{It is a priori unclear whether the purely-outgoing condition can be satisfied at any order $\mathfrak{n}$ by algebraically fixing two functions of integration. We check explicitly that this is true up to order $\mathfrak{n}=4$.} The remaining six are determined by the boundary conditions \eqref{eqn: fluid conf bdry 1} and \eqref{eqn: fluid conf bdry 2}. This leaves three functions of integration which are subject to \eqref{eqn: fluid P constraint}. In the next section, we identify them with fluid degrees of freedom living on $[g]_{mn}$.

\subsection{Identifying fluid variables} \label{sec: fluid ident}

Here we provide a prescription of fixing the metric solutions obtained from the previous section in terms of the fluid velocity and pressure. This will be done by recasting the conformal stress-tensor as the stress-tensor of a relativistic fluid.

To begin with, recall that a general relativistic fluid in local thermal equilibrium, defined on a three-dimensional manifold endowed with a fixed metric $g^{(\text{fluid})}_{mn}$, can be described by three independent variables: the fluid velocity field $u^m(x^m)$, satisfying $u^mu_m=-1$, and the local pressure $\mathcal{P}(x^m)$. Its stress-tensor is generically given by
\begin{equation}\label{eqn: gen fluid Tmn}
    T_{mn}^{(\text{fluid})}=\mathcal{P}g^{(\text{fluid})}_{mn} +\left(\mathcal{E}+\mathcal{P}\right)u_mu_n + \Pi_{mn} \, ,
\end{equation}
where $\mathcal{E}$ is the energy density related to $\mathcal{P}$ via the equation of state, and $\Pi_{mn}$ denotes the dissipative term containing derivatives of $\mathcal{P}$ and $u^m$. Imposing that $T_{mn}^{(\text{fluid})}$ is covariantly conserved with respect to the metric $g^{(\text{fluid})}_{mn}$ leads to a set of dynamical equations for $u^m$ and $\mathcal{P}$, referred to as fluid equations. As in any effective description, the fluid velocity $u^m$ is subject to field redefinitions. To fix this ambiguity, we work in the Landau gauge,
\begin{equation}\label{eqn: Landau gauge}
    u^m T_{mn}^{(\text{fluid})} = -\mathcal{E} u_n \, ,
\end{equation}
or equivalently $\Pi_{mn}u^m=0$.

Now we propose a prescription to identify the metric solutions around the empty dS cosmic patch to solutions of the fluid equations around a static fluid with a constant pressure. More specifically, we impose
\begin{equation}\label{eqn: fluid presc}
    T_{mn} =T^{(\text{fluid})}_{mn} \, , \qquad [g]_{mn} =g^{(\text{fluid})}_{mn} \,.
\end{equation}
The condition that $T_{mn}$ is traceless determines the fluid equation of state to be $\mathcal{E}=2\mathcal{P}$. Such an equation of state corresponds to an ideal three-dimensional conformal fluid.

Applying \eqref{eqn: fluid presc} to the stress-tensor \eqref{eqn: fluid Tmn 0th} leads to
\begin{equation}
    \mathcal{P}_{\text{dS}} = \frac{K}{24 \pi G_N} + \mathcal{O}((K\ell)^{-2}) \, , \qquad u^m_{\text{dS}}\partial_m = \partial_\tau \, ,
\end{equation}
which identifies the empty dS solution as a static fluid with constant pressure $\mathcal{P}_{\text{dS}}$. To use \eqref{eqn: fluid presc} on the general solutions, we parametrise the expansion of $\mathcal{P}$ and $u^m$ as
\begin{equation}\label{eqn: expanding P um}
\begin{cases}
    \mathcal{P}=\frac{K}{8 \pi G_N}\left(\frac{1}{3} + (K\ell)^{-1}p\right)+\mathcal{O}((K\ell)^{-2}) \, , \\
    u^m\partial_m = \partial_\tau + \frac{v^i \partial_i}{K\ell} + \frac{v^iv_i \partial_\tau}{2(K\ell)^2} + \mathcal{O}((K\ell)^{-3}) \, ,
\end{cases}
\end{equation}
where $p$ and $v^i$ are $x^m$-dependent. Inserting \eqref{eqn: expanding P um} into \eqref{eqn: gen fluid Tmn}, we obtain the $(K\ell)^{-1}$ expansion of the fluid stress-tensor up to $\Pi_{mn}$. Matching this with the conformal stress-tensor of the metric solutions at order $(K\ell)^{-\mathfrak{n}}$ leads to three equations which algebraically determine the remaining three functions of integration in terms of $v^i$ and $p$. As a result, the radial momentum constraint equations \eqref{eqn: fluid P constraint} become the fluid equations.

Generally, the expression of $\Pi_{mn}$ depends on the details of the system of interest. Using \eqref{eqn: expanding P um}, the general $\Pi_{mn}$ containing only first-order derivative terms and obeying \eqref{eqn: Landau gauge} can be decomposed into the trace and traceless parts as
\begin{equation}\label{eqn: Pi 1st order}
    \Pi_{ij} = - \zeta \frac{[g]_{ij} \nabla^k v_k}{K\ell} -\eta  \frac{\left(\nabla_i v_j+\nabla_j v_i - [g]_{ij}\nabla^k v_k \right)}{K\ell}  + \mathcal{O}((K\ell)^{-2}) \, , 
\end{equation}
and $\Pi_{\tau \tau}=\Pi_{\tau i}=0$, where $\nabla_i$ is the covariant derivative from $[g]_{ij}$. The transport coefficients $\zeta$ and $\eta$, referred to as the bulk and shear viscosity, are, in principle, arbitrary coefficients characterising the fluid. 

Below, we will see, via \eqref{eqn: fluid presc}, the Einstein field equations near the horizon obeying conformal boundary conditions completely determine $\Pi_{mn}$.

\subsection{Fluid/gravity correspondence}\label{sec: fluid result}

In this section we write the non-linear solution for the metric and the conformal stress tensor in terms of the fluid variables $(p,v^i)$, which parameterise the expansion of the pressure and the fluid velocity.

We start by writing the bulk metric in a perturbative expansion at large-$K\ell$,
\begin{equation}\label{eqn: fluid metric sol}
    ds^2 =  -\left(1-\frac{\rho}{\ell}\right) d\tau^2 + \ell^2d\Omega_2^2 + \frac{ds^2_{(1)}}{(K\ell)} + \frac{ds^2_{(2)}}{(K\ell)^2}+ \mathcal{O}((K\ell)^{-3}) \, ,
\end{equation}
where the leading correction is given by
\begin{eqnarray}
    ds^2_{(1)} = -d\tau d\rho +\frac{\rho}{\ell}p d\tau^2-\frac{2\rho}{\ell} v_i d\tau dx^i +2p \left(-d\tau^2+\ell^2d\Omega_2^2\right) \, .
\end{eqnarray}
The second order solution gives,
\begin{eqnarray}
    ds^2_{(2)} &=& v_i dx^id\rho - \frac{\rho}{\ell}\left(p^2-v^iv_i\right)d\tau^2 +\frac{\rho}{\ell} v_iv_jdx^idx^j -  \frac{2\rho}{\ell} p v_i d\tau dx^i -\rho\partial_\tau v_i d\tau dx^i\nonumber \\
    &&-\frac{\rho}{2} \nabla^iv_i d\tau^2+\frac{\rho}{4\ell}\left(8+\frac{\rho}{\ell}\right)d\tau^2 -\left(p^2 +2-\frac{\rho}{\ell}\right)\left(-d\tau^2+\ell^2 d\Omega_2^2\right) \,.
\end{eqnarray}
Note that the solution depends on $\rho$ polynomially. At the boundary $\rho=0$, the Weyl factor is determined by the pressure,
\begin{equation}
    e^{2\bomega} = 1 + \frac{2p}{(K\ell)} - \frac{\left(p^2+2\right)}{(K\ell)^2} + \mathcal{O}((K\ell)^{-3}) \,.
\end{equation}
The stress-tensor up to second order is
\begin{equation}\label{eqn: Tmn fluid result}
    T_{mn}dx^mdx^n =\frac{K
    }{24 \pi G_N    }\left(2d\tau^2+\ell^2d\Omega_2^2\right) + T_{mn}^{(1)}dx^mdx^n+ (K\ell)^{-1} T_{mn}^{(2)}dx^mdx^n+\mathcal{O}((K\ell)^{-2})\, ,
\end{equation}
where 
\begin{equation}
    8 \pi G_N T_{mn}^{(1)}dx^mdx^n =\frac{p}{\ell}\left(2d\tau^2+\ell^2d\Omega_2^2\right)-\frac{2}{\ell}v_id\tau dx^i \, ,
\end{equation}
and
\begin{equation}
        8 \pi G_N T_{mn}^{(2)}dx^mdx^n =-\frac{6p}{\ell} v_i d\tau dx^i+\frac{1}{\ell}v_iv_j dx^i dx^j+\frac{1}{\ell}v^iv_i d\tau^2
    - \frac{1}{2}\left(\nabla_i v_j+\nabla_j v_i - [g]_{ij}\nabla^k v_k \right)dx^i dx^j \, .
\end{equation}
It is straightforward to check that \eqref{eqn: Tmn fluid result} takes the form of the stress-tensor of a relativistic fluid, \eqref{eqn: gen fluid Tmn}, with the dissipative term  $\Pi_{mn}$ given by \eqref{eqn: Pi 1st order}. 

Importantly, and as opposed to the bulk viscosity  found in the usual membrane paradigm, we find the bulk viscosity $\zeta$ vanishes,
\begin{equation}
    \zeta =0 \,,
\end{equation} 
whilst the shear viscosity $\eta$ is given by 
\begin{equation}
    \eta = \frac{1}{16 \pi G_N} \,. 
\end{equation}
We can also write down the shear viscosity to entropy density ratio  $\tfrac{\eta}{s}=\tfrac{1}{4\pi}$, where the entropy density $s \equiv  \tfrac{1}{4G_N} $. 
This is the same as the result found in the membrane paradigm literature \cite{Damour:1979wya,Price:1986yy}, as well as the one stemming from the AdS/CFT correspondence \cite{Kovtun:2004de}.

The metric \eqref{eqn: fluid metric sol} solves the Einstein field equations provided \eqref{eqn: Tmn fluid result} is covariantly conserved. Equivalently, $(p,v^i)$ must satisfy the following fluid equations,
\begin{equation}
\begin{cases}
    \partial_\tau p &= -\frac{1}{2}\nabla_i v^i + (K\ell)^{-1} \left(-\frac{1}{2}v^i\nabla_i p -\frac{3}{2} p \nabla_i v^i\right) + \mathcal{O}((K\ell)^{-2}) \, , \\
    \partial_\tau v_i &= -\nabla_i p + (K\ell)^{-1}\left(3p \nabla_i p +\frac{\ell}{2}\left( \nabla^2v_i+R_i{}^jv_j\right) -v^j\nabla_j v_i + \frac{1}{2}v_i \nabla^jv_j\right) + \mathcal{O}((K\ell)^{-2})\, . \label{eqn: fluid eom}
\end{cases}
\end{equation}

Finally, for the perturbative expansion of \eqref{eqn: fluid metric sol} and \eqref{eqn: Tmn fluid result} to be reliable, we require the magnitude of $(p,v^i)$ and their time and spatial derivatives obey
\begin{equation}\label{eqn: fluid validity}
    |p| , \quad|\partial_\tau p |, \quad |\nabla_ip|, \quad |v^i|, \quad |\partial_\tau v^i|, \quad |\nabla_j v^i|  \ll K \ell \,.
\end{equation}
It follows that the solution space of \eqref{eqn: fluid eom} does not contain solutions with the time dependence $e^{-i\omega \tau}$ where $\omega \sim K\ell$. It would be interesting to have a non-linear fluid description that allows for such modes, along the lines of \cite{Knysh:2024asf}, and understanding the interplay of the fluid dynamical sector with the other dynamical modes. Perhaps, from an initial boundary value perspective, one way to dynamically isolate the fluid sector is by turning on a dynamical source term whose frequency scales in the same way as the fluid modes.

\subsection{Simple solutions to the fluid equations}

Finally, we present some simple examples of solutions to the fluid equations \eqref{eqn: fluid eom}, where we connect these non-linear solutions to the linearised modes found in section \ref{sec: lin fluid}.

\textbf{Stationary solutions.} We first consider $\tau$-independent solutions to \eqref{eqn: fluid eom}, i.e., $\partial_\tau p= \partial_\tau v^i=0$. Assuming spherical symmetry, the general solution is
\begin{equation} \label{eqn: stationary solution}
    p=c_{1}+ \frac{c_{2}}{K\ell} \, , \qquad v^i=0 \, ,
\end{equation}
for arbitrary constants $c_{1}$ and $c_{2}$. Plugging this into \eqref{eqn: fluid metric sol}, we find this describes the Schwarzschild-de-Sitter solution with a small black hole of mass $M$ in units of $\ell$,
\begin{equation}
    M \ell=-\frac{c_1}{K\ell} + \frac{1-c_1^2-2 c_2}{2(K\ell)^2} + \mathcal{O}((K\ell)^{-3})\,.
\end{equation}
It follows that the empty de Sitter solution is given by $c_1=0$ and $c_2=\tfrac{1}{2}$.

For non-spherically symmetric solutions, we find
\begin{equation}
    p=  -\frac{\ell^2c_{1}^2 \cos^2\theta}{2K\ell} \, , \qquad v^\phi = c_{1}+\frac{c_{2}}{K\ell} \, , \qquad v^\theta =0 \,,
\end{equation}
for arbitrary constants $c_{1}$ and $c_{2}$. This amounts to adding a small rotation to empty dS solution with angular velocity $\tfrac{c_{1}}{K\ell}+\tfrac{c_{2}}{(K\ell)^2}$.

\textbf{Linearised fluid modes.} Let us consider linearisation of \eqref{eqn: fluid eom} around the empty dS solution, \eqref{eqn: stationary solution} with $c_1=0$ and $c_2=\tfrac{1}{2}$. In particular, we consider the ansatz
\begin{equation}
    p = \frac{1}{2 K\ell}+ \delta p \, , \qquad v^i=\delta v^i \, ,
\end{equation}
for $x^m$-dependent $\delta p$ and $\delta v^i$. Inserting this to \eqref{eqn: fluid eom} and only keeping terms linear in $\delta p$ and $\delta v^i$, we find two sets of linearised modes. The first mode is the linearised sound mode,
\begin{equation}\label{eqn: fluid sound}
    \delta p=\left(\tilde c_1 + \frac{\tilde c_2}{K\ell} \right)e^{-i(\omega_1 + \tfrac{\omega_2}{K\ell}) \tau} \mathbb{S} \, , \qquad \delta v_i=\left(c_1 + \frac{c_2}{K\ell}\right) e^{-i(\omega_1 + \tfrac{\omega_2}{K\ell}) \tau} \partial_i\mathbb{S} \, ,
\end{equation}
where
\begin{equation}\label{eqn: fluid sound frequency}
    \tilde c_1=  i \omega_1 c_1 \, , \quad \tilde c_2= i (\omega_1 c_2-\omega_2 c_1) \, , \quad \omega_1^2\ell^2 =  \frac{l(l+1)}{2} \, , \quad \omega_2 \ell = - i \frac{l(l+1)-2}{4} \, ,
\end{equation}
for arbitrary constants $c_1$ and $c_2$. From this, we can identify the speed of sound as $c_s=\tfrac{1}{\sqrt{2}}$. The second mode is the linearised shear mode,
\begin{equation}\label{eqn: fluid shear}
    \delta p =0 \, , \qquad \delta v_i = \left(c_1 + \frac{c_2}{K\ell}\right) e^{-i(\omega_1 + \tfrac{\omega_2}{K\ell}) \tau} \mathbb{V}_i\, ,
\end{equation}
where
\begin{equation}\label{eqn: fluid shear frequency}
    \omega_1=0 \, , \qquad \omega_2 \ell= - i \frac{l(l+1)-2}{2} \, ,
\end{equation}
for arbitrary constants $c_1$ and $c_2$. Since $\mathbb{V}_i$ is divergenceless, this mode is incompressible, $\nabla^i \delta v_i=0$.

The linearised modes \eqref{eqn: fluid sound} and \eqref{eqn: fluid shear} reproduce the modes of \eqref{eq:shearmodes} and \eqref{cosmicsound} obtained using the Kodama-Ishibashi formalism upon identifying the time coordinate \eqref{eqn: fluid coord transf}, which in the large-$K\ell$ limit becomes
\begin{equation}
    t = K \ell \tau + \ell  \log \frac{2 K \ell}{\sqrt{1-\frac{\rho}{\ell}}} + \mathcal{O}((K\ell)^{-1})\,.
\end{equation}
The relative factor between the time coordinates $t$ and $\tau$ accounts for the red-shift factor of $\tfrac{1}{K\ell}$ in the frequencies \eqref{eqn: fluid sound frequency} and \eqref{eqn: fluid shear frequency} as compared to \eqref{eq:shearmodes} and \eqref{cosmicsound}. 

The sound and shear modes are only a subset of the admissible linearised modes obeying the conformal boundary conditions in the stretched horizon limit. But note that with respect to coordinate $\tau$, the frequencies of all the other modes found, such as \eqref{eq:cosmicplusIexp}, scale as $K\ell$. As such, they violate \eqref{eqn: fluid validity} and are outside the regime of validity of the large-$K\ell$ expansion presented in this section.

\textbf{Non-relativistic limit.} Lastly, we consider the non-relativistic limit of \eqref{eqn: fluid eom}. In particular, we consider the rescaled pressure $\tilde p$ and time $\tilde \tau$ defined as
\begin{equation}
    p = \frac{\tilde p}{K\ell}  \, , \qquad \partial_\tau = \frac{1}{K\ell}\partial_{\tilde\tau} \, .
\end{equation}
Note that in this non-relativistic limit, we do not rescale $v^i$ with $K\ell$, such that only the rescaled speed of sound goes to infinity. As a result, the leading order fluid equations become
\begin{equation}
    \nabla_i v^i=0 \, , \qquad \partial_{\tilde\tau} v_i = -\nabla_i \tilde p + \frac{\ell}{2}(\nabla^2 v_i + R_i{}^j v_j)-v^j \nabla_j v_i\,,
\label{eq:navierstokes}\end{equation}
which take the form of the incompressible Navier-Stokes equations on a round two-sphere of radius $\ell$, in agreement with \cite{Bredberg:2011xw,Anninos:2011zn}. On the gravity side, that is equivalent to suppressing scalar perturbations.


\section{Rindler dynamics} \label{sec: rindler}

In this section we study linearised dynamics around flat Rindler space. We show the precise stretched horizon limit in which the de Sitter results can be approximated by those in Rindler. Furthermore, in this limit, we show that all the phenomenology from both the pole and the cosmic patch dynamics can be obtained from Rindler dynamics. We end by generalising this result to other non-degenerate spherical horizons, with or without a cosmological constant, and discuss potential differences to distinguish them at the linearised level.

\subsection{Background solution}

Flat Rindler space is an exact solution to the Einstein equations with vanishing cosmological constant, and has the following line element
\beq ds^{2}=-\frac{z^{2}}{z_{0}^{2}}dt^{2}+dz^{2}+dx^{2}+dy^{2}\;,\label{eq:Rindmet}\eeq
for $(t,x,y)\in\mathbb{R}^{3}$. The Rindler horizon is located at $z=0$, and $z_{0}^{-1}>0$ denotes the proper acceleration of a uniformly accelerating observer. The parameter $z_0$ can always be absorbed by redefining the clock, so we will just use it as a book-keeping scale.

Let $z=\mathfrak{z}$ denote the location of a finite timelike boundary $\Gamma$. The induced metric, extrinsic curvature $\bar K_{mn}$ and its trace $K$ are given by, 
\beq ds^{2}|_{\Gamma}=\bar{g}_{mn}dx^{m}dx^{n}=-\frac{\mathfrak{z}^{2}}{z_{0}^{2}}dt^{2}+dx^{2}+dy^{2}\;,\quad \bar{K}_{mn}dx^{m}dx^{n}|_{\Gamma}=\sigma\frac{\mathfrak{z}}{z_{0}^{2}}dt^{2}\;, \quad  K=-\frac{\sigma}{\mathfrak{z}}\;,\label{eq:backgmn}\eeq
where indices $m,n$ range over coordinates $\{t,x,y\}$, and $\sigma=\pm 1$, depends on which region of space we consider. Even though there is not pole in Rindler space, in analogy to the de Sitter pole patch, we will call the region with $z\geq\mathfrak{z}$, the Rindler pole patch. In this case, $\sigma = +1$. We will call the region between the Rindler horizon and the boundary, $0\leq z \leq \mathfrak{z}$, the Rindler cosmic patch, which has $\sigma=-1$.

Note that the limit $\mathfrak{z}/z_0\to0$ coincides with $Kz_{0}\to-\sigma\infty$. 

For the Rindler background, the components of the conformal stress-tensor (\ref{eq: tmn}) are
\beq 
z_0^2 \bar{T}_{tt} = \frac{1}{12\pi G_{N} K}\;,\quad \bar{T}_{xx}=\bar{T}_{yy}=\frac{K}{24\pi G_{N}}\;,\label{eq:backSTRind}\eeq
such that, 
\beq (8\pi G_{N})^{2}\bar{T}^{mn}\bar{T}_{mn}|_{\Gamma}=\frac{2K^{2}}{3}\;,\label{eq:TsqmnRind}\eeq
for all $K$. This agrees with (\ref{eq: large K Tmn}) for the de Sitter static patch only in the de Sitter stretched horizon limit.

\subsection{Linearised dynamics}

We are interested in linearised metric perturbations about the Rindler background (\ref{eq:Rindmet}). For this we again  use the Kodama-Ishibashi formalism \cite{Kodama:2000fa}. Upon a convenient choice of gauge, the  metric perturbations $h_{\mu\nu}$  may be expanded in terms of scalar and vector harmonics $\mathbb{S}$ and $\mathbb{V}_{i}$ and master fields $\Phi^{(S/V)}$ with non-vanishing components (see appendix \ref{app:linpertRind} for details) 
\beq
\begin{split}
\begin{cases}
h_{mn}=\frac{k^{2}z_{0}^{2}}{2}\bar{g}_{mn}\Phi^{(S)}\mathbb{S}+z^{2}(\delta^{i}_{m}\delta^{t}_{n}+\delta^{i}_{n}\delta^{t}_{m})\partial_{z}\Phi^{(V)}\mathbb{V}_{i}\;,\\
h_{tz}=\frac{z_{0}^{2}}{2z}\left[z_{0}^{2}\partial_{t}^{3}+z\partial_{t}\partial_{z}+(k^{2}z^{2}-2)\partial_{t}\right]\Phi^{(S)}\mathbb{S}\;,\\
h_{zz}=z_{0}^{2}\left(\frac{k^{2}}{2}-\frac{z_{0}^{2}}{z^{2}}\partial_{t}^{2}+\frac{z_{0}^{2}}{z}\partial_{z}\partial_{t}^{2}+k^{2}z\partial_{z}\right)\Phi^{(S)}\mathbb{S}\;,\\
h_{zi}=\frac{z_{0}^{4}}{2z}\left[\partial_{t}^{2}-\frac{z}{z_{0}^{2}}\partial_{z}+\frac{k^{2}z^{2}}{z_{0}^{2}}\right]\Phi^{(S)}\partial_{i}\mathbb{S}+z_{0}^{2}\partial_{t}\Phi^{(V)}\mathbb{V}_{i}\;.
\end{cases}
\end{split}
\label{eq:metperRind}\eeq
The scalar and vector harmonics obey $(\partial_{x}^{2}+\partial_{y}^{2})\mathbb{S}=-k^{2}\mathbb{S}$ and $(\partial^{2}_{x}+\partial_{y}^{2})\mathbb{V}_{i}=-k^{2}\mathbb{V}_{i}$ for $k\in\mathbb{R}^{+}$ (with dimensions of inverse length), and $\partial_{i}\mathbb{V}^{i}=0$. For $k\neq0$, the master fields $\Phi^{(S/V)}$ satisfy the equation of motion 
\beq \left(-\frac{z_{0}^{2}}{z^{2}}\partial_{t}^{2}+\frac{1}{z}\partial_{z}+\partial_{z}^{2}-k^{2}\right)\Phi^{(S/V)}=0\;.\label{eq:scalEOMrind}\eeq
It is straightforward to verify that the Ricci scalar on the perturbed background $g_{\mu\nu}=\bar{g}_{\mu\nu}+\varepsilon h_{\mu\nu}$ vanishes at linear order in $\varepsilon$ upon implementing the master field equation of motion. 

At the boundary, the metric is taken to have the form
\beq ds^{2}|_{\Gamma}=e^{2\bomega(x^{m})}\left(-\frac{\mathfrak{z}^{2}}{z_{0}^{2}}dt^{2}+dx^{2}+dy^{2}\right)\;,\label{eq:indmetweylRind}\eeq
for unspecified Weyl factor $\bomega(x^{m})$.  The dynamics of the  Weyl factor is governed by (\ref{eq: 2.12}), which in this case becomes
\beq \mathcal{D}^{m}\mathcal{D}_{m}\bomega+\frac{1}{2}\mathcal{D}_{m}\bomega \mathcal{D}^{m}\bomega-16\pi^{2}G_{N}^{2}T^{mn}T_{mn}e^{-4\bomega}+\frac{K^{2}}{6}e^{2\bomega}=0\;.\label{eq:212Rind}\eeq
The background trivially satisfies (\ref{eq:212Rind}) via (\ref{eq:TsqmnRind}). Linearising around $\bomega=0$, we obtain,
\beq \mathcal{D}^{m}\mathcal{D}_{m}\delta\bomega+K^{2}\delta\bomega = 32\pi^{2}G_{N}^{2}\bar{T}^{mn}\delta T_{mn} \;.\label{eq:lin212Rindv2}\eeq

\noindent \textbf{Vector perturbations.} With our choice of gauge, the boundary condition $\delta K|_{\Gamma}=0$ is automatically satisfied. Insisting the conformal class of the metric is preserved imposes
\beq \left. \partial_{z}\Phi^{(V)}\right|_{\Gamma}=0\;.\label{eq:vecbcrind}\eeq
The Weyl factor is unaffected by the vector perturbations, and the only non-zero components of the linearised stress-tensor are
\beq 
\begin{split} 
\begin{cases}
8\pi G_{N}\delta T^{(V)}_{ti}=\left. -\frac{1}{6}\left(z\partial_{z}\Phi^{(V)}+3z^{2}k^{2}\Phi^{(V)}\right)\right|_{\Gamma}\mathbb{V}_{i}\;,\\
8\pi G_{N}\delta T^{(V)}_{ij}=\frac{z_{0}^{2}}{2}\partial_{t}\Phi^{(V)}|_{\Gamma}(\tilde{\nabla}_{i}\mathbb{V}_{j}+\tilde{\nabla}_{j}\mathbb{V}_{i})\;,
\end{cases}
\end{split}
\eeq
where we used the master field equation of motion (\ref{eq:scalEOMrind}).
It is then easy to verify $\bar{T}^{mn}\delta T_{mn}^{(V)}=0$,
such that the Weyl factor dynamics (\ref{eq:212Rind}) is self-consistent.

\vspace{2mm}

\noindent \textbf{Scalar perturbations.} With our choice of gauge, scalar perturbations keep the conformal class of the metric unchanged and imposing $\delta K|_{\Gamma}=0$ amounts to
\beq \left. \left(2\partial_{t}^{4}+\frac{2(2k^{2}z^{2}-1)}{z_{0}^{2}}\partial_{t}^{2}+\frac{k^{2}z^{2}}{z_{0}^{4}}(2k^{2}z^{2}-1)\right)\Phi^{(S)}-\frac{k^{2}z^{3}}{z_{0}^{4}}\partial_{z}\Phi^{(S)}\right|_{\Gamma}\;=0\;.\label{eq:deltaK0Rind}\eeq
Comparing to the components of the metric perturbation (\ref{eq:metperRind}), at the linear level the Weyl factor at the boundary is directly related to the master field, 
\beq \delta\bomega =\left. \frac{k^{2}z_{0}^{2}}{4}\Phi^{(S)}\mathbb{S} \right|_\Gamma \;.\label{eq:weylRindB}\eeq
At linearised order, the components of the conformal stress-tensor $\delta T_{mn}$ are  
\beq 
\begin{cases}
8\pi G_{N}\delta T_{tt}=-\frac{2\mathfrak{z}}{k^{2}z_{0}^{2}}(1-z_{0}^{2}\partial_{t}^{2}-k^{2}\mathfrak{z}^{2})\tilde{\nabla}^{2}\delta\bomega\;,\\
8\pi G_{N}\delta T_{ti}=-\frac{2}{2\mathfrak{z}k^{2}}(1-z_{0}^{2}\partial^{2}_{t}-k^{2}\mathfrak{z}^{2})\partial_{t}\tilde{\nabla}_{i}\delta\bomega\;,\\
 8\pi G_{N}\delta T_{ij}=\frac{2(1-z_{0}^{2}\partial^{2}_{t}-k^{2}\mathfrak{z}^{2})}{k^{4}\mathfrak{z}^{3}}\left[(k^{2}\mathfrak{z}^{2}+2z_{0}^{2}\partial_{t}^{2})\left(\tilde{\nabla}_{i}\tilde{\nabla}_{j}-\frac{1}{2}\tilde{g}_{ij}\tilde{\nabla}^{2}\right)-\frac{k^{2}\mathfrak{z}^{2}}{2}\tilde{g}_{ij}\tilde{\nabla}^{2}\right]\delta\bomega\;,   
\end{cases}
\label{eq:deltaTmnRindgen}\eeq
where we used the boundary condition  to eliminate $z$-derivatives on $\Phi$. We can easily recast the components in terms of the linearised Weyl factor (\ref{eq:weylRindB}), and it is quick to verify $\delta T_{mn}$ is  traceless, and transverse with respect to $\bar{g}^{mn}$. Further,
\beq 
\begin{split} 
\hspace{-4mm} 64\pi^{2}G_{N}^{2}\bar{T}^{mn}\delta T_{mn}&=\frac{\mathbb{S}z_{0}^{2}}{6}\left[-\frac{k^{2}(k^{2}z^{2}-2)}{z^{2}}+\frac{z_{0}^{2}(k^{2}z^{2}-2)}{z^{4}}\partial_{t}^{2}+\frac{2z_{0}^{4}}{z^{4}}\partial_{t}^{4}\right]\Phi^{(S)}-\frac{k^{2}z_{0}^{2}\mathbb{S}}{6z}\partial_{z}\Phi^{(S)}\biggr|_{\Gamma}\;.
\end{split}
\label{eq:TdeltaT}\eeq 
As in the de Sitter context, we can either substitute the Weyl factor (\ref{eq:weylRindB}) and (\ref{eq:TdeltaT}) into the Weyl equation of motion (\ref{eq:lin212Rindv2}) to recover boundary condition (\ref{eq:deltaK0Rind}), or we can use (\ref{eq:weylRindB}) and (\ref{eq:deltaK0Rind}) to recast (\ref{eq:TdeltaT}) in terms of $\delta\bomega$, self-consistently recovering the equation of motion (\ref{eq:lin212Rindv2}). 

\noindent \textbf{Physical diffeomorphisms.} The Kodama-Ishibashi formalism only captures gauge-invariant bulk modes. In Rindler, this analysis is only valid for $k\neq0$. In addition to the bulk modes, there can also be physical diffeomorphisms. We verified that at the linearised level there are no such physical diffeomorphisms, except in the $k=0$ case, that we describe below.

For this translationally invariant sector, there exists physical diffeomorphisms which lead to a non-vanishing Weyl factor. The conformal boundary conditions impose that the Weyl factor induced by such diffeomorphisms is subject to a non-linear equation analogous to \eqref{eqn: brane eqn} for the de Sitter case. In particular, written in terms of $E_\text{conf}$, we find 
\begin{equation}
\label{eqn: brane eqn Rindler}
    \partial_t^2\bomega =  - \frac{1}{2}(\partial_t \bomega)^2-24 \pi^2G_N^2 E_\text{conf}^2e^{-4\bomega} + K^2\frac{\mathfrak{z}^2e^{2\bomega}}{6z_0^2} \,.
\end{equation}
Here, $E_\text{conf}$ is defined as the energy density 
which is given by
\begin{equation}
    E_\text{conf} = \frac{K\mathfrak{z}e^{3\bomega}\mp 3 \sigma e^{2\bomega}z_0\partial_t \bomega}{12 \pi G_N z_0} \, .
\end{equation}
The $\mp$ arises due to the fact we have two possible solutions with our choice of embedding. 
The conformal stress-tensor is then
\begin{equation}
    T_{mn} dx^m dx^n = \frac{E_\text{conf}z_0}{2\mathfrak{z}}\left(2\frac{\mathfrak{z}^2}{z_0^2}dt^2+dx^2+dy^2\right) \, . 
\end{equation}

Linearising the Weyl equation of motion (\ref{eqn: brane eqn Rindler}) about the $\bomega=0$ background yields 
\beq   \partial_t^2\delta\bomega = \mp \frac{1}{z_{0}}\partial_{t}\delta\bomega\;.
\label{eq:linweylrind}\eeq
The solutions to this equation are $\delta\bomega(t)=e^{-i\omega^{(0)}t}$ with either $\omega^{(0)}z_{0}=i$ or $\omega^{(0)}z_{0}=-i$. Notably, their superposition is not a solution to (\ref{eq:linweylrind}). This behaviour is compatible with taking the stretched horizon limit of the de Sitter Weyl equation of motion (\ref{eqn: brane eqn}), and then performing the linearisation. This is distinct from linearising (\ref{eqn: brane eqn}) and then taking the stretched horizon limit.

\subsubsection{The stretched horizon limit} 

At this point, we are in position of making a direct comparison with the stretched horizon limit of the dS static patch.

First, recall the static patch metric \eqref{eq:dSmetric}. Consider the following change of coordinates, where we take the de Sitter radial coordinate $r$ to 
\begin{equation}
    r= \ell - \frac{z^2}{2\ell} \left(1-\frac{\r^2}{\ell^2}\right) = \ell - \frac{z^2}{2\ell} \left( |K\ell|^{-2} + \mathcal{O}(|K\ell|^{-4}) \right) \,. 
\end{equation}
With this change of coordinates, in the stretched horizon limit, the de Sitter static patch metric becomes, 
\begin{equation}
    ds^2 =\frac{1}{|K\ell|^2}\bigg(dz^2 -\frac{z^2}{\ell^2}dt^2 +|K\ell|^2\ell^2 d\Omega_2^2\bigg)+\mathcal{O}(|K\ell|^{-4})\;.
\end{equation}
The radius of the $S^2$ is $|K\ell|\ell$, so in the $|K\ell|\rightarrow \infty$
limit it can be approximated by two-dimensional flat space. Identifying $\ell\leftrightarrow z_0$, we note that to leading order in the large-$|K\ell|$ expansion, the bulk metric looks conformal to flat Rindler.

In fact, in the $|K\ell|\rightarrow \infty$ limit, the eigenfunctions and eigenvalues of the Laplacian operator on $S^2$ become the ones of two-dimensional flat space. In particular, in this limit, we can identify the Rindler momentum $k$ with 
\begin{align}\label{eq:idkl}
    k z_0 \leftrightarrow \frac{\sqrt{l(l+1)}}{|K\ell|} \,.
\end{align}
Note this means that finite momentum $k z_0$ in Rindler will agree with the de Sitter result in the stretched horizon limit, provided we also take $l\rightarrow\infty$ with the above ratio fixed.

Now we can consider the master field bulk equation of motion \eqref{eq:scalEOMrind}. The matching with the de Sitter bulk equation of motion \eqref{eq: master equation} can be seen explicitly by considering the Fourier decomposition of the master fields and changing the radial coordinate from $r$ to $z$. Using \eqref{eq:idkl}, it follows that the de Sitter master field equation matches the Rindler master field equation at leading order in the large-$|K\ell|$ expansion,
\begin{equation}
     \nabla^{2}_{\text{dS}}\Phi(r,t)-\frac{l(l+1)}{r^2}\Phi(r,t)=|K\ell|^2\bigg(\nabla^{2}_{\text{Rind}}\Phi(z,t) -k^2 \Phi(z,t)\bigg)+\mathcal{O}(1) \,,
\end{equation}
where $\nabla^{2}_{\text{Rind}}$ is the Laplacian on two-dimensional Rindler space. Similarly, changing the radial coordinate from $r$ to $z$ in the de Sitter conformal boundary conditions and expanding in large $|K\ell|$ with \eqref{eq:idkl} fixed leads, at leading order, to the same conformal boundary conditions as in Rindler. In the vector sector, the derivative term dominates in the large-$|K\ell|$ limit and we recover \eqref{eq:vecbcrind}, up to subleading corrections. In the scalar sector, we recover \eqref{eq:deltaK0Rind}.

Hence, we have shown that at leading order in the large-$|K\ell|$ expansion, with $\tfrac{l(l+1)}{(K\ell)^2}$ fixed, the de Sitter linearised dynamics with conformal boundary conditions matches the analogue Rindler problem.

\subsection{Pole and cosmic patch}

We proceed in exactly the same way as we did with the de Sitter perturbations. We first identify bulk solutions with regularity properties that resemble those of the pole and the cosmic patch of de Sitter.

As before, we assume a Fourier decomposition for our master fields, $\Phi^{(S/V)}(t,z)=\mathfrak{R}\, e^{-i\omega t}\phi^{(S/V)}(z)$. Then, for each allowed frequency, the master field equation becomes,
\beq 0=\left(\frac{z_{0}^{2}\omega^{2}}{z^{2}}+\frac{1}{z}\partial_{z}-\partial^{2}_{z}-k^{2}\right)\phi^{(S/V)}\;,\label{eq:mastfieldnoteom}\eeq
which has the general solution
\beq \phi^{(S/V)}(z)=c_{1}J_{iz_{0}\omega}(-ikz)+c_{2}Y_{iz_{0}\omega}(-ikz)\;,\label{eq:genphiRindscal}\eeq
for constant coefficients $c_{1,2}$ and $J_{n}(x)$ and $Y_{n}(x)$ respectively denote Bessel functions of the first and second kind.

\textbf{Pole patch.} In de Sitter, the pole patch solution for the master field was chosen by imposing regularity at the pole $r=0$. In the case of Rindler, there is no equivalent requirement, since there is no pole. Nevertheless, we can select the solution that decreases as $z$ increases towards $z=z_0$. The linear combination of the Bessel functions with this property is the modified Bessel function of the second kind, 
\beq \phi^{(S/V)}_{\text{pole}}(z)=K_{iz_{0}\omega}(kz),\label{eq:polsolrind}\eeq
which is exponentially decaying when we increase $z$.

\textbf{Cosmic patch.} For the cosmic patch, similar to de Sitter, we choose master field solutions that are outgoing at the Rindler horizon $z=0$.
The linear combination of Bessel functions which captures the outgoing solution is
\beq \phi^{(S/V)}_{\text{cosmic}}(z)=J_{-iz_{0}\omega}(-ikz)\;.\label{eq:cpsolphi}\eeq

Next, we need to impose the conformal boundary conditions for both the vector and the scalar sector. Notice that for both patches, both boundary conditions, \eqref{eq:vecbcrind} and \eqref{eq:deltaK0Rind}, depend on the Rindler momentum only through the following dimensionless combination,
\begin{equation}
    \kappa \equiv k \mathfrak{z} \,.
\end{equation}

As in de Sitter, we can study the solutions perturbatively, both in the limit $\kappa \to \infty$, which corresponds to the large angular momentum limit in the static patch, and in the $\kappa \to 0$ limit, which corresponds to the large mean curvature limit. 

Given the previous discussion, we stress that {all} the phenomenology of modes that we found in the de Sitter case can be reproduced perturbatively to all orders in the $\kappa$-expansion in Rindler. Moreover, the solutions will quantitatively agree up to numerical coefficients that are order one in large $\L$. Given the detailed presentation in the previous sections, in the following, we will be brief. All perturbative results have been checked against the numerical evaluations and they agree in the appropriate limits. The interested reader can find even more details in appendix \ref{app:Rindmodes}.

\subsection{Pole patch modes}

\subsubsection{Scalar sector} We start the discussion of the pole patch modes in the scalar sector. Imposing (\ref{eq:deltaK0Rind}) for the master field (\ref{eq:polsolrind}), we obtain that scalar perturbations obey 
\beq 
\begin{split}
\left(
2\kappa^{4} + 2z_{0}^2 \omega^2 + 2z_{0}^4 \omega^4 
- \kappa^{2} \left(1 + 4z_{0}^2 \omega^2 \right)
\right)
K_{i z_{0} \omega}(\kappa)+ \frac{1}{2} \kappa^{3} \left(
K_{-1 + i z_{0} \omega}(\kappa) + K_{1 + i z_{0} \omega}(\kappa)
\right) = 0 \;.
\end{split}
\label{eq:CBCpoleRind}\eeq
Note that this equation has the same properties as its analogue in the de Sitter pole patch. Namely, if $\omega$ is a solution, so are $-\omega$ and minus the complex conjugate of $\omega$. Hence, solutions for $\omega$ will be symmetric in the complex plane both with respect to the real and imaginary axes.

\textbf{Large spatial momentum.} We first take the large spatial momentum limit, $\kappa \to \infty$. 
The master field equation in this limit can be analyzed using the WKB approximation, similar to section \ref{sec: large l}. We find that the master field is given by an Airy function at leading order and that it localizes within a width $\kappa^{-2/3}$ near the boundary. We can solve analytically order by order the equation together with the conformal boundary condition \eqref{eq:CBCpoleRind} and obtain,
\beq \omega_{(\kappa)} z_{0}= \kappa+ \frac{\nu}{2^{1/3}}\kappa^{1/3}+2^{1/3}\frac{36 + 89 \nu^3 + 16 \nu^6}
{60  \, \nu \left(-7 + 16 \nu^3\right)}\kappa^{-1/3} + \mathcal{O}(\kappa^{-1}) \;,\label{eqn: large kappa scalar pole}\eeq
with $\nu$ the solution to the same equation as in the case of the de Sitter pole patch, namely \eqref{eqn: Airy bdry eqn}. The (same) solutions with positive imaginary part are present. These frequencies exhibit the same structure as the equivalent de Sitter pole patch ones, see \eqref{eqn: omega L sca strch pole}.

\textbf{Large mean curvature.} In the opposite limit, we take $\kappa \to 0$, to obtain
\begin{align}\label{rindlerlimitdK}
    (\omega z_0)^2 ((\omega z_0)^2+1)\phi^{(S)}(\kappa) =0.
\end{align}
Just like in the case of de Sitter, the solutions to leading order are $\omega z_0 = 0,\pm i$ and a tower of modes originating from the vanishing of the Rindler master field at the boundary. Expanding around each, we obtain,

\begin{itemize}
    \item \textbf{Soft modes} with the following allowed frequencies 
    \begin{align}
      \omega_{(\pm i)} z_{0} =\pm i \mp i\kappa^{2}+\mathcal{O}(\kappa^{4}\log(\kappa)) \,.
\end{align}
    The corresponding master field diverges at the boundary in the same way as for the dS soft modes, justifying the choice of Bessel function in \eqref{eq:polsolrind}. These modes have radial profiles localised near the boundary similar to their de Sitter counterparts, and their associated Weyl factor grows/decreases exponentially with time to leading order.
    \item \textbf{Gapless modes} with the following allowed frequencies 
    \begin{align}
    \omega z_0&=\pm \frac{\kappa}{\sqrt{2}} +\mathcal{O}(\kappa\log^{-1}\kappa) \,.
\end{align}
    These modes are always real, and their profile localises close to the boundary.
    \item \textbf{Tower of real modes} with allowed frequencies that satisfy the following equation at leading order in the $\kappa$ expansion
    \beq
\omega_{n}z_{0}=\frac{i}{2\log(\kappa/2)}\left[\log\left(-\frac{\Gamma(-i\omega_{n}z_{0})}{\Gamma(i\omega_{n}z_{0})}\right)+2\pi i n\right]\;,\; n\in\mathbb{Z}/\{0\}\;. \label{eq:realtowscalRind_2}
\eeq
These are further divided into low frequency modes, which have a radial profile that localises logarithmically at small-$\kappa$, and a family of high frequency modes, whose radial profile has support over all the Rindler pole patch.
\end{itemize}

\subsubsection{Vector sector} In the vector sector, imposing the boundary condition (\ref{eq:vecbcrind}) on (\ref{eq:polsolrind}) yields
\begin{equation} 
    k(K_{1+iz_{0}\omega}(\kappa)+ K_{-1+iz_{0}\omega}(\kappa))=0\;.\label{eq:CBCpoleRindvec}
\end{equation} 
Similar to the scalar sector, taking $\omega\to-\omega$ or $\omega\to-\omega^*$ preserves this equation, implying that solutions for $\omega$ are symmetric in the complex plane with respect to both real and imaginary axes.

\textbf{Large momentum.} The large (spatial) momentum limit, $\kappa \to \infty$, can be obtained in a similar way as the scalar perturbations, with the master field given by an Airy function localized within a width $\kappa^{-2/3}$ near the boundary. The frequency which solves \eqref{eq:CBCpoleRindvec} in the $\kappa^{-1}$ expansion is
\begin{equation}
    \omega z_0 = \kappa + \frac{\nu}{2^{1/3}}\kappa^{1/3}+2^{1/3}\frac{6+\nu^3}{60 \nu}\kappa^{-1/3}+\mathcal{O}(\kappa^{-1})\,\label{eqn: large kappa vector pole}
\end{equation}
with $\nu$ satisfying \eqref{eqn: Airy prime bdry cond}, which in particular does not have any complex solution. These frequencies exhibit the same structure as those of the de Sitter vector pole patch, see \eqref{eqn: omega L vec strch pole}.

\textbf{Large mean curvature.} In the opposite limit, $\kappa \to 0$, the frequency solutions to \eqref{eq:CBCpoleRindvec} only give a {tower of real modes}. These are frequencies that satisfy the following equation at leading order in the small-$\kappa$ expansion
    \beq
\omega_{n}z_{0}=\frac{i}{2\log(\kappa/2)}\left[\log\left(\frac{\Gamma(-i\omega_{n}z_{0})}{\Gamma(i\omega_{n}z_{0})}\right)+2\pi i n\right]\;,\;  n\in\mathbb{Z}/\{0\}\;.
\label{eq:realtowvectorRind}\eeq
This tower also is split into low and high frequency modes in the same way as the scalar tower of real modes.

\subsubsection{Connection to quasinormal modes}
Similar to the de Sitter case, in the strict $\kappa\rightarrow 0$ limit, it is natural to inquire whether the Rindler quasinormal modes can be recovered from our analysis of the Rindler pole patch normal modes. For both scalar and vector sectors, the Weyl factor of the normal modes vanishes at the boundary, replicating the Dirichlet problem. Following the remarks from section \ref{sectionQNM}, we inspect the ratio between the coefficients of the outgoing and ingoing components of the pole patch solutions \eqref{scatteringphase}. We find
 \begin{align}
        \bigg(\frac{A^{\text{out}}}{A^{\text{in}}}\bigg)_{\hspace{-1mm}S}^{\hspace{-1mm}\text{Rind}}&=  \frac{2^{2 i\omega z_0} \, \Gamma( i\omega \ell) }
{\Gamma(- i\omega\ell)}\;,\quad  \bigg( \frac{A^{\text{out}}}{A^{\text{in}}}\bigg)_{\hspace{-1mm} V}^{\hspace{-1mm}\text{Rind}}= \frac{2^{2 i\omega z_0} \, \Gamma(1 + i\omega \ell)}
{\Gamma(1 - i\omega\ell)}\;.
    \end{align}
The zeros and poles are at $\mp i n$. Comparing with \eqref{QNMssmatrix}, we notice that dividing the de Sitter ratio by the Rindler ratio, we find a quantity whose poles are only the de Sitter quasinormal models \cite{Law:2022zdq}.

\subsection{Cosmic patch modes}

Finally we turn to the cosmic patch master field solutions (\ref{eq:cpsolphi}). Scalar perturbations in the cosmic patch obey (\ref{eq:deltaK0Rind}),
\beq (2\kappa^{4}+2z_{0}^{2}\omega^{2}(1+z_{0}^{2}\omega^{2})+\kappa^{2}[-1+z_{0}\omega(i-4z_{0}\omega)])J_{-iz_{0}\omega}(-i\kappa)-i\kappa^{3}J_{1-iz_{0}\omega}(-i\kappa) = 0 \;,\eeq
while the vector perturbations satisfy (\ref{eq:vecbcrind}),
 \beq J_{-1-iz_{0}\omega}(-i\kappa)-J_{1-iz_{0}\omega}(-i\kappa) = 0\;.\eeq

Below we briefly report on the allowed frequencies in each case, which coincide which those of the dS cosmic patch analysis in the appropriate limit. In this Rindler regime, it is still subtle to take the strict $\kappa \to 0$ limit. Note that the radial profile takes a very simple form, that can be non-trivial even in the strict limit. In the Rindler cosmic patch, modes do not particularly localise close to the boundary, but they have support from the boundary to the Rindler horizon, $0 \leq z \leq \mathfrak{z}$.

In the cosmic patch, a summary of the allowed modes and their respective radial profile can be found in table \ref{table rindler}.

\begin{center}
\begin{tcolorbox}[
    tab2,
    tabularx={l||l||l}, 
    title=Cosmic patch modes in the Rindler stretched horizon limit,
    boxrule=2pt,
    width=15.6cm,
    code={\setstretch{1.7}}
]

\multicolumn{3}{c}{\textbf{Soft scalar modes}} \\
\hline\hline
\textbf{Type} & \textbf{Frequency} & \textbf{Master field} \\
\hline\hline
$\omega_{(+i)}$ & $\omega z_0 = i -i\frac{\kappa^2}{2} + \mathcal{O}(\kappa^4)$ & $ \phi^{(S)}(z) = \frac{z}{\mathfrak{z}} + \mathcal{O}(\kappa^2)$ \\
$\omega_{(-i)}$ & $\omega z_0 =-i + i\frac{3\pm \sqrt{17}}{8}\kappa^2+\mathcal{O}(\kappa^4)$ & $\phi^{(S)}(z)= \frac{-1\pm\sqrt{17}}{4}\frac{\mathfrak{z}}{z} +\frac{5\mp \sqrt{17}}{4}\frac{z}{\mathfrak{z}} + \mathcal{O}(\kappa^2)$ \\
\hline\hline

\multicolumn{3}{c}{\textbf{Fluid modes}} \\
\hline\hline
Shear  & $\omega z_0 = -\frac{i}{2}\kappa^2 + \mathcal{O}(\kappa^{4})$ & $\phi^{(V)}(z) = \left(\frac{\mathfrak{z}}{z}\right)^{\frac{\kappa^2} {2}}+\mathcal{O}(\kappa^2)$ \\
Sound  & $\omega z_0 = \pm\frac{1}{\sqrt{2}}\kappa-\frac{i}{4}\kappa^2 + \mathcal{O}(\kappa^{3})$ & $\phi^{(S)}(z) = \left(\frac{z}{\mathfrak{z}}\right)^{\mp \frac{i\kappa} {\sqrt{2}}}+\mathcal{O}(\kappa^2\log \kappa)$ \\
\hline\hline

\multicolumn{3}{c}{\textbf{Quasi-normal like modes}} \\
\hline\hline
Scalar & $\omega z_0 = -in + \frac{i (-1)^n \kappa^{2n}}{2^{2n}\Gamma(n+1)\Gamma(n)} + \mathcal{O}(\kappa^{2n+2})$ & $ \phi^{(S)}(z) = \left(\frac{z}{\mathfrak{z}}\right)^{-n}-\left(\frac{z}{\mathfrak{z}}\right)^{n} + \mathcal{O}(\kappa^{2})$ \\
Vector & $\omega z_0 = -in + \frac{i(-1)^{n+1}\kappa^{2n}}{2^{2n}\Gamma(n+1)\Gamma(n)} + \mathcal{O}(\kappa^{2n+2})$ & $ \phi^{(V)}(z) = \left(\frac{z}{\mathfrak{z}}\right)^{-n}+\left(\frac{z}{\mathfrak{z}}\right)^{n} + \mathcal{O}(\kappa^{2})$ \\

\end{tcolorbox}
\end{center}

\begin{center}
\begin{minipage}[c]{\textwidth}
\captionof{table}{Allowed modes in the stretched horizon limit of the Rindler cosmic patch. For the quasinormal mode-like scalar modes, $n \geq 2$, while for the quasinormal mode-like vector modes, $n\geq1$.} \label{table rindler}
\end{minipage}
\end{center}

Other details, such as the leading correction to the Weyl factor at the boundary and/or the conformal stress tensor, can be seen directly from the de Sitter analysis in section \ref{sec: large K cosmic}. They coincide after making the identification (\ref{eq:idkl}) and further taking the large-$l$ limit of the de Sitter expressions.

These modes are all obtained in the small-$\kappa$ expansion. In the cosmic patch there are also large momentum modes that essentially coincide with the pole patch modes. Namely, \eqref{eqn: large kappa scalar pole} for the scalar perturbations and \eqref{eqn: large kappa vector pole} for the vector perturbations, with $\nu$ now given by \eqref{eqn: Airy bdry eqn 2} and \eqref{eqn: Airy prime bdry cond 2}, respectively.

\subsection{Horizon disparity?} 

The analysis of linearised perturbations about flat Rindler space reveals all of the modes found both in the pole and the cosmic patch analyses in the de Sitter stretched horizon limit (including the angular momentum modes) can be matched one-to-one to equivalent Rindler modes. Discrepancies only appear in numerical factors that are suppressed in the large  momentum limit.

We finish this section by showing that, in fact, this is a universal aspect of non-degenerate horizons in general relativity, at least for horizon patches. For pole patches, depending on the asymptotics of the spacetime, further boundary conditions might be needed. We leave this as an interesting problem for the future.\footnote{Consider, for instance,  the problem in asymptotically AdS spacetimes. Here we could impose a Fefferman-Graham type expansion close to the conformal boundary of AdS, supplemented by conformal boundary conditions at a boundary parametrically close to an AdS black hole. These configurations are not in the standard AdS/CFT dictionary.}

For now, let us consider a horizon patch in a background of the form 
\begin{equation}
    ds^2 = - f(r) dt^2 + \frac{dr^2}{f(r)}+r^2 d\Omega_2^2 \, , 
    \label{eq: general background metric copy}
\end{equation}
where we assume that there exists an $r_{\text{h}}$ such that $f(r_\text{h})=0$ but $\partial_rf(r_\text{h})\neq0$. The horizon patch is defined as the bulk region between a boundary at $r=\mathfrak{r}$ and the horizon. 
We can similarly redefine the radial coordinate 
\begin{align}
r=r_{\text{h}}-\frac{z^2}{2 r_{\text{h}}}\bigg(1-\frac{\mathfrak{r}^2}{r_{\text{h}}^2}\bigg)=r_{\text{h}} + \frac{z^2}{4 }\partial_r f(r_{\text{h}}) (K r_{\text{h}})^{-2} + \mathcal{O}((K r_{\text{h}})^{-4})\;.  
\end{align}
Then the bulk metric at large-$|K r_{\text{h}}|$ looks like 
\begin{align}
    ds^2&=(K r_{\text{h}})^{-2}\bigg( -\frac{z^2 |\partial_r f(r_{\text{h}})|^2}{4} dt^2+ dz^2 +r_{\text{h}}^2 |K r_{\text{h}}|^2 d\Omega_2^2 \bigg) + \mathcal{O}((K r_{\text{h}})^{-4})\;.
\end{align}
Introducing 
\begin{equation}\label{eqn: iden z0}
    z_0=\frac{2}{|\partial_r f(r_{\text{h}})|}\, ,
\end{equation}
and replacing the two-sphere $S^2$ with infinite radius in the large-$K r_{\text{h}}$ limit with two-dimensional flat space gives
\begin{align}
    ds^2&=(K r_{\text{h}})^{-2}\bigg( -\frac{z^2 }{z_0^2} dt^2+ dz^2 +dx^2 +dy^2 \bigg) + \mathcal{O}((K r_{\text{h}})^{-4})\;.
\end{align}
At the level of the eigenvalues of the Laplacian operator, the limit in which we approximate $S^2$ with the two-dimensional flat space corresponds to the limit $l(l+1)\rightarrow \infty$ and $|K r_{\text{h}}| \rightarrow \infty$ while keeping
\begin{align}\label{identifRind}
   k z_0 = \frac{\sqrt{l(l+1)}}{|K r_{\text{h}}|}\frac{2}{r_{\text{h}}|\partial_r f(r_{\text{h}})|} \,
\end{align}
fixed. Like we saw for de Sitter, this fixed quantity maps to the Rindler momentum.

Furthermore, similar to the de Sitter case, the metric identification implies 
\begin{align}
    \nabla^{2}\Phi(r,t)= (K r_{\text{h}})^2\nabla^{2}_{\text{Rind}}\Phi(z,t) +\mathcal{O}(1)\;,
\end{align}
where $ \nabla^{2}$ is the Laplacian on the two-dimensional $(t,r)$-sector of the metric
\eqref{eq: general background metric copy}. The generic form of the master field equation on \eqref{eq: general background metric copy} is
\begin{align}
    \nabla^{2}\Phi(r,t)=V_{\text{eff}}(r)\Phi(r,t)\;,
\end{align}
where the expression for $V_\text{eff}(r)$ depends on the particular $f(r)$ considered and is given in Eqs. (3.2)--(3.7) of \cite{Kodama:2003jz}. It is easy to see that by changing the radial coordinate from $r$ to $z$ and taking the $|K r_{\text{h}}|\rightarrow \infty$ limit with $k z_0$ fixed, the leading order term in the potential is always
\begin{align}
    V_{\text{eff}}(z)&=k^2 (K r_{\text{h}})^2+\mathcal{O}(1) \,.
\end{align}
Subleading terms depend on the specific background metric. Hence, for arbitrary static backgrounds, the master field equation in a near-horizon expansion, maps, to leading order, to flat Rindler with $k z_0$ fixed.

In fact, the leading large $K \rh$-limit of the vector and scalar boundary conditions around any static and spherically symmetric Einstein metric are given by those of Rindler, \eqref{eq:vecbcrind} and \eqref{eq:deltaK0Rind}, upon following the identifications of $z_0$  and $k$, \eqref{eqn: iden z0} and \eqref{identifRind}. The first deviation away from Rindler is suppressed by a factor of $\tfrac{1}{(K\rh)^2}$.

Thus, we expect the same type of analysis to hold for all spherically symmetric, non-degenerate horizons. Similar considerations apply to the large angular momentum modes, see appendix \ref{app: large l generic}. Interestingly, the fluid dynamical picture of a conformal fluid with zero bulk viscosity and shear viscosity to entropy density ratio   $\tfrac{\eta}{s} = \tfrac{1}{4\pi}$ holds independently of the black hole type and/or the sign of the cosmological constant. It does not require an AdS$_4$ asymptotic boundary. It would be interesting to see how this picture gets modified when the horizon becomes (near-)extremal  \cite{Banihashemi:2025qqi, Galante:2025tnt}. We leave this for the future.

\begin{center}
\pgfornament[height=5pt, color=black]{83}
\end{center}
\vspace{5pt} 

To contrast the above universality, we end the section with an attempt to recognise the differences between different types of horizons in the linearised gravity analysis. As mentioned, to leading order in this double, small $\kappa$ expansion, the Rindler horizon will capture all the phenomenology, so the differences will be hidden in subleading contributions.

For this analysis, we will restrict to the soft modes. In the cosmic patch of de Sitter, we found (\ref{eq:cosmicplusIexp}), that we can conveniently write  as
\begin{equation}
    \frac{\omega^{(\text{dS}_4)} \r}{\sqrt{f(\r)}} = \frac{2\pi  i }{\b} \left(1-\frac{l(l+1)-2}{2 K^2\ell^2} + \mathcal{O}((K\ell)^{-4})\right) \,,\label{eqn: +i dS}
\end{equation}
for inverse conformal temperature $\tilde{\beta}$ at the boundary.
To compare with a black hole, in principle, we need to solve the full problem, and given the less symmetric background, we do not have access to analytic results. However, perturbatively around the soft mode frequency,  $\omega_{(+i)}$, it is possible to solve the linearised problem with conformal boundary conditions in more general backgrounds. For instance, for the de Sitter black hole, taking $K r_\text{h} \to +\infty$, we obtain 
\begin{equation}\label{bhsoft}
    \frac{\omega \r}{\sqrt{f(\r)}} = \frac{2\pi  i }{\b} \left(1-\frac{l(l+1)+\rh \partial_r f(\rh)}{2 K^2r_\text{h}^2} + \mathcal{O}((Kr_\text{h})^{-4})\right) \,,
\end{equation}
where $r_\text{h}$ is the radius of the horizon which replaces the de Sitter horizon in the cosmic patch analysis. This result is valid both for cosmic, $r_\text{h}>\tfrac{\ell}{\sqrt{3}}$, and black hole patches, $0<r_\text{h}<\tfrac{\ell}{\sqrt{3}}$, of Schwarzschild-de Sitter solutions (see \cite{Anninos:2024wpy} for details of the definition of cosmic and black hole patches and their inverse conformal temperature $\b$). Note that when setting $r_\text{h} = \ell$, we recover \eqref{eqn: +i dS}. In the flat space limit, we further obtain
\begin{equation}
    \frac{\omega^{(\text{BH})} \r}{\sqrt{f(\r)}} = \frac{2\pi  i }{\b} \left(1-\frac{l(l+1)+1}{2 K^2r_\text{h}^2} + \mathcal{O}((K r_\text{h})^{-4})\right) \,.
\end{equation}
As expected, the leading large-$l$ terms agree in all three formulas, but the order one term is different. This difference gets magnified for the spherically symmetric mode, where the cosmic patch horizon receives a positive correction, while the black hole patch will have a negative one.

It would be interesting to explore whether these subtle linearised differences between cosmic and black-hole horizons are  connected to the more significant signatures that emerge at the non-linear level, such as the Gao–Wald effect \cite{Gao:2000ga}.

\noindent\section*{Acknowledgments}

We are grateful to Michael Anderson, Mike Blake, Shira Chapman, Grigorios Fournodavlos, Christoph Kehle, Maria Knysh, Albert Law, Rob Myers, David Ramirez, Edgar Shaghoulian, Eva Silverstein, Toby Wiseman, and Themistocles Zikopoulos for useful discussions. We would also like to thank all the participants of the workshop \href{https://scgp.stonybrook.edu/archives/45387}{Timelike Boundaries in Classical and Quantum Gravity} for stimulating discussions and the Simons Center for Geometry and Physics at Stony Brook University for its hospitality. D.A. and A.S. are funded by the Royal Society under the grant ``Concrete Calculables in Quantum de Sitter". The research of D.A. is also supported in part by KU Leuven grant C16/25/010. The work of D.A.G. is funded by UKRI Stephen Hawking Fellowship EP/W005530/1 ``Quantum Emergence of an Expanding Universe". S.G. is funded by a Royal Society Newton International Fellowship NIF/R1/241888. C.M. is funded
by STFC under grant number ST/X508470/1. D.A., D.A.G., and A.S. are further supported by STFC consolidated grant ST/X000753/1.

\appendix

\section{WKB approximation at large angular momentum} \label{app: WKB}
In this appendix we review the WKB approximation applied specifically to the master field dynamics in the large angular momentum limit. For detailed reviews of the subject, we refer to \cite{Berry:1972na,Knoll:1975zs}.
Recall that after a judicious change of variables, the master field equation of motion can be recast as an effective 1D Schr\"{o}dinger problem \eqref{eqn: Meqn dS large l},
\begin{equation}\label{eqn: Meqn dS large l copy}
    \left(-\frac{1}{\mathfrak{L}^2}\partial_{r^*}^2 + V_{\text{eff}}(r^*)\right)\phi (r^*) = \frac{\omega^2}{\mathfrak{L}^2}\phi(r^*) \, , \qquad V_{\text{eff}}(r^*) = \frac{1}{\ell^2 \sinh^2\frac{r^*}{\ell}} \, .
\end{equation} 
Here we will only be interested in the bulk solution (without specifying boundary conditions), and therefore we will drop the superscript specifying whether the master field refers to the scalar or vector sector perturbations.  As in the main text, the turning point is defined as
\begin{equation}\label{eqn: turning point copy}
    \frac{\omega^2}{\mathfrak{L}^2}=V_\text{eff}(r^*_\text{t}) \, .
\end{equation}
In the following, we solve \eqref{eqn: Meqn dS large l copy} in the large-$\mathfrak{L}$ limit while keeping $\tfrac{\omega^2}{\mathfrak{L}^2}$ finite. 

\textbf{Solutions far from the turning point.} We first consider the following solution ansatz
\begin{equation}\label{eqn: ansatz WKB}
    \phi(r^*)= c\exp\left( i \mathfrak{L} \int^{r^*} dx \,\sigma(x)\right) \, , \qquad \sigma(r^*) = \sum_{n=0}^\infty \mathfrak{L}^{-n}\sigma_n(r^*) \, ,
\end{equation}
where $c$ is an arbitrary constant, which can be used to define the lower bound of integration. Plugging this ansatz into \eqref{eqn: Meqn dS large l copy}, we obtain a set of equations at each order of $\mathfrak{L}$. Upon solving these, we obtain coefficients $\sigma_n$ in terms of $\tfrac{\omega^2}{\mathfrak{L}^2}$, $V_\text{eff} (r^*)$, and its derivatives. To leading order, we obtain
\begin{equation}
    \sigma_0^2 (r^*) + V_\text{eff} (r^*) = \frac{\omega^2}{\mathfrak{L}^2}\, ,
\end{equation}
having two branches of solutions, 
\begin{equation}\label{eqn: sigma0}
    \sigma_0^{(\pm)}(r^*) = \pm \sqrt{V_\text{eff}(r^*_\text{t})-V_\text{eff}(r^*)} \, ,
\end{equation}
for turning point $r^{*}_{\text{t}}$.
The first sub-leading equation upon inserting the ansatz \eqref{eqn: ansatz WKB} into \eqref{eqn: Meqn dS large l copy} gives
\begin{equation}
    i \partial_{r^*}\sigma_0 (r^*) - 2 \sigma_0 (r^*) \sigma_1 (r^*) =0 \, .
\end{equation}
Using \eqref{eqn: sigma0}, this results in
\begin{equation}\label{eqn: sigma1}
    \sigma_1^{(\pm)}(r^*) = i\partial_{r^*}\log \left|V_\text{eff}(r^*_\text{t})-V_\text{eff}(r^*)\right|^{1/4} \, .
\end{equation}
Inserting \eqref{eqn: sigma0} and \eqref{eqn: sigma1} into \eqref{eqn: ansatz WKB}, we reproduce the solution \eqref{eqn: WKB sol 1} in the main text,
\begin{equation}\label{eqn: WKB sol 1 copy}
    \phi(r^*) = c_1 \frac{\exp\left(i \mathfrak{L} \int^{r^*} dx\, \sigma_0^{+}(x)\right)}{\left|V_\text{eff}(r^*)-V_\text{eff}(r^*_\text{t})\right|^{1/4}}+c_2 \frac{\exp\left(i \mathfrak{L} \int^{r^*} dx\, \sigma_0^{-}(x)\right)}{\left|V_\text{eff}(r^*)-V_\text{eff}(r^*_\text{t})\right|^{1/4}} + \mathcal{O}(\mathfrak{L}^{-1}) \, .
\end{equation}

This solution can be systematically corrected by including contributions with $\sigma_n$ for $n\geq2$. From \eqref{eqn: WKB sol 1 copy}, we find that $\sigma_0^{(\pm)}$ in \eqref{eqn: sigma0} determines the leading phase of the solution. For real $\sigma^{(\pm)}$, it implies the solution oscillates rapidly with wavelength of the order $\mathfrak{L}^{-1}$. The constants $c_{1,2}$ can be selected such that the solution obeys particular behaviour.

For instance, the solution that is regular near the origin $r^*=0$ is given by
\begin{equation}\label{eqn: wkb pole sol}
    \phi(r^*) = c_1 \frac{\exp\left(\mathfrak{L} \int^{r^*} dx\, \sqrt{V_\text{eff}(x)-V_\text{eff}(r^*_\text{t})}\right)}{\left|V_\text{eff}(r^*)-V_\text{eff}(r^*_\text{t})\right|^{1/4}} +\mathcal{O}(\mathfrak{L}^{-1})\, .
\end{equation}
To see this, we note that near the origin, the potential can be approximated by $V_\text{eff}(r^*)=\tfrac{1}{r^{*2}}$, which dominates $V_\text{eff}(r^*_\text{t})$. Hence, as $r^*\to0$,
\begin{equation}
    \phi(r^*) \to c_1 \sqrt{r^*}\text{exp}\left(\mathfrak{L} \log r^*\right) = c_1 r^{*(\mathfrak{L}+\frac{1}{2})} \, ,
\end{equation}
which is regular at the origin.
Similarly, the solution that is purely-outgoing as $r^*\to +\infty$ is
\begin{equation}
    \phi(r^*) = c_1 \frac{\exp\left(i\mathfrak{L} \int^{r^*} dx\, \sqrt{V_\text{eff}(r^*_\text{t})-V_\text{eff}(x)}\right)}{\left|V_\text{eff}(r^*)-V_\text{eff}(r^*_\text{t})\right|^{1/4}} +\mathcal{O}(\mathfrak{L}^{-1})\, .
\end{equation}
This is because, as $r^*\to+\infty$, the potential  $V_\text{eff}(r^*)$ is sub-dominant compared to $V_\text{eff}(r^*_\text{t})$. Hence, by re-expressing $r^*_\text{t}$ in terms of the frequency \eqref{eqn: turning point copy}, we find that as $r^*\to+\infty$,
\begin{equation}
    \phi(r^*) \to c_1 \left|\frac{\mathfrak{L}}{\omega}\right|^{1/2}\text{exp}\left(+i \omega r^*\right) \, .
\end{equation}

For our large-$\L$ analysis it is important to establish the range of validity of this solution. Using \eqref{eqn: sigma1}, it is clear that for this perturbative analysis to be valid we require
\begin{equation}
    |\sigma_0(r^*)|^2 \gg |\mathfrak{L}^{-1}\partial_{r^*}\sigma_0(r^*)| \, .
\end{equation}
Using \eqref{eqn: sigma0} and expanding the potential near the turning point $r^*_\text{t}$, this condition becomes
\begin{equation}
    \left|\partial_{r^*}V_\text{eff}(r^*_\text{t})(r^*-r^*_\text{t})\right| \gg \left|\mathfrak{L}^{-1}\sqrt{\frac{\partial_{r^*}V_\text{eff}(r^*_\text{t})}{r^*-r^*_\text{t}}}\right| \, .
\end{equation}
Now, provided $\partial_{r^*}V_\text{eff}(r^*_\text{t})$ does not scale with $\mathfrak{L}$ (which is the case for the dS perturbations we consider), the validity condition imposes a lower bound on the radial coordinate,
\begin{equation}
    |r^*-r^*_\text{t}|\gg \mathfrak{L}^{-2/3} \, .
\end{equation}
Thus, in the near turning point region $|r^*-r^*_\text{t}|\sim \mathfrak{L}^{-2/3}$, the perturbative solution \eqref{eqn: ansatz WKB} breaks down and must be replaced with a different ansatz.

\textbf{Solutions near the turning point.} Let us consider an ansatz valid in the region near the turning point, i.e.,  $|r^*-r^*_\text{t}|\sim \mathfrak{L}^{-2/3}$. To do so, it is convenient to define a radial coordinate which is finite over this region. Let $\tilde\xi$ be the dimensionless near turning point coordinate defined as
\begin{equation}\label{eqn: xi coord def}
    \tilde\xi \equiv \left(-
    \mathfrak{L}^2 \partial_{r^*}V_\text{eff}(r^*_{\text{t}})\right)^{1/3}\left(r^*_{\text{t}}-r^*\right) \, ,
\end{equation}
where we choose a real cubic root. Note this is similar but not equal to the $\xi$ coordinate defined in the main text. For the dS background, the potential is monotonically decreasing, $\partial_{r^*}V_\text{eff}(\r^*) < 0$, and thereby $\tilde \xi $ increases (decreases) toward the origin (horizon). Consider the following ansatz for the master scalar field,
\begin{equation}\label{eqn: WKB ansatz 2}
    \phi(r) = \sum_{n=0}^\infty \mathfrak{L}^{-n}\phi_n(\tilde \xi) \, .
\end{equation}
Inserting \eqref{eqn: xi coord def} and \eqref{eqn: WKB ansatz 2} into the master field equation of motion \eqref{eqn: Meqn dS large l copy} and taking $\mathfrak{L}\to\infty$ while keeping $\tfrac{\omega}{\mathfrak{L}}$ finite, we can systematically obtain differential equations for the coefficients $\phi_n(\xi)$. The leading equation becomes,
\begin{equation}
    \left(\partial_{\tilde \xi}^2 - \tilde\xi\right) \phi_0 (\tilde\xi) = 0 \, .
\end{equation}
Solutions to this equation are superpositions of Airy functions $\text{Ai} (\tilde\xi)$ and $\text{Bi}(\tilde\xi)$, with general solution
\begin{equation}\label{eqn: WKB sol near copy}
    \phi_0(\tilde\xi)= c_3 \text{Ai}(\tilde\xi) + c_4 \text{Bi}(\tilde\xi) \, ,
\end{equation}
where $c_3$ and $c_4$ are constants of integration. Re-expressing $\tilde\xi$ in terms of $r^*$ using \eqref{eqn: xi coord def}, the solution \eqref{eqn: WKB sol near copy} gives the leading term of \eqref{eqn: WKB sol 3}.

\textbf{Matching condition.} To obtain the global solution, it is important that the far and near turning point solutions, \eqref{eqn: WKB sol 1 copy} and \eqref{eqn: WKB sol near copy}, are smoothly connected. We thus match the constants of integration of \eqref{eqn: WKB sol 1 copy} in the limit $|r^*-r^*_\text{t}|\to 0$ and \eqref{eqn: WKB sol near copy} in the limit $|\tilde \xi |\to \infty$. For instance, the solution which is regular at the origin, \eqref{eqn: wkb pole sol}, is well-approximated in the far region by
\begin{equation}\label{eqn: large L pole far}
    \phi(r^*) = \begin{cases}
        c_1\frac{\exp\left(\mathfrak{L} \int^{r^*}_{r^*_\text{t}} dx\, \sqrt{V_\text{eff}(x)-V_\text{eff}(r^*_\text{t})}\right)}{\left|V_\text{eff}(r^*)-V_\text{eff}(r^*_\text{t})\right|^{1/4}} \, , & 0<r^*<r^*_\text{t} \, , \\
        c_1\frac{\exp\left(i\mathfrak{L} \int^{r^*}_{r^*_\text{t}} dx\, \sqrt{V_\text{eff}(r^*_\text{t})-V_\text{eff}(x)}-i\frac{\pi}{4}\right)}{\left|V_\text{eff}(r^*)-V_\text{eff}(r^*_\text{t})\right|^{1/4}}+c_1\frac{\exp\left(-i\mathfrak{L} \int^{r^*}_{r^*_\text{t}} dx\, \sqrt{V_\text{eff}(r^*_\text{t})-V_\text{eff}(x)}+i\frac{\pi}{4}\right)}{\left|V_\text{eff}(r^*)-V_\text{eff}(r^*_\text{t})\right|^{1/4}}\, , & r^*_\text{t}<r^* \, ,
    \end{cases}
\end{equation}
while in the near turning point region, it is approximated by
\begin{equation}\label{eqn: large L pole near}
    \phi(r^*)=c_1 2 \sqrt{\pi}\left(\frac{\mathfrak{L}}{-\partial_{r^*}V_\text{eff}(r^*_\text{t})}\right)^{1/6}\text{Ai}\left(\left(-\mathfrak{L}^2\partial_{r^*}V_{\text{eff}}(r^*_\text{t})\right)^{1/3}(r^*_\text{t}-r^*)\right) \, .
\end{equation}
From this, it can be seen that upon crossing the turning point, the solution becomes rapidly oscillating with phase difference between the incoming and outgoing solution as $\tfrac{\pi}{2}$.

A similar analysis can be done for the purely-outgoing solution. This results in the approximation of $\phi(r^*)$ as
\begin{equation}\label{eqn: large L cosmic far}
    \phi(r^*) = \begin{cases}
        c_1\frac{\exp\left(-\mathfrak{L} \int^{r^*}_{r^*_\text{t}} dx\, \sqrt{V_\text{eff}(x)-V_\text{eff}(r^*_\text{t})}-i\frac{\pi}{4}\right)}{\left|V_\text{eff}(r^*)-V_\text{eff}(r^*_\text{t})\right|^{1/4}} \, , & 0<r^*<r^*_\text{t} \, , \\
        c_1\frac{\exp\left(i\mathfrak{L} \int^{r^*}_{r^*_\text{t}} dx\, \sqrt{V_\text{eff}(r^*_\text{t})-V_\text{eff}(x)}\right)}{\left|V_\text{eff}(r^*)-V_\text{eff}(r^*_\text{t})\right|^{1/4}}\, , & r^*_\text{t}<r^* \, ,
    \end{cases}
\end{equation}
with near turning point behaviour
\begin{equation}\label{eqn: large L cosmic near}
    \phi(r^*)=c_1 2 \sqrt{\pi}e^{-\frac{\pi i}{12}}\left(\frac{\mathfrak{L}}{-\partial_{r^*}V_\text{eff}(r^*_\text{t})}\right)^{1/6}\text{Ai}\left(e^{\frac{2\pi i}{3}}\left(-\mathfrak{L}^2\partial_{r^*}V_{\text{eff}}(r^*_\text{t})\right)^{1/3}(r^*_\text{t}-r^*)\right) \, .
\end{equation}
As the solution crosses the turning point it exponentially grows toward the origin.

\subsection{Master field when the boundary is close to the turning point} \label{sec: derive Ai(xi-nu)}

Here we provide a derivation of \eqref{eqn: WKB sol 3} and its first correction, which describes the large $\mathfrak{L}$ solution to \eqref{eqn: Meqn dS large l} in the region near both the turning point and the boundary. 

We recall that the near-boundary radial coordinate is defined as \eqref{eqn: near bdry xi},
\begin{equation}\label{eqn: near bdry xi copy}
    \xi \equiv \left(-
    \mathfrak{L}^2 \partial_{r^*}V_\text{eff}(\r^*)\right)^{1/3}\left(\r^*-r^*\right)\,,
\end{equation}
where the cubic root is chosen such that $\xi \in \mathbb{R}$ and defined so $\xi=0$ at the boundary. We consider the expansion for the frequencies in \eqref{eqn: ansatz near Rt},
\begin{equation}\label{eqn: ansatz near Rt copy}
    \omega = \mathfrak{L}\sqrt{V_\text{eff}(\r^*)}+\mathfrak{L} \sum_{n=1}^\infty \omega_n \mathfrak{L}^{-2n/3}\, ,
\end{equation}
where the $\omega_n$ are independent of $r^*$ and $\mathfrak{L}$. Plugging in \eqref{eqn: near bdry xi copy} and \eqref{eqn: ansatz near Rt copy} into \eqref{eqn: Meqn dS large l} and series expanding $V_\text{eff}(r^*)$ near $\r^*$, we obtain a differential equation for $\phi(\xi)$ organized in powers of $\mathfrak{L}^{-2/3}$,
\begin{equation}
    \left(-\left(\frac{-\partial_{r^*}V_\text{eff}(\r^*)}{\mathfrak{L}}\right)^{2/3}\partial_\xi^2 + \sum_{n=0}^\infty \frac{(-\xi)^n}{(-\mathfrak{L}^2\partial_{r^*}V_\text{eff}(\r^*))^{n/3}}\frac{\partial_{r^*}^nV_\text{eff}(\r^*)}{n!}\right)\phi(\xi)= \left(\sqrt{V_\text{eff}(\r^*)} + \sum_{n=1}^\infty \omega_n \mathfrak{L}^{-2n/3}\right)^2 \phi(\xi)\,.
\end{equation}
To perturbatively solve this equation, we consider the solution ansatz
\begin{equation}
    \phi(\xi) = \sum_{n=0}^\infty \phi_n(\xi) \mathfrak{L}^{-2n/3} \, .
\end{equation}
As such, the leading equation is given by
\begin{equation}
    \left(-\partial_\xi^2 + \xi-\nu \right) \phi_0(\xi) = 0 \, , \qquad \nu \equiv \frac{2 \sqrt{V_\text{eff}(\r^*)} }{\left(-\partial_{r^*}V_\text{eff}(\r^*)\right)^{2/3}}\omega_1 \,,
\end{equation}
where $\nu$ is defined in \eqref{eqn: WKB sol 3}. The solution to this equation is given by a linear combination of the Airy functions,
\begin{equation}\label{eqn: appen dum sol1}
    \phi_0(\xi)  = c_1\text{Ai}(\xi-\nu) + c_2 \text{Bi}(\xi-\nu) \, ,
\end{equation}
reproducing the leading term of \eqref{eqn: WKB sol 3}.

The first sub-leading equation reads
\begin{equation}
    \left(-\partial_\xi^2 + \xi-\nu\right)\phi_1(\xi) =\left(\mathfrak{a}_1+\mathfrak{a}_2\xi^2\right)\phi_0(\xi) \, , \qquad \begin{cases}
        \mathfrak{a}_1 =\frac{(-\partial_{r^*} V_\text{eff}(\r^*))^{2/3}\nu^2}{4 V_\text{eff}(\r^*)}+\frac{2\sqrt{V_\text{eff}(\r^*)}\omega_2}{(-\partial_{r^*} V_\text{eff}(\r^*))^{2/3}} \, , \\
        \mathfrak{a}_2=-\frac{\partial_{r^*}^2 V_\text{eff}(\r^*)}{2 (-\partial_{r^*} V_\text{eff}(\r^*))^{4/3}} \, .
    \end{cases}
\end{equation}
Using \eqref{eqn: appen dum sol1}, the general solution $\phi_1(\xi)$ is given by
\begin{equation}\label{eqn: appen dum sol2}
    \phi_1(\xi) = \frac{\mathfrak{a}_1}{5}\xi \phi_0(\xi)-\left(\mathfrak{a}_1+\mathfrak{a}_2\xi^2+\frac{4\mathfrak{a}_2}{15}\left(\nu-\xi\right)\left(2\nu+3\xi\right)\right)\partial_\xi \phi_0(\xi)+c_3\text{Ai}(\xi-\nu) + c_4 \text{Bi}(\xi-\nu)\, .
\end{equation}
The constants of integration $c_3$ and $c_4$ can be absorbed into the definitions of $c_1$ and $c_2$. The inhomogeneous part of $\phi_1(\xi)$ is given by polynomial functions multiplying $\phi_0(\xi)$ and $\partial_\xi \phi_0(\xi)$. This implies that $\phi_1(\xi)$ exhibits the similar asymptotic behaviour as $\phi_0(\xi)$.

\textbf{Pole patch solutions.} To obtain the pole patch solution of \eqref{eqn: appen dum sol1} and \eqref{eqn: appen dum sol2}, we require that these solutions smoothly connect to \eqref{eqn: large L pole far} as $|\xi| \to \infty$. In particular, this leads to the condition that $\phi_0(\xi)$ decays exponentially as $\xi \to + \infty$. The solution $\phi_0(\xi)$ which satisfies this condition is
\begin{equation}
    \phi_0(\xi) = c_1 \text{Ai}(\xi-\nu) \, ,
\end{equation}
for some arbitrary constant $c_1$.

\textbf{Cosmic patch solutions.} Similarly, the cosmic patch solution of \eqref{eqn: appen dum sol1} and \eqref{eqn: appen dum sol2} can be obtained by requiring that the solution smoothly connect to \eqref{eqn: large L cosmic far} as $|\xi| \to \infty$. This means, in particular, that $\phi_0(\xi)$ is purely-outgoing as $\xi \to - \infty$. The solution $\phi_0(\xi)$ which satisfies this condition is given by
\begin{equation}
    \phi_0(\xi) = c_1 \text{Ai}(e^{\frac{2\pi i}{3}}(\xi-\nu)) \, ,
\end{equation}
where $c_1$ is some other arbitrary constant.

\section{Large angular momentum modes in generic backgrounds} \label{app: large l generic}

In this appendix, we show that the large-$\L$ behaviour found in section \ref{sec: large l} can be easily generalised to perturbations about other spherically symmetric backgrounds. To this end, consider the following Einstein backgrounds,
\begin{equation}
    ds^2 = - f(r) dt^2 + \frac{dr^2}{f(r)}+r^2 d\Omega_2^2 \, , 
    \label{eq: general background metric}
\end{equation}
which we assume it is a solution to Einstein gravity with positive, negative or vanishing cosmological constant. Here, for simplicity, we assume backgrounds with a non-degenerate horizon, such that there exists a $\rh$ with $f(\rh) = 0$ but $\partial_r f(\rh) \neq 0$. It is interesting to generalise the following arguments to (near-)extremal horizons as well \cite{Banihashemi:2025qqi,Galante:2025tnt}. 

For this subsection, we focus on the horizon patch, that is, the bulk region of interest lies between a boundary at  $r=\r$ and the horizon. For the background (\ref{eq: general background metric}), the extrinsic curvature at $r=\r$ can be written as 
\begin{equation}
\bar{K}_{mn} dx^m dx^n|_{\Gamma} =\sigma \mathfrak{r}\sqrt{f(\mathfrak{r})}\bigg( -\frac{\partial_r f(\mathfrak{r})}{2\mathfrak{r}}dt^2 +d\Omega_2^2\bigg) \Rightarrow K =\sigma \frac{4 f(\mathfrak{r}) + \mathfrak{r} \partial_{r}f(\mathfrak{r})}{2 \mathfrak{r} \sqrt{f(\mathfrak{r})}}   \,.    
\end{equation}
Consequently, the conformal stress-tensor  for the background geometry obeys
\begin{equation}
    64 \pi^2 G_N^2 \bar{T}_{mn} \bar{T}^{mn} = \frac{\left(\mathfrak{r} \partial_{r}f(\mathfrak{r})-2 f(\mathfrak{r})\right)^2}{6 \mathfrak{r}^2 f(\mathfrak{r})} \,.
    \label{eq: tt general bh}
\end{equation}

Note that these generalised geometries also have a stretched horizon limit, in which $\r \to \rh$. In that case $|K\r| \to \infty$, as 
 \begin{align}
      K^2\r^2 = \frac{\r^2 \partial_{r}f(\rh)}{4 (\r-\rh)} +\mathcal{O}(1) \,.
 \end{align}

We are interested in perturbations around backgrounds (\ref{eq: general background metric}). Though less symmetric than the dS$_4$ static patch, it is still possible to perform a Kodama-Ishibashi decomposition of the metric perturbations given the spherical symmetry of the compact directions. The technical difference in these more general cases is that the equations of motion for the master fields are not analytically solvable in general. 
Nevertheless, following an analogous decomposition to that in \eqref{eqn: spherical l>2 ansatz}, we find the master equation generalises to 
\begin{align} \label{generalmasterfieldeq}
     -\partial_t^2 \Phi^{(S/V)} +f(r)\big(f(r) \partial_r^2 \Phi^{(S/V)} + \partial_{r}f(r) \partial_r  \Phi^{(S/V)}\big) &= V_\text{eff}(r)  \Phi^{(S/V)}.
\end{align}
The expression for $V_\text{eff}(r)$ depends on the particular $f(r)$ considered (such as $M,\Lambda$, etc.), and is given in Eqs. (3.2) -- (3.7) in \cite{Kodama:2003jz}. We note that, in the large angular momentum limit, $V_\text{eff}(r)$ simplifies to 
\begin{equation}
    V_\text{eff}(r) = \frac{\mathfrak{L}^2f(r)}{r^2}+ \mathcal{O}(\mathfrak{L}^0) \, .
\end{equation}
Here we focus on the case of four-dimensional spacetimes with spherical symmetry but these arguments can likely be generalised to other dimensions and/or other background topologies.\footnote{For a $(d+1)$-dimensional spherically symmetric background, the trace of the extrinsic curvature at $r=\r$ is
\begin{equation}
      {K}=\sigma\frac{2(d-1)f(\mathfrak{r})+\mathfrak{r}\partial_{r}f(\mathfrak{r})}{2\mathfrak{r}\sqrt{f(\mathfrak{r})}} \rightarrow 64 \pi^2 G_N^2 \bar{T}_{mn} \bar{T}^{mn} = \frac{(d-1)}{4d}\frac{\left(\mathfrak{r} \partial_{r}f(\mathfrak{r})-2 f(\mathfrak{r})\right)^2}{ \mathfrak{r}^2 f(\mathfrak{r})} \,.
\end{equation}
To leading order in the large$-\L$ limit, it is still true that $V_\text{eff}(r) = \frac{\mathfrak{L}^2f(r)}{r^2}$. The allowed frequencies in higher dimensions generalise to $ \frac{\omega \r}{\sqrt{f(\r)}} = \mathfrak{L} + \nu \left(\frac{d}{d-1}32 \pi^2 G_N^2\r^2 \bar{T}_{mn}\bar{T}^{mn}\right)^{1/3}\mathfrak{L}^{1/3} + \mathcal{O}(\mathfrak{L}^{-1/3}) \, ,$ where $\nu$ now is a different constant that will come from solving the boundary conditions in the different sectors.
}

One can solve \eqref{generalmasterfieldeq} in the large-$\L$ limit, using a WKB approximation analogous to the one shown in this section. The first step is to use a tortoise coordinate to write the equation in Schrodinger form. By doing this, depending on $f(r)$, it might happen that the new effective potential $V_{\text{eff}}$ is not monotonic. This happens, for instance, for the Schwarzschild background. For the rest of this section, we will assume that the boundary sits at an $\r$ close enough to the horizon such that in the region between $\r$ and the horizon, $V_{\text{eff}}$ is indeed monotonic. We leave the more general case for future work.

Here we just report the final result. In both vector and scalar perturbations, the allowed frequencies in the large-$\L$ limit take the form, 
\begin{equation}
    \frac{\omega \r}{\sqrt{f(\r)}} = \mathfrak{L} + \nu \left(48 \pi^2 G_N^2\r^2 \bar{T}_{mn}\bar{T}^{mn}\right)^{1/3}\mathfrak{L}^{1/3} + \mathcal{O}(\mathfrak{L}^{-1/3}) \, ,
\end{equation}
where, remarkably, $\nu$ is the same as in the previous subsection. Namely, the allowed $\nu$ are given by solutions of \eqref{eqn: Airy prime bdry cond 2} for vector and \eqref{eqn: complex nu 2} for scalar perturbations. Here $\bar{T}_{mn} \bar{T}^{mn}$ is given by \eqref{eq: tt general bh}.\footnote{We note that $\bar T_{mn} \bar T^{mn} \sim \partial_r V_\text{eff}(\r)$. By varying $\r$ such that it approaches the local extremum of $V_\text{eff}$, the coefficient of $\mathfrak{L}^{1/3}$ tends to zero. At this point, the potential ceases to be monotonic, and further analysis is required.} As before, $\Re(\omega) \to -\Re(\omega)$ is also an allowed frequency.

Interestingly, in the stretched horizon limit, where we take $|K\r|\to \infty$ after taking the large-$\L$ limit we obtain, 
\begin{equation}\label{eq: general large l frequency}
    \omega = \frac{|\partial_{r}f(\rh)|}{2}\frac{\mathfrak{L}}{|K\rh|} + \nu \frac{|\partial_{r}f(\rh)|}{2}\left(\frac{\mathfrak{L}}{2|K\rh|}\right)^{1/3} + \mathcal{O}(\mathfrak{L}^{-1/3})
\end{equation}
where we only kept the leading term in the large $|K\r|$ expansion. Note that in this double limit the perturbation parameters $\L$ and $|K \rh|$ are actually organised in terms of $\L/|K \rh|$. When we take both $\L$ and $|K \rh|$ to be large, the limit analysed corresponds to taking the limit, while further requiring $\L/|K \rh| \gg 1$. The result \eqref{eq: general large l frequency} agrees with \eqref{eq: frequencies cosmic large l large K} when $r_{\text{h}} \to \ell$ and $f(r) = 1-\tfrac{r^2}{\ell^2}$. 
This analysis also agrees with the Rindler result, after suitable identifications, see section \ref{sec: rindler}.

\section{Conformal stress tensor with off-shell frequencies} \label{app: cancellation}

\subsection{Soft modes}

Here we compute the first non-trivial correction to the conformal stress-tensor (for scalar metric perturbations) for the soft modes in the pole patch in the stretched horizon limit. See section \ref{sssec:confstpmiPP}. 
In principle, this correction comes from the $\omega_1$ correction in the frequency expansion \eqref{eq: frequency expansion}. To this end, it is convenient to split the linearised stress-tensor (\ref{eq: tmn}) into two contributions
\begin{equation}
    8\pi G_N \delta T_{mn} = \delta T^{(1)}_{mn} + \delta T^{(2)}_{mn} \,,
\end{equation}
where
\begin{equation}
    \begin{cases}
       \delta T^{(1)}_{mn} = \delta \bomega \, \bar{T}_{mn} \,, \\
        \delta T^{(2)}_{mn} = \delta \left(-K_{mn} +\frac{K}{3} g_{mn} \right)\,.
    \end{cases}
\end{equation}
Using the Weyl factor \eqref{eq: weyl w=i}, we find the only non-trivial components of $\delta T_{mn}^{(1)}$ are given by\footnote{We emphasise this relation between the components of $\delta T_{mn}^{(1)}$ is only valid to leading order in the large-$K\ell$ expansion.}
\begin{equation}
    \ell \delta T_{tt}^{(1)} =\frac{2}{\ell} (K\ell)^{-2} \delta T_{\theta\theta}^{(1)} =\frac{2 (K\ell)^{-2} }{\ell\sin^2 \theta}\delta T_{\varphi\varphi}^{(1)} = - e^{\pm t/\ell} \left(\tfrac{1}{2} (l^2+l)-1\right) \frac{2^{l+1}  \Gamma \left(l+\frac{3}{2}\right) \mathbb{S}}{3 \sqrt{\pi } \Gamma (l+2)} + \mathcal{O} ((K \ell)^{-2}\log |K\ell|) \,,
\end{equation}
inheriting the traceless property from the background stress tensor. We observe that this contribution of the stress tensor does not depend explicitly on $\omega_1$.  
The second contribution, however, does explicitly depend on $\omega_1$ and contributes to the same leading order. A direct calculation gives,
\begin{equation}
  \delta T_{tt}^{(2)} = -e^{\pm t/\ell}  \left(\tfrac{3}{2} (l^2 + l) - 1+ 2\omega_1 \ell\right) \frac{2^{l+1}  \Gamma \left(l+\frac{3}{2}\right) \mathbb{S}}{3  \ell\sqrt{\pi } \Gamma (l+2)} + \mathcal{O} ((K \ell)^{-2}\log |K\ell| ) \,.
\end{equation}
The full correction to the conformal stress tensor is therefore
\begin{equation}
   8\pi G_N \delta T_{tt} = -e^{\pm t/\ell}  ( (l^2 + l -1)+ \omega_1 \ell) \frac{2^{l+2} \Gamma \left(l+\frac{3}{2}\right) \mathbb{S}}{3 \ell\sqrt{\pi } \Gamma (l+2)}  + \mathcal{O} ((K \ell)^{-2}\log |K\ell| ) \,.
\end{equation}
It exactly vanishes when $\omega_1\ell$ is given by the allowed frequency given in \eqref{eq: subleading frequencies pole patch}, making $\delta T_{tt} = \mathcal{O} \left((K \ell)^{-2} \log |K\ell|\right)$.

At next order, a similar cancellation occurs. That is, the contribution to $\delta T_{tt}^{(1)}$ will explicitly depend on $\omega_1$, but not on $\omega_2$. Instead, $\delta T_{tt}^{(2)}$ depends on both $\omega_1$ and $\omega_2$, in such a way that the sum $\delta T_{tt}^{(1)} + \delta T_{tt}^{(2)}$ vanishes to this order when $\omega_2$ takes its on-shell value. The first non-trivial contribution to $\delta T_{tt}$ then enters at order $(K\ell)^{-2}$, given by
\begin{equation}
    \delta T_{tt} = -e^{\pm t/\ell}  ( l+2) \frac{2^{l} \Gamma \left(l+\frac{3}{2}\right) \mathbb{S}}{\ell\sqrt{\pi } \Gamma (l-1)} (K\ell)^{-2} + \mathcal{O} ((K\ell)^{-4}\log |K\ell|)  \,,
\end{equation}
consistent with \eqref{deltaTcomponents} in the main text.
It would be interesting to have a physical understanding for the origin of this exact cancellation. 

The other diagonal components are non-vanishing to leading order (here we directly set $\omega_1 \ell$ to its on-shell value),
\begin{equation}
  8\pi G_N \delta  T_{\theta \theta} = - 8\pi G_N \frac{\delta T_{\varphi\varphi}}{\sin^2\theta} = e^{\pm t/\ell}\frac{ 2^l \Gamma \left(l+\frac{3}{2}\right)  \ell \left(l (l+1) \mathbb{S}+2 \cot{\theta} \partial_\theta \mathbb{S}\right)}{\sqrt{\pi } \Gamma (l+2)} (K\ell)^2 +\mathcal{O} (\log|K\ell|) \,,
  \label{eq: delta tmn theta theta}
\end{equation}
which ensures the correction to the conformal stress tensor is traceless to this order, and consistent with \eqref{deltaTcomponents}. As mentioned in the main text, to this order, not only is $\delta T_{mn}$ traceless with respect to $g_{mn}$, but $\delta T_{ij}$ is also traceless with respect to the two-dimensional metric $\tilde{g}_{ij}$.

\subsection{Large angular momentum modes}
We now turn to the large angular momentum modes and compute the components of the conformal stress-tensor for $\omega=\omega_0 l +\omega_1 l^{1/3}+\mathcal{O}(1)$. The background stress-tensor does not depend on $l$, so the leading non-vanishing components of $\delta T_{mn}^{(1)}$ are $\mathcal{O}(l^2)$ in the large-$l$ limit. By explicit computation, we note that they are subleading compared to $\delta T_{mn}^{(2)}$. In fact, we find that
\begin{align}
\begin{cases}
     8\pi G_N\delta T_{tt}
    &= \frac{\ell
 \left(1 - \frac{\mathfrak{r}^{\,2}}{\ell^{2}} - \mathfrak{r}^{\,2}\omega_{0}^{2}\right)
       \left(1 - \frac{\mathfrak{r}^{\,2}}{\ell^{2}} + 2\mathfrak{r}^{\,2}\omega_{0}^{2}\right)\Phi(\mathfrak{r})\mathbb{S}}
{6\,\mathfrak{r}^{\,2}\,\sqrt{1 - (\mathfrak{r}/\ell)^{2}}} l^4+ \mathcal{O}(l^{10/3}), \\
   8\pi G_N\delta T_{\theta\theta} &=\frac{
\ell\left( 1 - \frac{\mathfrak{r}^{2}}{\ell^{2}} - \mathfrak{r}^{2}\omega_{0}^{2} \right)
\left( 2 - \frac{2\mathfrak{r}^{2}}{\ell^{2}}\, \mathfrak{r}^{2}\omega_{0}^{2} \right)
\Phi(\mathfrak{r})\mathbb{S} }
{6\left( 1 - \frac{\mathfrak{r}^{2}}{\ell^{2}} \right)^{3/2}}l^4 +\mathcal{O}(l^{10/3}),\\
    8\pi G_N \delta T_{\varphi\varphi} &=\frac{\ell
\left( 1 - \frac{\mathfrak{r}^{2}}{\ell^{2}} - \mathfrak{r}^{2}\omega_{0}^{2} \right)^{2}
\sin^{2}\!\theta \Phi(\mathfrak{r})\mathbb{S}}
{6 \left( 1 - \frac{\mathfrak{r}^{2}}{\ell^{2}} \right)^{3/2}}l^4 +\mathcal{O}(l^{10/3}).
\end{cases}
\end{align}
Here $\Phi(\mathfrak{r})=\mathfrak{R}e^{-i\omega t}\text{Ai}(-\nu)$. We notice all the components vanish to leading order when $\omega_0 = \sqrt{f(\r)}/\r$, its on-shell value.

\section{Radial profiles of pole patch soft modes} 
\label{app: radial profile}

In this appendix we present a more detailed analysis of the radial profiles of the soft modes in the pole patch in the stretched horizon limit. To do so, we find it convenient to introduce the following dimensionless radial coordinate,
\begin{align}\label{rhocoordinate}
   \rho & =\frac{\ell-r}{\ell-\mathfrak{r}} \,,
\end{align}
such that $\rho=1$ at the boundary and diverges as $|K\ell|^2$ close to the origin. Note this is a different coordinate from the $\rho$-coordinate introduced in section \ref{sec: non-linear fluid}.

\begin{figure}[t!]
    \centering
    
    \begin{subfigure}[t]{0.48\textwidth}
        \centering
        \includegraphics[width=\textwidth]{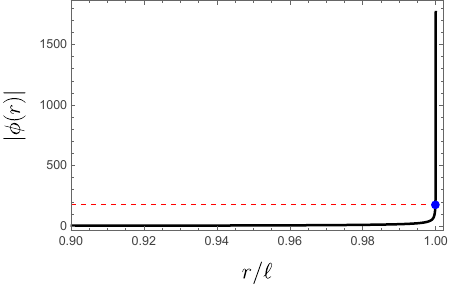}
        \caption{Radial profile.}
        \label{fig:omiradprofapp}
    \end{subfigure}
    \hfill
    \begin{subfigure}[t]{0.48\textwidth}
        \centering
        \begin{picture}(0,0)
        \put(110,79){\rotatebox{-12}{\small $\sim \log|K\ell|^{-1}$}} 
    \end{picture}%
        \includegraphics[width=1.03\textwidth]{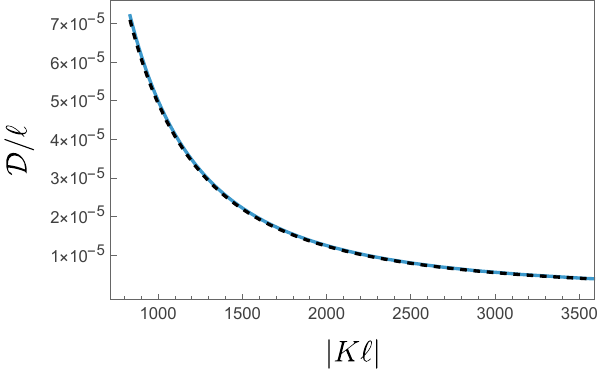}
        \caption{Width.}
        \label{fig:widthPPlgK}
    \end{subfigure}
    \caption{Master field radial profile and width for $\omega_{(\pm i)}$ modes. Here $l=2$. \emph{Left.} Localisation of master field radial profile in pole patch in the stretched horizon limit (solid black curve; here $|K\ell|\approx 707.106$, and $\omega\ell\approx 0.999992 i$).  The master field decays to (blue dot) $\alpha=\frac{1}{10}$ of its boundary value at $r_{\ast}/\ell=0.999899$. \emph{Right.} The width $\mathcal{D}/\ell=(\mathfrak{r}-r_*)/\ell$ as a function of $|K\ell|$ for $\alpha=\frac{1}{10}$. The light blue line represents the numerical values. The dashed black line represents the analytical approximation \eqref{pmimodeswidth} with the first two terms.}
    \label{fig:bulkprofsandDomrealvecn2xxxx}
\end{figure}

For the soft modes, the radial part of the master field is, as a function of coordinate $\rho$ 
\begin{align}
    \phi(\rho)&=\frac{2^{\,1 + l}\,\Gamma\!\left(\tfrac{3}{2} + l\right)}
     {\sqrt{\pi}\,\,\Gamma(2 + l)\sqrt{\rho}}|K\ell|
\;-\;
\frac{2^{\,l}\,
\big[-2 + (l^2+l)(2 + \rho)\big]\,
\Gamma\!\left(\tfrac{3}{2} + l\right)\,
}
{\,\sqrt{\pi}\,\sqrt{\rho}\,\Gamma(2 + l)}\frac{\log|K\ell|}{|K\ell|} +\mathcal{O}(|K\ell|^{-1})\;.
\label{eq:radprofomirho}\end{align}
As explained in the main text, a way to characterise the localisation of the radial profile of these modes in the large-$K\ell$ limit, which can be easily compared with numerical results, is to introduce a parameter $0<\alpha< 1$ and consider the position $\rho_*$ at which the master field decays to $|\phi(\rho_*)|=\alpha |\phi(1)|$ (see figure \ref{fig:omiradprofapp} for an illustration). We define the width $\mathcal{D}$ as the distance between this position and the boundary\footnote{The difference with respect to the numerical analysis from the main text is that here $\mathcal{D}$ is measured in the radial coordinate $r$ of the de Sitter static patch, while $D$ in the main text is the proper distance $s$ away from the boundary. As such, the powers of $|K\ell|$ appearing in the expansions of $\mathcal{D}$ and $D$ differ. For the analytic expansions in this appendix, we find it more convenient to use coordinates $r$ and $\rho$.}
\begin{align} \label{defwidth}
   \mathcal{D}\defeq \mathfrak{r}-r_*=(\ell-\mathfrak{r})(\rho_*-1)\;.
\end{align}
Solving for $\rho_*$ order by order in the large-$|K\ell|$ limit, we find
\begin{align}\label{pmimodeswidth}
    \mathcal{D}&=\frac{1-\alpha^2}{2\alpha^2}\ell|K\ell|^{-2}-\frac{l(l+1)(1-\alpha^2)}{2\alpha^4}\ell |K\ell|^{-4}\log|K\ell|+ \mathcal{O}(|K\ell|^{-4})\;,
\end{align}
which goes to 0 in the strict large-$|K\ell|$ limit. We compare this expansion with numerical results in figure \ref{fig:widthPPlgK}.

\begin{figure}[t!]
\centering
 \includegraphics[width=.5\textwidth]{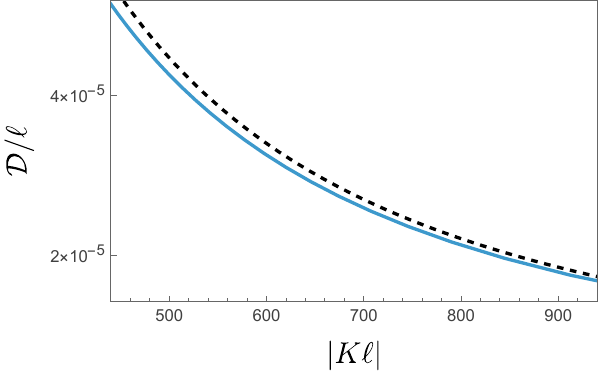}
\caption{The width $\mathcal{D}/\ell=(\mathfrak{r}-r_*)/\ell$ as a function of $|K\ell|$ for $l=2$, $\omega\ell=+i$ modes and $\alpha=|K\ell|^{-1/4}$. The light blue line represents the numerical values. The dashed black line represents the analytical approximation \eqref{width Kdepeta}. As $|K\ell|$ increases, the analytical approximation fits better the numerics.}
\label{fig:bulkradPPlgK}
\end{figure}

For the perturbative expansion to be valid we need to choose the parameter $\alpha$ to obey
\begin{align}
    \frac{1}{\alpha^2}\ll \frac{|K\ell|^2}{l(l+1)\log|K\ell|}\;.
\end{align}
We can choose $\alpha$ such that it is solely determined by $|K\ell|$. For example, $\alpha=|K\ell|^{-1/4}$ satisfies the condition above provided that $\frac{|K\ell|^{3/2}}{\log|K\ell|}\gg l(l+1)$. In this case, 
\begin{align}\label{width Kdepeta}
    \mathcal{D}&= \frac{\ell}{2|K\ell|^{3/2}}+\mathcal{O}(|K\ell|^{-2})\;,
\end{align}
such that the master field drops to $|K\ell|^{-1/4}$ of its maximal value within a region of width $\frac{\ell}{2|K\ell|^{3/2}}$ near the boundary, see figure \ref{fig:bulkradPPlgK}.

\section{Differential operator $\mathbb{H}$}\label{sec: H}

In this appendix we provide details on the differential operator $\mathbb{H}$ \eqref{eqn: fluid Gmn} arising from the Einstein tensor components $G^{(\mathfrak{n})}_{mn}=0$, solving for its zero modes. Explicitly, $\mathbb{H}$ acting on the metric perturbation $h_{mn}$ is given by
\begin{equation}
    \begin{cases}
        \mathbb{H}\,h_{\tau \tau} = \frac{1}{2}(1-\frac{\rho}{\ell}) \left(\ell\partial_\rho - 2 \left(1-\frac{\rho}{\ell}\right)\ell^2\partial_\rho^2\right)  [g]^{ij}h_{ij} \,, \\
        \mathbb{H}\,h_{\tau i}=  -\left(1-\frac{\rho}{\ell}\right)\ell^2\partial_\rho^2 h_{\tau i} \, ,\\
        \mathbb{H}\,h_{ij}= \left(\ell\partial_\rho -\left(1-\frac{\rho}{\ell}\right)\ell^2\partial_\rho^2\right)\left(h_{ij}- [g]_{ij}  [g]^{kl}h_{kl}\right) -  [g]_{ij} \ell^2\partial_\rho^2 h_{\tau\tau} \, ,
    \end{cases}
\end{equation}
where $[g]_{mn}$ is the conformal representative defined in \eqref{eqn: conf rep}.

\textbf{Zero modes of $\mathbb{H}$}. Now we consider the zero modes of $\mathbb{H}$,
\begin{equation}
    \mathbb{H} h_{mn} = 0\, .
\end{equation}
This amounts to solving \eqref{eqn: fluid Gmn} at first order when the source term is absent. Such $h_{mn}$ can be obtained analytically, 
\begin{equation}
    \begin{cases}
        h_{\tau\tau}=c_1\left(\left(1-\frac{\rho}{\ell}\right)^{3/2}-1\right)+c_2 +c_3 \frac{\rho}{\ell} \, , \\
        h_{\tau \theta}=h_{\theta \tau}= c_4+c_5 \frac{\rho}{\ell} \, , \\
        h_{\tau \phi}=h_{\phi \tau}= c_6+c_7 \frac{\rho}{\ell} \, , \\
        h_{\theta \theta} = \ell^2\left(6c_1\left(\sqrt{1-\frac{\rho}{\ell}}-1\right)+c_8  \right) + \left(c_{9}+c_{10}\log(1-\frac{\rho}{\ell})\right) \, , \\
        h_{\phi \phi} = \ell^2\sin^2\theta\left(6c_1\left(\sqrt{1-\frac{\rho}{\ell}}-1\right)+c_8  \right) - \sin^2\theta\left(c_{9}+c_{10}\log(1-\frac{\rho}{\ell})\right)\, ,\\
        h_{\theta \phi}=h_{\phi \theta} = \left(c_{11}+c_{12}\log(1-\frac{\rho}{\ell})\right) \, ,
    \end{cases}
\end{equation}
where $c_i=c_i(\tau,\theta,\phi)$ for $i=1,2,\dots,12$ are functions of integration. 

\section{Linearised perturbations about Rindler} \label{app:linpertRind}

Here we provide details describing the linearised metric perturbations about flat Rindler space. We apply the Kodama-Ishibashi formalism \cite{Kodama:2000fa}. The unperturbed Rindler background (\ref{eq:Rindmet}) has the form
\beq d\bar{s}^{2}=g_{ab}(y)dy^{a}dy^{b}+r^{2}(y)\sigma_{ij}(x)dx^{i}dx^{j}\;,\label{eq:genbackapp}\eeq
where $g_{ab}$ is the two-dimensional flat space metric in Rindler coordinates, i.e.,  $g_{ab}dy^{a}dy^{b}=-z^{2}z_{0}^{-2}dt^{2}+dz^{2}$, the warp factor is $r=1$, and $\sigma_{ij}dx^{i}dx^{j}=dx^{2}+dy^{2}$. In particular, we are interested in gauge-invariant scalar and vector sector metric perturbations.

\subsection*{Scalar perturbations}

A scalar perturbation $h^{(S)}_{\mu\nu}$ to the generic metric (\ref{eq:genbackapp})  may be expanded in terms of the scalar harmonics with components
\beq h^{(S)}_{ab}=f_{ab}\mathbb{S}\;,\quad  h^{(S)}_{ai}=f_{a}\mathbb{S}_{i}\;,\quad h^{(S)}_{ij}=2(\gamma\sigma_{ij}\mathbb{S}+H_{T}\mathbb{S}_{ij})\;.
\label{eq:scalpertcomps}\eeq
Here $\mathbb{S}$ is the scalar harmonic function satisfying 
\beq 0=(\tilde{\nabla}^{2}+k^{2})\mathbb{S}=(\partial^{2}_{x}+\partial^{2}_{y}+k^{2})\mathbb{S}\;,\eeq
with $k\in\mathbb{R}^{+}$ and $\tilde{\nabla}^{2}$ being the D'Alembertian on the space with metric $\sigma_{ij}$. Meanwhile, $\mathbb{S}_{i}=-\frac{1}{k}\tilde{\nabla}_{i}\mathbb{S}$ and $\mathbb{S}_{ij}=\frac{1}{k^{2}}\tilde{\nabla}_{i}\tilde{\nabla}_{j}\mathbb{S}+\frac{1}{n}\sigma_{ij}\mathbb{S}$ are scalar-type harmonic vectors and tensors, respectively, (see Eqs. (39) -- (44) of \cite{Kodama:2000fa}), where $\tilde{\nabla}_{i}$ refers to the covariant derivative compatible with metric $\sigma_{ij}$. Presently, $\tilde{\nabla}_{i}=\partial_{i}$, and the dimension $n$ of the transverse space with metric $\sigma_{ij}$ has $n=2$. We will demand that $h_{ti}=0$ and $h_{ij}\propto \sigma_{ij}$, such that $f_{t}=H_{T}=0$. Consequently, 
the metric components (\ref{eq:scalpertcomps}) become
\beq h^{(S)}_{ab}=f_{ab}\mathbb{S}\;,\quad  h^{(S)}_{ai}=-\frac{1}{k}f_{a}\tilde{\nabla}_{i}\mathbb{S}\;,\quad h^{(S)}_{ij}=2\gamma\sigma_{ij}\mathbb{S}\;.
\label{eq:scalpertcomps2app}\eeq
Finally, $f_{ab}$ and $f_{a}$ are related to gauge invariant quantities 
\beq 
\begin{split} 
&F=\gamma+\frac{1}{n}H_{T}+\frac{1}{r}(D^{a}r)X_{a}\;,\\
&F_{ab}=f_{ab}+D_{a}X_{b}+D_{b}X_{a}=f_{ab}+(\partial_{a}X_{b}+\partial_{b}X_{a}-2\Gamma^{c}_{\;ab}X_{c})\;,
\end{split}
\label{eq:gaugeinvarquants}\eeq
for $X_{a}\equiv \frac{r}{k}\left(f_{a}+\frac{r}{k}D_{a}H_{T}\right)$, and $D_{a}$ is the covariant derivative with respect to the metric $g_{ab}$.

 For linearised perturbations about a vacuum background, the linearised Einstein's equations yield (see Eqs. (64) and (66) of \cite{Kodama:2000fa})
\beq F^{a}_{\;a}=-2(n-2)F\;,\quad D_{b}(r^{n-2}F^{b}_{\;a})=2D_{a}(r^{n-2}F)\;.\eeq
 The general solutions to these equations may be cast in terms of a single master field $\Phi^{(S)}$, leading to (for $n=2$ and $\lambda=0$)
\beq F=\frac{1}{4}\nabla^{2}\Phi^{(S)}\;,\quad F_{ab}=D_{a}D_{b}\Phi^{(S)}-\frac{1}{2}g_{ab}\nabla^{2} \Phi^{(S)}\;,\label{eq:gaugeinvquants2}\eeq
for $\nabla^{2}\equiv g^{ab}D_{a}D_{b}=-\frac{z_{0}^{2}}{z^{2}}\partial_{t}^{2}+\frac{1}{z}\partial_{z}+\partial_{z}^{2}$. Clearly we see $F^{a}_{a}=0$, and $F_{ab}$ has components
\beq 
\begin{split}
&F_{tt}=\frac{1}{2}\frac{z^{2}}{z_{0}^{2}}\left(\frac{z_{0}^{2}}{z^{2}}\partial_{t}^{2}-\frac{1}{z}\partial_{z}+\partial_{z}^{2}\right)\Phi^{(S)}\;,\\
&F_{tz}=\left(\partial_{t}\partial_{z}-\frac{1}{z}\partial_{t}\right)\Phi^{(S)}\;,\\
&F_{zz}=\frac{1}{2}\left(\frac{z_{0}^{2}}{z^{2}}\partial_{t}^{2}-\frac{1}{z}\partial_{z}+\partial^{2}_{z}\right)\Phi^{(S)}\;.
\end{split}
\eeq
 Using gauge invariant quantities (\ref{eq:gaugeinvarquants})
and implementing $f_{t}=H_{T}=0$ (such that $X_{t}=0$ and $F=\gamma$)
we also have
\beq
\begin{split}
&F_{tt}=f_{tt}-\frac{2z}{kz_{0}^{2}}f_{z}\;,\\
&F_{tz}=f_{tz}+\frac{1}{k}\partial_{t}f_{z}\;,\\
&F_{zz}=f_{zz}+\frac{2}{k}\partial_{z}f_{z}\;.
\end{split}
\eeq
Rearranging the first equation to solve for $f_{z}$ gives
\beq f_{z}=-\frac{kz_{0}^{2}}{2z}(F_{tt}-f_{tt})\;,\quad f_{tz}=F_{tz}+\frac{z_{0}^{2}}{2z}\partial_{t}(F_{tt}-f_{tt})\;,\quad f_{zz}=F_{zz}+z_{0}^{2}\partial_{z}\left[\frac{1}{z}(F_{tt}-f_{tt})\right]\;.\eeq
Further, we have the gauge condition $2\gamma+z_{0}^{2}z^{-2}f_{tt}=0$.
Altogether, the metric components become (\ref{eq:scalpertcomps2app})
\beq 
\begin{cases}
h_{tt}=-2\gamma \frac{z^{2}}{z_{0}^{2}}\mathbb{S}\;,\\
h_{tz}=\left[F_{tz}+\frac{z_{0}^{2}}{2z}\partial_{t}\left(F_{tt}+2\gamma\frac{z^{2}}{z_{0}^{2}}\right)\right]\mathbb{S}\;,\\
h_{zz}=\biggr\{F_{zz}+z_{0}^{2}\partial_{z}\left[\frac{1}{z}\left(F_{tt}+2\gamma\frac{z^{2}}{z_{0}^{2}}\right)\right]\biggr\}\mathbb{S}\;,\\
h_{ti}=0\;,\\
h_{zi}=\frac{z_{0}^{2}}{2z}\left(F_{tt}+2\gamma\frac{z^{2}}{z_{0}^{2}}\right)\partial_{i}\mathbb{S}\;,\\
h_{ij}=2\gamma\sigma_{ij}\mathbb{S}\;.
\end{cases}
\label{eq:compsintmedrind}\eeq

From Eqs. (156-157) of \cite{Kodama:2000fa}, we also have (for $n=2$, $\lambda=0$, $r=1$, and vanishing constant sectional curvature $K=0$)
\beq 0=D_{a}D_{b}\left(\nabla^{2}\Phi-k^{2}\Phi\right)\;. \eeq
For modes with $k\neq0$, we have that the master field for the scalar perturbation satisfies
$\nabla^{2}\Phi=k^{2}\Phi$. Combining this with (\ref{eq:gaugeinvquants2}), the gauge invariant quantities (\ref{eq:gaugeinvarquants}) obey (for $r=1$)
\beq
\begin{split}
&\gamma=F=\frac{1}{4}\nabla^{2}\Phi=\frac{k^{2}}{4}\Phi\;.
\end{split}
\eeq
We further use the master field equation to express the metric components (\ref{eq:scalpertcomps2app}) without higher derivatives with respect to $z$. Specifically, note
\beq
\begin{cases} 
F_{tt}=\frac{z^{2}}{z_{0}^{2}}\left(\frac{z_{0}^{2}}{z^{2}}\partial_{t}^{2}-\frac{1}{z}\partial_{z}+\frac{k^{2}}{2}\right)\Phi\;,\\
F_{tt}+2\gamma \frac{z^{2}}{z_{0}^{2}}=\left[\partial_{t}^{2}-\frac{z}{z_{0}^{2}}\partial_{z}+\frac{k^{2}z^{2}}{z_{0}^{2}}\right]\Phi\;,\\
F_{zz}=\left(\frac{z_{0}^{2}}{z^{2}}\partial_{t}^{2}-\frac{1}{z}\partial_{z}+\frac{k^{2}}{2}\right)\Phi\;.
\end{cases}
\eeq
With some further massaging the components (\ref{eq:compsintmedrind}) take the form presented in the main text.

\subsection*{Vector perturbations}

The vector sector of perturbations has metric components 
\beq h^{(V)}_{ab}=0\;,\quad h^{(V)}_{ai}=f_{a}\mathbb{V}_{i}\;,\quad h^{(V)}_{ij}=2H_{T}\mathbb{V}_{ij}\;,\eeq
where $f_{a}$ and $H_{T}$, are in principle different from those appearing in the scalar perturbations (\ref{eq:scalpertcomps}). Here we will likewise impose $H_{T}=0$. Further, here the vector harmonic $\mathbb{V}_{i}$ and vector-type harmonic tensor $\mathbb{V}_{ij}$ obey
\beq (\tilde{\nabla}^{2}+k^{2})\mathbb{V}_{i}=0\;,\quad \tilde{\nabla}_{i}\mathbb{V}^{i}=0\;,\quad \mathbb{V}_{ij}=-\frac{1}{2k}(\tilde{\nabla}_{i}\mathbb{V}_{j}+\tilde{\nabla}_{j}\mathbb{V}_{i})\;.\eeq

As shown in Eq. (139) of \cite{Kodama:2000fa}, for a vector perturbation on a vacuum bulk spacetime, the quantity $f_{a}$ may be cast in terms of a single master field $\Phi^{(V)}$,
\beq f^{a}=\epsilon^{ab}D_{b}\Phi^{(V)}\;.\eeq 
Thus, we conclude the only components of the vector sector metric perturbations are
\beq h_{ti}^{(V)}=z^{2}z_{0}^{-2}\partial_{z}\Phi^{(V)}\mathbb{V}_{i}\;,\quad h^{(V)}_{zi}=\partial_{t}\Phi^{(V)}\mathbb{V}_{i}\;,\eeq
where our convention is $\epsilon^{tz}=-1$.

\section{Rindler modes in the pole patch stretched horizon limit} \label{app:Rindmodes}

In this appendix, we present details of the linearised dynamics for Rindler pole patch in the stretched horizon limit. In the scalar sector, we find soft modes, gapless modes and an infinite set of real modes, coming from the solutions of \eqref{rindlerlimitdK} at leading order in the large-$|K\ell|$ expansion.

\subsection{Soft and growing modes} 
Solving $\delta K|_{\Gamma}=0$ order by order in the $\kappa\rightarrow 0$ limit, we find 
\beq \omega z_{0} =\pm i \mp i\kappa^{2}\pm i\kappa^{4}\log(\kappa)+...\label{eq:freqexpRind}\eeq
It is straightforward to verify this frequency expansion has the same form as the allowed frequencies in the dS pole patch (\ref{eq: frequency expansion}).

Regarding the master field for $\omega z_{0}=\pm i$, we can expand it about $kz=0$ to obtain
\beq \phi_{\text{pole}}^{(S/V)}|_{\omega z_{0}=\pm i}\approx \frac{1}{kz}+\frac{kz}{2}\log(kz)+...\;,\label{eq:asympexppoleRind}\eeq
where the ellipsis has, for current purposes, unimportant subleading terms. The asymptotic expansion (\ref{eq:asympexppoleRind}) matches the form of the dS$_{4}$ radial profile (cf. Eq. \ref{eq:radprofomirho}), which justifies our choice \eqref{eq:idkl}.

The linearised Weyl factor for the scalar perturbations (\ref{eq:weylRindB}) for small $\kappa$ goes as
 \beq
 \begin{split}
\delta\bomega|_{\Gamma}&\approx e^{\pm t/z_{0}}\mathbb{S}\left(\frac{k^{2}z_{0}^{2}}{4\kappa}+\frac{k^{2}z_{0}^{2}}{8}\kappa \log(\kappa)+\mathcal{O}(\kappa)\right)\;.
\end{split}
\eeq 
Via our identification \eqref{eq:idkl}, we see find the same scaling with extrinsic curvature as in de Sitter (\ref{eq: weyl w=i}). Further, upon normalising the Rindler profile, the coefficients of the leading terms in the dS and Rindler precisely match.

Next, it is straightforward to show that the stress-tensor components (\ref{eq:deltaTmnRindgen}) have the same scaling behaviour as the those in the dS pole patch (\ref{deltaTcomponents}). For example, 
\beq
\begin{split} 
8\pi G_{N}\delta T_{tt}&=\frac{e^{-it\omega}k^{2}\mathfrak{z}(1-k^{2}\mathfrak{z}^{2}+z_{0}^{2}\omega^{2})}{2}\mathbb{S}K_{iz_{0}\omega}(\kappa)\approx \mathcal{O}(\kappa^{2})\;,
\end{split}
\eeq
where to arrive to the last equality we substituted in the frequency expansion (\ref{eq:freqexpRind}) and performed a small $\kappa$ expansion for fixed $k$. Similarly, the spatial $ij$-components diverge as 
\beq \delta T_{ij}=\mathcal{O}(\kappa^{-2})\;.\eeq
Combined, we find the $\delta T_{tt}$ and $\delta T_{ij}$ components of the stress-tensor are consistent with the structure for de Sitter (\ref{eq: leading tmn w=i}). A difference here is that we find $\delta T_{ti}$ components diverge like $\mathcal{O}(\kappa^{-1})$.

Further, we find (\ref{eq:TdeltaT}) 
\beq
\begin{split}\label{TdTRindlerpole}
32\pi^{2}G_{N}^{2}\bar{T}^{mn}\delta T_{mn}=\frac{k^{2}(1-\kappa^{2}+z_{0}^{2}\omega^{2})}{\kappa^{2}}\delta\bomega\;,
\end{split}
\eeq
such that for $\omega z_{0}=\pm i$, this is $\mathcal{O}(\kappa^{-1})$. 

Let us now comment on the linearised Weyl equation \eqref{eq:lin212Rindv2}, which for $\delta\bomega(t,x,y)=\delta\bomega(t)\mathbb{S}(x,y)$, becomes
\beq \frac{k^{2}z_{0}^{2}}{\kappa^{2}}\left[-\delta\bomega''(t)+\frac{(1-\kappa^{2})}{z_{0}^{2}}\delta\bomega\right]=32\pi^{2}G^{2}_{N}\bar{T}^{mn}\delta T_{mn}\;.\label{eq:weyleomsimpRind}\eeq

 In the strict stretched horizon limit, there is no contribution from the $\bar{T}^{mn}\delta T_{mn}$ on the right-hand side and the equation of motion simplifies to $-z_{0}^{2}\delta\bomega''(t)+\delta\bomega(t)=0$. Solutions are of the form $\delta\bomega(t)=N_{\bomega}e^{\pm t/z_{0}}$, recovering the $\omega z_{0}=\pm i$ modes. As for the de Sitter pole patch, the first correction in the frequency expansion (\ref{eq:freqexpRind}) is a consequence of a contribution from $\bar{T}^{mn}\delta T_{mn}$, which can be easily obtained from \eqref{TdTRindlerpole}.

\subsection{Gapless modes}
For the gapless $\omega z_0=0$ modes, the expansion of \eqref{eq:CBCpoleRind} yields
\beq 
\omega z_{0}=\pm \kappa\sqrt{\frac{1}{2}\left(1+\frac{1}{\log(\kappa/2)+\gamma}\right)}+\mathcal{O}(\kappa^2)\;.\eeq
where $\gamma$ is the Euler-Mascheroni constant and the subleading order represents $\kappa^2$ times an infinite expansion in $\log\kappa$. Further expanding the leading order term, we find a perfect match to the de Sitter result \eqref{eq:omega0expan}
\begin{align}
    \omega z_0&=\pm \frac{\kappa}{\sqrt{2}} +\mathcal{O}(\kappa\log^{-1}\kappa)
\end{align}
Since the conformal stress-tensor for these modes satisfies
\beq
\begin{split}\label{TdTRindlerpole_z0}
32\pi^{2}G_{N}^{2}\bar{T}^{mn}\delta T_{mn}=\bigg(\frac{k^{2}}{\kappa^{2}}+\mathcal{O}(\kappa^2)\bigg)\delta\bomega\;,
\end{split}
\eeq
it follows that we need to consider this contribution in the Weyl mode equation \eqref{eq:weyleomsimpRind} in order to obtain the correct leading order time-independent Weyl factor.

\subsection{Normal scalar modes}

The tower of real modes comes from the vanishing of the Rindler master field at the boundary,
\begin{align}
 \left(\frac{\kappa}{2}\right)^{-2i\omega z_{0}}=-\frac{\Gamma(-i\omega z_{0})}{\Gamma(i\omega z_{0})}.
\end{align}
Equivalently, we can put this equation in a similar form to \eqref{pole tower eq},
\beq
\omega_{n}z_{0}=\frac{i}{2\log(\kappa/2)}\left[\log\left(-\frac{\Gamma(-i\omega_{n}z_{0})}{\Gamma(i\omega_{n}z_{0})}\right)+2\pi i n\right]\;,\; n\in\mathbb{Z}/\{0\} \label{eq:realtowscalRind}
\eeq
Further, taking  $\omega z_0$ small, assuming a low frequency ansatz valid for $n\ll |\log\kappa|$
\begin{align}
    \omega_n z_0&= -\frac{\pi n}{\log\kappa}\bigg(a_0+ \frac{a_1}{\log\kappa}+ \frac{a_2}{\log^2\kappa}+\frac{a_3}{\log^3\kappa}+\mathcal{O}(\log^{-4}\kappa)\bigg)\;,
\end{align}
we find the following coefficients
\begin{align}
\begin{cases}
    a_0=1,\\a_1=\log 2-\gamma,\\a_2=(\log 2-\gamma)^2, \\ a_3=\bigl(\log 2-\gamma\bigr)^3
+ \frac{n^2 \pi^2}{3}\zeta(3) \,.
\end{cases}
\end{align}
The difference between neighboring modes is easily found to be
\beq \Delta\omega \approx\frac{\pi}{z_{0}|\log\kappa|}\;,\eeq
which goes to 0 in the stretched horizon limit. Just like in the de Sitter case, there exists also a high frequency regime $n\gg |\log\kappa|$, for which one can use the Stirling approximation of the Gamma functions in order to obtain an approximation for the allowed frequencies.

\subsection{Normal vector modes}
Finally, substituting in the Fourier decomposition of the vector master fields into boundary conditions (\ref{eq:vecbcrind}) shows that vector perturbations must satisfy
\beq 0=K_{-1+iz_{0}\omega}(\kappa)+K_{1+iz_{0}\omega}(\kappa)\;.\eeq
In the small $\kappa$ limit we obtain an equation for an infinite tower of real modes, 
\begin{align}
\omega_{n}z_{0}=\frac{i}{2\log(\kappa/2)}\left[\log\left(\frac{\Gamma(-i\omega_{n}z_{0})}{\Gamma(i\omega_{n}z_{0})}\right)+2\pi i n\right]\;,\;  n\in\mathbb{Z}/\{0\}.
\end{align}
This is reminiscent to the tower of real vector modes found for dS (\ref{vectorpoleeqq}) and analytic approximations for the solutions can be found in a similar way, for both low frequencies and high frequencies.

\bibliography{bdryrefs}

\end{document}